\documentclass[12pt]{article}

\usepackage{epstopdf,amsfonts,amsmath,amssymb}
\usepackage{graphicx}
\DeclareGraphicsExtensions{.eps}

\newcommand\encadremath[1]{\vbox{\hrule\hbox{\vrule\kern8pt
\vbox{\kern8pt \hbox{$\displaystyle #1$}\kern8pt}
\kern8pt\vrule}\hrule}}
\def\enca#1{\vbox{\hrule\hbox{
\vrule\kern8pt\vbox{\kern8pt \hbox{$\displaystyle #1$}
\kern8pt} \kern8pt\vrule}\hrule}}

\newcommand\framefig[1]{
\begin{figure}[bth]
\hrule\hbox{\vrule\kern8pt
\vbox{\kern8pt \vbox{
\begin{center}
{#1}
\end{center}
}\kern8pt}
\kern8pt\vrule}\hrule
\end{figure}
}

\newcommand\figureframex[3]{
\begin{figure}[bth]
\hrule\hbox{\vrule\kern8pt
\vbox{\kern8pt \vbox{
\begin{center}
{\mbox{\epsfxsize=#1.truecm\epsfbox{#2}}}
\end{center}
\caption{#3}
}\kern8pt}
\kern8pt\vrule}\hrule
\end{figure}
}
\newcommand\figureframey[3]{
\begin{figure}[bth]
\hrule\hbox{\vrule\kern8pt
\vbox{\kern8pt \vbox{
\begin{center}
{\mbox{\epsfysize=#1.truecm\epsfbox{#2}}}
\end{center}
\caption{#3}
}\kern8pt}
\kern8pt\vrule}\hrule
\end{figure}
}

\makeatletter
\@addtoreset{equation}{section}
\makeatother
\newtheorem{theorem}{Theorem}[section]

\newtheorem{remark}{Remark}[section]
\newtheorem{proposition}{Proposition}[section]
\newtheorem{lemma}{Lemma}[section]
\newtheorem{corollary}{Corollary}[section]
\newtheorem{definition}{Definition}[section]
\def\br{\begin{remark}\rm\small}
\def\er{\end{remark}}
\def\bt{\begin{theorem}}
\def\et{\end{theorem}}
\def\bd{\begin{definition}}
\def\ed{\end{definition}}
\def\bp{\begin{proposition}}
\def\ep{\end{proposition}}
\def\bl{\begin{lemma}}
\def\el{\end{lemma}}
\def\bc{\begin{corollary}}
\def\ec{\end{corollary}}
\def\beaq{\begin{eqnarray}}
\def\eeaq{\end{eqnarray}}
\newcommand{\proof}{{\noindent \bf proof:}$\quad$ }
\newcommand{\eproof}{ $\square$ }

\newcommand{\be}{\begin{equation}}
\newcommand{\ee}{\end{equation}}
\newcommand{\beq}{\begin{equation}}
\newcommand{\eeq}{\end{equation}}
\newcommand{\bea}{\begin{eqnarray}}
\newcommand{\eea}{\end{eqnarray}}

\newcommand{\Tr}{{\rm Tr}\,}

\newcommand{\ii}{{\rm i}\,}

\newcommand{\Tau}{{\cal T}}

\newcommand{\spcurve}{{\cal S}}
\newcommand{\curve}{{\Sigma}}
\newcommand{\curverond}{\overset{\circ}{\curve}}
\newcommand{\genus}{{\mathfrak g}}

\newcommand{\xirond}{\overset{\circ}{\xi}}
\newcommand{\acycle}{{\cal A}}
\newcommand{\bcycle}{{\cal B}}

\newcommand{\modsp}{{\mathcal M}}
\newcommand{\modspmero}[1]{\modsp_{{#1}+\text{mero}}}

\newcommand{\Lieg}{{\mathfrak g}}
\newcommand{\Lieh}{{\mathfrak h}}
\newcommand{\weil}{{\mathfrak w}}
\newcommand{\Ker}{{\rm Ker\ }}
\newcommand{\Img}{{\rm Im\ }}

\newcommand{\x}{{\rm x}}
\newcommand{\y}{{\rm y}}

\newcommand{\td}{\tilde}
\newcommand{\Res}{\mathop{\,\rm Res\,}}
\usepackage[pdftex]{hyperref}
\hypersetup{colorlinks,urlcolor=magenta,citecolor=red,linkcolor=blue,filecolor=black}

\begin{document}

\sloppy

\pagestyle{empty}
\hfill IPhT-T17/086, CRM-3360
\addtolength{\baselineskip}{0.20\baselineskip}
\begin{center}
\vspace{26pt}
{\large \bf {The Geometry of integrable systems from topological recursion.
\\ 
Tau functions and homology of Spectral curves. Perturbative definition.}}
\newline
\vspace{26pt}

{\sl B.\ Eynard}\hspace*{0.05cm}

\vspace{6pt}
Institut de Physique Th\'{e}orique de Saclay,\\
F-91191 Gif-sur-Yvette Cedex, France.\\
CRM, Centre de recherches math\'ematiques  de Montr\'eal,\\
Universit\'e de Montr\'eal, QC, Canada.
\end{center}
\begin{center}
\includegraphics[scale=0.3]{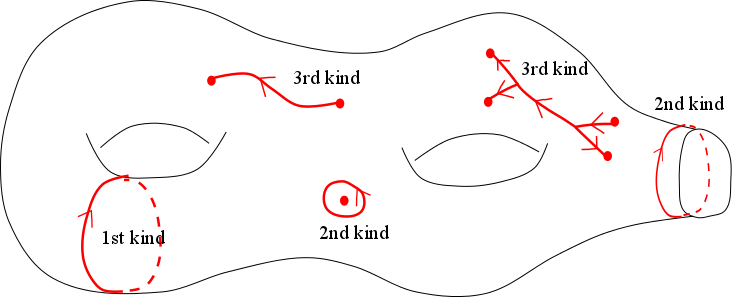}
\end{center}
\vspace{20pt}
\begin{center}
{\bf Abstract}
\end{center}

We describe a geometric method to construct, from an object called a "spectral curve",  
 an integrable system, and in particular its Tau function, Baker-Akhiezer functions and "current amplitudes", all having an interpretation as CFT conformal blocks.
 The construction identifies Hamiltonians with cycles on the curve, and times with periods (integrals of forms over cycles). 
 All the integrable structure is formulated in terms of homology of contours, the phase space is a space of cycles where the symplectic form is the intersection, the Hirota operator is a degree 2 second-kind cycle, a Sato shift is an addition of a 3rd kind cycle. 
In this setting the Hirota equations can be illustrated as merging 3rd kind cycles (monopoles) yielding a 2nd kind cycle (dipole).
This article is also a preparation of a series of 3:

1) here: classical case, perturbative: the spectral curve is a ramified cover of a base Riemann surface -- with some additional structure -- and the integrable system is defined as a formal power series of a small "dispersion" parameter $\epsilon$.

2) In \cite{BeEy19} we defined the  dispersive classical case, non perturbative: the spectral curve is defined not as a ramified cover (which would be a bundle with discrete fiber), but as a vector bundle -- whose dispersionless limit consists in chosing a finite set of vectors in each fiber. 

3) In preparation, and based on \cite{BEOPF11, CEM11}: non-commutative case, and perturbative. The spectral curve is here a "non-commutative" surface, whose geometry will be defined in lecture III.

4) the full non-commutative dispersionless theory is currently under development.


%

\vspace{0.5cm}

\vspace{26pt}
\pagestyle{plain}
\setcounter{page}{1}


\tableofcontents

\section{Introduction}

Integrable systems are a corner stone in classical mechanics and dynamical systems, for 2 reasons: one is that they are ubiquitous in physics, they are often the simplest toy models, like the hydrogen atom or the motion of a single planet, and also because  integrable systems are the dynamical systems that can be solved exactly, in some sense they are the contrary of chaotic systems. 
An important property, is that up to a "good" change of variable, they can be brought to linear motion at constant velocity, however, all the complexity is hidden in finding the good change of variables.

Many classical books and lectures exist on integrable systems and we refer to \cite{BBT}.
It has been understood that geometry plays an important role in integrable systems.
The Kyoto school constructions (Sato, Hirota, Miwa--Jimbo-Ueno-Takasaki \cite{Sato2, Hirota, JMI,JMII,JMIII}) associate a "\textbf{Tau--function}" to an integrable system. The Tau function plays the role of a partition function in statistical physics, it encodes most of the properties of the system and a large part of integrable system theory consists in computing the Tau function. Tau functions enjoy many beautiful mathematical properties, they obey Hirota equations, they have some modular and automorphicity properties.

The Lax--pair method \cite{LaxLev, BBT} allows to encode most of the integrable system property into an operator --the Lax operator-- and into its eigenvalues locus -- the "\textbf{spectral curve}".
In the simplest case of "isospectral" integrable systems, also called finite--gap solution, there is a method to recover the Lax pair and find the Tau--function and all properties of the integrable system, from the spectral curve's geometry, this is known as the geometric reconstruction method \cite{Dub81,RevRusInt,ItsMatv,Kri77,Kri92,Kri92b}. 

\medskip

Here we shall start from a spectral curve.
This lecture is largely based on the method presented in \cite{BE10}, with many updates and additions.

To a spectral curve $\spcurve$ (defined below), we shall associate a  "Tau-function" and a family of "$n$-points amplitudes", that we shall denote -- mimicking CFT notations:
\begin{eqnarray}
\Tau(\spcurve) = W_0(\spcurve) &=& \left< \mathcal V(\spcurve) \right> \cr
\hat W_n(\spcurve;X_1,\dots,X_n) &=& \left< \mathcal V(\spcurve)  J(X_1) \dots J(X_n)\right>.
\label{CFTnote1}
\end{eqnarray}

$\bullet$ {\bf CFT notations}:
The right hand side is a mere notation, borrowed from CFT (Conformal Field Theory), the bracket notation $\langle.\rangle$ is named "amplitude" or "conformal block amplitude". $\mathcal V(\spcurve)$ will be called the "generalized vertex operator" associated to the spectral curve $\spcurve$, and $J(X_i)$ will be called a "current operator" at point $X_i$ of the spectral curve.
A point $X_i$ of a spectral curve is actually a pair $X_i=(x_i,Y_k(x_i))$ of a point $x_i$ on the base curve, and the $k^{\rm th}$ value $Y_k(x_i)$ of a multivalued function $Y(x)$, meaning that the currents are multivalued functions of $x_i$, or monovalued functions of $X_i$. $J(X_i)$ can be seen as a vector with components $J_k(x_i)=J((x_i,Y_k(x_i)))$ for $k=1,\dots,\text{rk}$ with rk the rank of a vector space to which it belongs (often this will be a Cartan or Lie algebra).
We shall show that our definition of currents will agree with Sugawara currents \cite{sugawara} in CFT.

$\bullet$ {\bf Integrability}:
Usually \cite{BBT} Tau-functions were defined as functions of "times" $\Tau(t_1,t_2,t_3,\dots)$ required to obey some equations with respect to changing the times, either differential equations $\partial_{t_i} \log\Tau = \dots$, or finite shifts $\Tau(t_k+u_k) =\dots$\ . The operators $\partial_{t_i}$ are associated to Hamiltonians, required to commute.
Here $\Tau(\spcurve)$ will be defined for each spectral curve, point-wise in the moduli space of spectral curves, and deformations equations will arise as consequences, not as definitions.

Times will be viewed as local coordinates in the moduli space of spectral curves, and time deformations $\partial_{t_k}$ belong to the tangent space of the moduli space. We shall see that the tangent space is isomorphic to the space of meromorphic differential forms on the spectral curve, and through form-cycle duality it can be identified with the space of cycles (in fact generalized cycles) on the spectral curve: times can thus be seen as coordinates in the homology space (space of cycles). This will allow to re-define the Tau function as a function of cycles, and define $\partial_\gamma \log\Tau=\dots$ or $\Tau(\spcurve+\gamma)=\dots$ where $\gamma$ is a cycle.

This definition using cycles is 

- geometric,

- intrinsic: independent of a choice of coordinates (the times),

- since cycles are rigid (like integers), they don't deform, therefore all deformations are much easier when written in terms of cycles. Somehow all "complicated" expressions in integrable systems come from a choice of coordinates\footnote{It is well known that in "action--angle" coordinates, every integrable system is a linear motion at constant velocity. The complication is only in finding these coordinates.}.

- cycles are equipped with a symplectic structure: the intersection, and by pushforward, this gives a symplectic form on the tangent space, where it then coincides with the Goldman symplectic structure. 
In fact, there is another symplectic structure emerging from the complex structure of the spectral curve.
The interplay between the 2 symplectic structures provides locally an hyperK\"ahler structure on the moduli space of spectral curves.

- There is an integer symplectic lattice in the space of cycles, giving rise to modular properties.

\section{Spectral curves}

We shall define several notions of spectral curves, classical, quantum, non--commutative...
In this article we start with the simplest, based on Riemann surfaces.
We refer to classical textbooks on Riemann surfaces in particular  \cite{Fay,MumTata,farkas}.

\subsection{Classical spectral curves}

For an open Riemann surface $\curve$, we denote $\mathfrak M^1(\curve)$ the $\mathbb C$--vector space (infinite dimension) of meromorphic 1-forms on $\curve$.
It is usually decomposed into 3 parts: 1st kind forms have no poles, 3rd kind forms are meromorphic forms with only simple poles, and 2nd kind forms have poles of degree $\geq 2$.

\bd[Spectral curve]
A spectral curve data $(\curve,\x,\y,B)$ is the data of:
\begin{itemize}

\item a differentiable orientable open surface $\curve$, not necessarily connected\footnote{For example, $\curve$ can be a disjoint union of discs, this was called a local spectral curve in \cite{eynclasses2,DOSS}. },  with possible boundaries $\partial\curve$ = finite union of topological circles.


\item a $C^\infty$ map $\x:\curve\to\curverond$, from $\curve$ to a Riemann surface $\curverond$ (called the base), such that the boundaries of $\curve$ are mapped to boundaries of $\curverond$.
The pull back of the complex structure of $\curverond$ induces a complex structure on $\curve$, and thus in which $\x$ is analytic, and with which $\curve$ can be seen as a Riemann surface.

\item a locally holomorphic (with respect to the complex structure above) 1-form $\y$ on $\curve$ (locally meromorphic means  having at most a finite number of poles in any compact subset of $\curve$. It allows all kinds of essential singularities at the boundary of $\curve$.)
The map $z\mapsto (\x(z),\x_*\y(z))$ is a locally meromorphic immersion of $\curve$ into the total cotangent bundle of $\curverond$:
\bea
\curve &\hookrightarrow & T^*\curverond  \cr
& \x \searrow & \downarrow \cr
& & \curverond
\eea

\item a meromorphic 1-1 symmetric bilinear differential on $\curve\times \curve$, having a double pole on the $\x$-diagonal divisor $\text{Diag}_\x = \{ (p,q) | \ \x(p)=\x(q) \}\subset \curve\times\curve $, and without residue
\beq
B \in H_0(\curve\times\curve, K_\curve \boxtimes K_\curve(2\text{Diag}_\x)^{\rm Sym})
\eeq
Near $\text{Diag}_\x$, writting $G_p$ the group of permutations of  $\x^{-1}(\x(p))=\sigma(p) $, we require that  it locally behaves as
\beq\label{normalizedBkappa}
B(p,q) \mathop{\sim}_{q\to \sigma(p)} \left( \frac{\kappa_{\sigma}}{2\ (\x(p)-\x(q))^2}  + O(1) \right) \ d\x(p) d\x(q)
\eeq
where $\sigma\in G_p $.

Cases where $G_p$ is a Weil group and $\kappa$ the Cartan matrix are interesting.
However, from now on we shall choose -- unless stated otherwise --
\beq\label{normalizedB}
\kappa_\sigma = 2 \ \delta_{\sigma,{\rm Id}},
\eeq
so that $B$ has a double pole only on the diagonal of $\curve$, normalized to 1.


\end{itemize}

We define the category $\mathbb S$, whose objects are spectral curve datas, and whose morphisms are defined as follows:

we say that there is a morphism between two spectral curve datas $(\curve,\x,\y,B)$ and $(\tilde\curve,\tilde\x,\tilde\y,\tilde B)$, if they have the same base curve and there is a $C^\infty$ map such that $\phi^*\tilde\x = \x$, $\phi^*\tilde \y=\y$, $\phi^* \tilde B = B$.
Two spectral curve datas are isomorphic if there is a morphism from the 1st to the second, and a morphism from the second to the first, whose composition is the identity morphism.

In particular if two spectral curve datas are isomorphic, this implies that $\curve$ and $\tilde \curve$ are isomorphic as topological surfaces and as Riemann surfaces.

Spectral curves are diffeomorphism classes:
\beq
\spcurve=[(\curve,\x,\y,B)].
\eeq
We call $\modsp$ the moduli space of spectral curves, i.e. $\mathbb S/\text{isomorphisms}$.

 \ed

\bd[$\mathbb C^*$ rescaling]
We define a $\mathbb C^*$ rescaling of spectral curve datas, i.e. multiplication of a spectral curve data $\spcurve=(\curve,\x,\y,B)$ by a non-zero scalar $\lambda\in \mathbb C^*$ as:
\beq
\lambda (\curve,\x,\y,B) = (\curve,\x,\lambda \y,B).
\eeq
This rescaling obviously descends to diffeomorphisms equivalence classes, i.e. to spectral curves.

\ed
People often denote the scaling parameter as $1/\lambda=\hbar,\epsilon,t/N,g_s,\sqrt{-\epsilon_1 \epsilon_2}$... depending on the context.
The large $\lambda$ limit is sometimes called the semi-classical limit or the heavy limit.
Below, we shall write $1/\lambda=\epsilon$ and call $\epsilon$ the \textbf{dispersion parameter}\index{dispersion}.

\subsubsection{Some geometric properties of spectral curves}

\begin{itemize}

\item {\bf Degree:}  is the generic number of preimages of a point: 
$\deg \x=\#\x^{-1}(x)$ for a generic $x\in\curverond$.
We shall most often (unless stated) consider spectral curves with finite degree, and with degree $\geq 2$.

\item {\bf Branchpoints:}
The points $a$ of $\curve$ where $\x$ is not locally invertible are called ramification points, and their images $\x(a)$ on $\curverond$ are called branchpoints. If $a$ is a ramification point, then $b_a={\rm order}_a (\x-\x(a))-1$ is called the order of the branchpoint. For generic branchpoints, the order is 1.
Let $\mathcal R=\sum_a b_a[a] $ be the ramification points divisor.

\end{itemize}

\subsection{The moduli space of spectral curves}


%
Let $\modsp$ denote the moduli space of spectral curves.
For the moment we don't have a topology or any structure on it.

It is sometimes interesting to consider sub-spaces, the most familiar being:

\begin{itemize}

\item  The space $\modsp_{\text{KP}}$ of algebraic spectral curves, where $\curverond=\mathbb C P^1=\bar{\mathbb C}$ is the Riemann sphere, $\curve$ icompact without boundaries, and $\x$ and $\y$  meromorphic, so there is a polynomial relationship $P(\x, y)=0$ (where $y=\y/d\x$).
This is the space of multicomponent {\bf KP} (Kadamtsev-Petiashvili)\index{KP} systems -- the number of components being the number of poles of $\y$ and $\x$.

\item  The space $\modsp_{\text{KdV}}\subset\modsp_{\text{KP}}$ of hyperelliptical algebraic spectral curves with $\deg\x=2$,  with a quadratic polynomial relationship $y^2=P(\x)$.
This is the space of {\bf KdV} (Kortweg de Vries) systems\index{KdV}. 

\item The space $\modsp_{\text{Toda}}$ of Toda spectral curves  concerns the case where $\curverond=\bar{\mathbb C}$, $\curve$ is compact and $d\x$ and $d(\y/d\x)$ are meromorphic 1-forms, this in particular allows $\x$ and $\y$ to have logarithmic singularities. It contains KP and KdV.

\item The space of spectral curves with a symmetry group.
In particular the space $\modsp_{\text{Hitchin}(G)}$ of spectral curves coming from a {\bf Hitchin system} with Lie group $G$.
Given a Lie group $G$ and its Lie algebra $\Lieg$, and a principal $G$ bundle $\mathcal E\to \curverond$ over a compact base Riemann surface $\curverond$, and a Higgs field:  a $\Lieg$-valued 1-form $\Phi\in H^0(\curverond, {\rm End}\mathcal E \otimes K_{\curverond} )$. The spectral curve is the eigenvalue locus of  $\{(x,\y) | \det(\y \text{Id}-\rho(\Phi(x)))=0\} $ (with $\rho$ a faithful representation  $\Lieg \to \Lieg l_n(\mathbb C)$), which defines an immersion of $\curve$ into the total cotangent space of $\curverond$
\bea
\curve & \hookrightarrow & T^*\curverond \cr
& \searrow & \downarrow \cr
& & \curverond
\eea
The 1-form $\y$ on $\curve$ is the Liouville form: the restriction of the tautological form of $T^*\curverond$ to $\curve$.

\item The  space $\modsp_{\text{Fuchsian}}$ of {\bf Fuchsian} spectral curves, same as above, but we allow $\Phi$ to have $N$ simple poles, $z_1,\dots,z_N$.
The Liouville 1-form $\y$ has then $N\times \dim\rho$ simple poles on $\curve$, whose residues are the eigenvalues of $\rho(\alpha_i)$ with $\alpha_i \in \Lieh/\weil$ (Cartan algebra quotiented by Weil group) the radial part of $\Res_{z_i} \Phi$, $\alpha_i$ is called the "charge" at $z_i$.

\item We can then enlarge Hitchin systems to meromorphic $\Phi$ having higher order poles.
 In some sense, a higher order pole can be reached as a limit (often very singular) of coalescing simple poles.
 
 \item Since any Lie group $G$ can be a subgroup of some $GL_n(\mathbb C)$, the $G$ Hitchin systems subspace is a subspace of the $GL_n(\mathbb C)$ Hitchin systems subspace.

\end{itemize}

\framefig{
\includegraphics[scale=0.4]{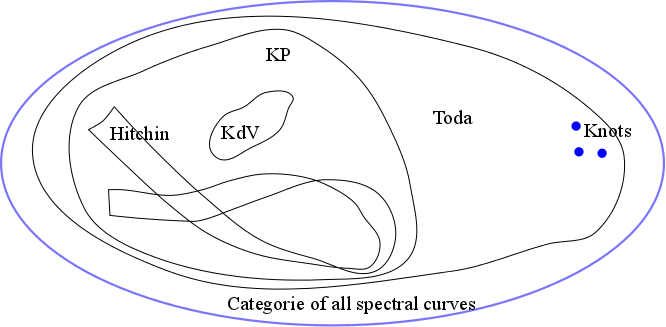}
\caption{The space of all spectral curves contains many classical subspaces.
Some subspaces can be manifolds or orbifolds with singularities.
The winding and self intersecting picture, illustrates the idea that certain subspaces, especially those with symmetry groups, can have a non--trivial topology, singular-points, self--intersections, be non--contractible, and other exotic features.
We also illustrate that there is a spectral curve associated to a knot (called its $A$-polynomial), but knot spectral curves are undeformable, they are somehow "isolated" points.
}
}

The idea is to define our Tau function and amplitudes {\bf objectwise}, independently of any subspace to which the considered spectral curve may belong, and independently of its possible deformations in a neighbourhood, and independently of any topology or structure of the space: just objectwise.

\subsection{Invariants of spectral curves}

Without explaining how to compute them, we recall that there is a family of "invariants" associated to any spectral curve:

\bd[EO invariants \cite{EO07}]

One associates to a spectral curve $\spcurve$, a double sequence $\omega_{g,n}(\spcurve)$ indexed by two non--negative integers $g,n$ such that $(g,n)\neq (0,0)$, of symmetric $n-$forms  on $\curve^n$.
For $n=0$, these are scalar and often denoted $\omega_{g,0}(\spcurve)=\mathcal F_g(\spcurve)$. We have a map defined objectwise:
\beq
\spcurve \mapsto
\{\omega_{g,n}(\spcurve) \}_{g,n}.
\eeq
For $(g,n)=(0,1)$ and $(0,2)$ these are, by definition:
\beq
\omega_{0,1}(\spcurve)=\y
\qquad , \qquad
\omega_{0,2}(\spcurve)=B.
\eeq
The invariants $\omega_{g,n}(\spcurve)$ with $2g-2+n>0$ are called "stable" and the ones with $2g-2+n<0$ are called "unstable". The only unstable ones are $(0,1),(0,2),(1,0)$.

Let us also mention the $(0,3)$ invariant  in the case $\curverond=\mathbb CP^1$:
\beq
\omega_{0,3}(\spcurve;z_1,z_2,z_3) = \sum_{a\in \mathcal R} \Res_{z\to a} \frac{B(z,z_1)B(z,z_2)B(z,z_3)}{d\x(z) \ d\left(\frac{\y(z)}{d\x(z)}\right) }
\eeq

\ed

These invariants were defined in \cite{EO07} initially only for algebraic spectral curves, and only those having only simple ramification points.
But in fact the definition extends to the whole space of spectral curves, the case of higher order ramification points being defined in \cite{BouEyn13}, and the algebraicity being needed nowhere in the definitions.

The case of $\mathcal F_0=\omega_{0,0} $ will be discussed below in section \ref{sec:F0}.
We just mention that all stable invariants $\omega_{g,n}(\spcurve)$ are defined by a recursion on $2g-2+n$, and involve residues at ramification points, and we refer to \cite{EO07} for details, and to \cite{KS17,ABCD17} for a recent algebraic reformulation.

\bt[Homogeneity \cite{EO07}]
\label{prop:homogeneity}
 If $(g,n)\neq (1,0)$, the invariant $\omega_{g,n}$ is homogeneous of degree $2-2g-n$:
\beq
\omega_{g,n}(\lambda\spcurve) = \lambda^{2-2g-n} \omega_{g,n}(\spcurve).
\eeq
And
\beq
\omega_{1,0}(\lambda\spcurve) = \omega_{1,0}(\spcurve) + \frac{\deg \mathcal R}{24}\ \ln\lambda
\eeq
where $\deg\mathcal R$ is the number (with multiplicity) of ramification points.

\et

\bt[Analytic properties \cite{EO07}]
\label{prop:polesWgn}
If $n\geq 1$ and $(g,n)\neq (0,1),(0,2)$, then $\omega_{g,n}(\spcurve)$:
\begin{itemize}
\item is a symmetric tensor of $1$-form on $\curve^n$,
\item has poles only at ramification points, and without residue,
\end{itemize}
\beq
\omega_{g,n}(\spcurve) \in H^0(\curve^n ,{\rm{Sym}}(K_{\curve}(*\mathcal  R)^{\boxtimes n}) )
\eeq
where $\mathcal R$ is the divisor of ramification points, $K_{\curve}$ is the cannonical bundle, $\boxtimes$ means a tensor product of 1-forms in each variable, and $*$ means poles of any degree.

\et

More properties of the EO invariants --all proved in \cite{EO07}-- will appear along this text, and shall be introduced when needed.

\subsection{Form--cycle dualities}

Usually Tau-functions are defined as functions of "times" \cite{BBT}, and we shall argue that times are local coordinates on $\mathcal M$,  said otherwise, they are coordinates in the tangent space of $\mathcal M$, and we shall argue that they are thus coordinates
in the space of meromorphic differential forms $\mathfrak M^1(\curve)$, and using form-cycle dualities, they are coordinates in the space of cycles, so that eventually we shall consider the Tau function as a function on the space of cycles.
Let us first study the space of cycles.

\subsubsection{Space of cycles}
\label{sec:cycles}

In all this section, $\spcurve$ is kept fixed, and in particular the base curve $\curverond$ is kept fixed, and we choose once for all an atlas of charts of $\curverond$, each chart being an open subset of $\mathbb C$, so that locally in each chart there is a well defined local coordinate, in other words we do as if $\x(z)\in \mathbb C$ locally. Also, for each boundary $\overset{\circ}{b}$ of $\curverond$, oriented so that the surface lies at its right, we define once for all a map $\xirond_{\overset{\circ}{b}}:\overset{\circ}{b} \to S^1$ that extends analytically to a neighborhood of $\overset{\circ}{b}$.

\framefig{
\includegraphics[scale=0.4]{cycles.png}
\caption{Cycles of 1st kind are usual non-contractible cycles.
Cycles of 2nd kind consist of a small circle around a point, weighted by a local meromorphic function, or cycles around a boundary weighted by a local holomorphic function.
Cycles of 3rd kind are chains whose boundaries are degree zero divisors. The addition of 3rd kind cycles is not commutative, an order must be choosen, encoded as an oriented graph.}
}

$\bullet$ {\bf 1st kind cycles:}

Let $H_1(\curve,\mathbb Z)$ (resp. $H_1(\curve,\mathbb C)$) the integer (resp. complex) homology space of $\curve$, i.e. the space of integer (resp. complex) linear combinations of homotopy classes of closed Jordan arcs on $\curve$.

There is the usual Poincarr\'e pairing between homology cycle $\gamma$ and 1-form $\omega$:
\beq
<\gamma,\omega> = \oint_\gamma \omega,
\eeq
which shows that cycles can be viewed as elements of the dual of holomorphic 1-forms: 
\beq
H_1(\curve,\mathbb C) \subset \mathfrak M^1(\curve)^\vee.
\eeq
These constitute "1st kind cycles", and we keep in mind that the space of 1st kind cycles contains an integer lattice $ H_1(\curve,\mathbb Z)\subset H_1(\curve,\mathbb C)$.

If $\curve$ is compact and orientable, of some genus $\genus$, then $\dim H_1(\curve,\mathbb C)=2\genus$, otherwise it is often infinite dimensional.

\br
Notice that $H_1(\curve,\mathbb Z)$ and $H_1(\curve,\mathbb C)$ don't depend on $\curverond$, $\x$, $\y$ or $B$, they depend only on the topology of $\curve$.
\er

We shall now enlarge the homology space of cycles, by considering a larger subset of $\mathfrak M^1(\curve)^\vee$.
In section \ref{sec:formcycleB} below we shall provide an intrinsic definition of our space of generalized cycles, as the subset of the dual $\mathfrak M^1(\curve)^\vee$ whose pairing with $B$ is a meromorphic 1-form.
But for the moment let us construct it explicitly with a basis as follows

$\bullet$ {\bf 2nd kind cycles:}
First we shall consider "meromorphic currents" around marked points or boundaries, denoted:
$\mathcal C_p. f$ where $\mathcal C_p$ is a "small" counterclockwise circle\footnote{Here "small" circle means the inductive limit of a family of circles around $p$. In other words a circle closer to $p$ than any other special point that is considered.} around a point $p$, or around a boundary $p$ of $\curve$, and $f$ is a function holomorphic in a neighborhood of $\mathcal C_p$.
If $p$ is a point, then $f$ is required to be meromorphic in a neighborhood of $p$ with a possible pole at $p$ (of any degree).
$\gamma = [\mathcal C_p. f]$ is the equivalence class modulo small homotopic deformations of $\mathcal C_p$ together with analytic continuation of $f$, and we define homology classes as integer (resp. complex) linear combinations of these.

If $\omega$ is a meromorphic 1-form, and $\gamma=\mathcal C_p.f$, the following pairing is well defined, and denoted 
\beq
<\gamma,\omega> = \oint_\gamma \omega = \oint_{\mathcal C_p.f} \omega = \oint_{\mathcal C_p} f \omega \qquad (=  2\pi\ii \Res_p f\omega\quad \text{if}\, p \,\, \text{is a point}).
\eeq
These meromorphic currents are called "2nd kind"  cycles, they generate the space of second kind cycles, which is always infinite dimensional, because the degree of poles can be as high as desired.

There is also a lattice in it, indeed, if $p$ is a point, a basis of the functions $f$ meromorphic in a neighbourhood of $p$ is
\beq
\{\xi_p^k \}_{k\in \mathbb Z }
\qquad , \qquad
\xi_p=(\x-\x(p))^{1/{\rm order}_\x(p)}.
\eeq
The set of cycles generated by
\bea\label{def:M12ndpm}
\acycle_{p,k} = \mathcal C_p .\xi_p^k \ , \ p\in \curve, \ k\geq 0
\cr
\bcycle_{p,k} =  \frac{1}{2\pi\ii} \mathcal C_p . \frac{\xi_p^{-k}}{-k} \ , \ p\in \curve, \ k\geq 1
\eea
defines a space of second kind cycles, that contains an integer lattice
\beq
\sum_{p\in \curve}\sum_{k\geq 0} \acycle_{p,k}\mathbb Z
\oplus \sum_{p\in \curve}\sum_{k\geq 1}   \bcycle_{p,k} \mathbb Z.
\eeq

$\bullet$ \textbf{boundary 2nd kind cycles:}

A boundary $b$ of $\curve$ -- oriented such that the surface lies on its right -- is mapped by $\x$ to a boundary $\overset\circ{b}$ of $\curverond$, with winding $d_b\in \mathbb Z_+$, and there is a map $\xi_{\overset\circ{b}}: \overset\circ{b} \to S^1 $  such that a neighborhood of $\overset\circ{b}$ is mapped to the exterior of the unit disc.
We define the map $\xi_b: b \to S^1$
\beq
\xi_b(z)  = \left(\xi_{\overset\circ{b}}(\x(z))\right)^{1/d_b}.
\eeq
The following set of cycles generate an integer lattice in the space of second kind cycles:
\bea\label{def:M12ndpmbnd}
\acycle_{b,k} = \mathcal C_b . \xi_b^k \ , \ k\geq 0
\cr
\bcycle_{b,k} =  \frac{1}{2\pi\ii} \mathcal C_b \frac{\xi_b^{-k}}{-k} \ ,  \ k\geq 1 .
\eea

\br
Notice that all integer 2nd kind cycles depend on  a choice of coordinate on $\curverond$ and on its boundaries. However, the space of complex 2nd kind cycles is independent, indeed a change of local coordinate amounts to linearly combining elements of the basis.
Later, we shall consider only deformations at fixed $\curverond$ so that the integer basis will be kept fixed.
\er

$\bullet$ {\bf 3rd kind cycles:}
Then we shall also consider open chains $\gamma=\gamma_{q\to p}$ with boundary $\partial\gamma = [p]-[q]$ a divisor of degree $0$.
We can define the pairing, for meromorphic 1-forms $\omega$ that have no pole at $\partial\gamma$:
\beq
\oint_\gamma \omega = <\gamma,\omega> = \int_{\gamma_{q\to p}}  \omega.
\eeq
For 1-forms $\omega$ that have poles at $p$ or $q$, there is a way to "regularize" the integral of $\omega$ along $\gamma_{q\to p}$ by subtracting the poles. In order to lighten the presentation the precise definition of $\oint_\gamma \omega $ is provided in appendix \ref{app:defint3rdkind}.
\br
The space of 3rd kind cycles is independent of a choice of charts and coordinates in $\curverond$.
\er

$\bullet$ \textbf{boundary 3rd kind cycles:}

We also define boundary chains as follows, for a boundary $b$, we have a map $\xi_b: b\to S^1$, and we choose a point $p_0\in b$ such that $\xi_p(p_0)=1$, and define:
\beq
\bcycle_{b,p_0,0} = \frac{1}{2\pi\ii}  \gamma_{p_0\to p_0} .  \ln {\xi_b}
\eeq
with the log cut on $\mathbb R_+$.
Notice that if $b$ winds $d_p$ times around $\x(b)$, then there exist $d_p$ possible choices for $p_0$, and if $\tilde p_0$ is another choice we have
\beq
\bcycle_{b,\tilde p_0,0} = \bcycle_{b,p_0,0} + \gamma_{p_0\to \tilde p_0},
\eeq
where $\gamma_{p_0\to \tilde p_0}$ is the oriented  boundary arc between the 2 points.

Again, one can consider linear combinations of chains, with either integer, or complex coefficients, so that there is a lattice also in the space of 3rd kind cycles.

\smallskip
\br\label{rem:chains}
When we consider 2 chains $\gamma_{q\to p}$ and $\gamma'_{q'\to p'}$, the chain $\gamma'_{q'\to p'}$ is defined in the relative homology of $\curve-\gamma_{q\to p}$, i.e. we can consider chains $\gamma'_{q'\to p'}$ only relatively to the homology of $\gamma_{q\to p}$.
In other words, the addition of 3rd kind cycles is {\bf not commutative}, we should always tell in what order they are added.

Most of the time this non-commutativity will turn out to be  irrelevant, except in multiple integrals that involve integration of forms having poles at coinciding points, in particular for double integrals of $B$ as we shall see later.
\er

\bd[Space of (generalized) cyles]
We define $\mathfrak M_1(\curve)$ as the set of linear combinations of 1st, 2nd, 3rd kind cycles.
It is a $\mathbb C$--vector space of infinite dimension.

At fixed $\curverond$, it contains an integer lattice $\mathfrak M_1(\curve,\mathbb Z)$ as the set of integer linear combinations of integer cycles.

\ed

\br\label{rem:otherdefofM1}
In section \ref{sec:formcycleB} below we shall see an alternative --more intrinsic-- definition for the space of generalized cycles as $\mathfrak M_1(\curve)=\hat B^{-1}(\mathfrak M^1(\curve))$, i.e. the space of elements of the dual $\mathfrak M^1(\curve)^\vee$ whose pairing with $B$ is a meromorphic 1-form.
\er

\subsubsection{Intersection and symplectic structure}

On 1st kind cycles is defined the intersection $\gamma \cap \gamma'=-\gamma'\cap \gamma$, as the algebraic counting of oriented crossings of transverse Jordan arcs representative of the cycles.

We shall extend it to 2nd kind and 3rd kind cycles in the following way

\bd[Intersection]

We define the intersection form as an antisymmetric bilinear form on $\mathfrak M_1(\curve)\times \mathfrak M_1(\curve)\to \mathbb C$ (resp. $\mathfrak M_1(\curve,\mathbb Z)\times \mathfrak M_1(\curve,\mathbb Z)\to\mathbb Z$), as follows.

We define the intersection of 1st kind cycles as the usual crossing number, and also the intersection of 1st kind and 3rd kind as the usual crossing number of a cycle with a chain in the relative homology.
For the intersection of two 3rd kind paths with distinct boundaries we define the intersection as the crossing number, and if the boundaries are not distinct the crossing number multiplied by $\frac12$, for example if both paths end  at $p$
\beq
\gamma_{q\to p} \cap \gamma'_{q'\to p} = \pm\ \frac12  
\eeq
where $\pm 1$ is, like in usual crossing numbers, the respective orientation of the second Jordan arc with respect to the first one.

Then we define the other intersections by the following table (completed by antisymmetry):

\begin{tabular}{|l||l|l|l|}
\hline
kind & 1st & 3rd &  2nd \cr
& $\gamma'$ & $\gamma'_{q'\to p'}$ &  $\mathcal C'_{p'}.f'$\cr
\hline
\hline
$\gamma$ & $\gamma\cap \gamma'$ &  $\gamma \cap \gamma'_{q'\to p'}$ &  $\gamma \cap \mathcal C_{p'}.f'=0$ \cr
\hline
$\gamma_{q\to p}$ &  &  $ \begin{array}{l} \gamma_{q\to p} \cap \gamma'_{q'\to p'}\cr \pm \frac{\delta_{p,p'}}{2} \pm \frac{\delta_{q,q'}}{2} \cr
\pm \frac{\delta_{p,q'}}{2} \pm \frac{\delta_{q,p'}}{2}  \end{array}$  &  $\gamma_{q\to p} \cap \mathcal C_{p'}.f'=\left\{\begin{array}{l}
 0\,\,\,\,{\rm if}\,\,f'(p')=\infty \cr (\delta_{q,p'}  -\delta_{p,p'})f'(p')  
 \end{array}\right.$ \cr
\hline
$\mathcal C_p.f$ &  &  & $\delta_{p,p'} \oint_{\mathcal C_p} f \ df'$ \cr
\hline 
\end{tabular}

\ed
Notice that it indeed takes integer values on  $\mathfrak M_1(\curve,\mathbb Z)\times \mathfrak M_1(\curve,\mathbb Z)$.
In particular, with the basis \eqref{def:M12ndpm}
\beq
\acycle_{p,k} \cap \bcycle_{p',k'} = \delta_{p,p'} \delta_{k,k'}.
\eeq
Except 3rd kind cycles that can have half--integer intersections.
This is illustrated in fig. \ref{fig:intersections}
\framefig{
\includegraphics[scale=0.5]{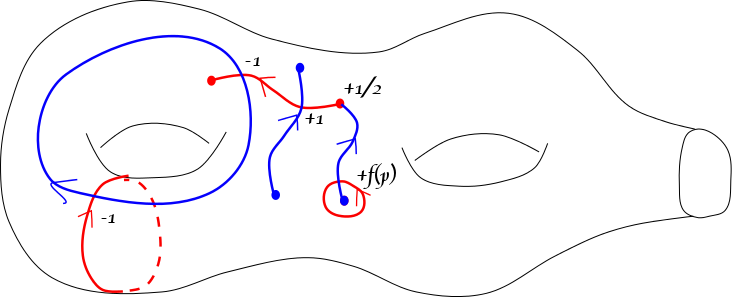}
\caption{Intersections of cycles blue $\cap$ red.\label{fig:intersections}}
}

There is an alternative way to define or compute intersections:

\bp\label{prop:Binter}
We have
\beq\label{eq:Binter}
\oint_{\gamma} \oint_{\gamma'} B - \oint_{\gamma'} \oint_{\gamma} B = 2\pi\ii\ \gamma\cap \gamma'.
\eeq
As a consequence:
\beq
\mathcal Q(\gamma,\gamma') = \frac{1}{2\pi\ii} \oint_{\gamma} \oint_{\gamma'} B - \frac12 \  \gamma\cap \gamma'
\eeq
is a symmetric bilinear form on $\mathfrak M_1(\curve)\times \mathfrak M_1(\curve)$.
\ep

\begin{proof}
It is easy to verify that it holds for every pair of cycles in the basis used to define $\mathfrak M_1(\curve)$.

Let us recall the proof for elements of $H_1(\curve,\mathbb Z)$.
Choose some transversally crossing Jordan arcs representatives. Away from intersection points the order of integrations can be exchanged.
In a neighborhood of intersection points, we have the behaviour \eqref{normalizedBkappa}, $B(z,z')\sim \frac{\kappa}{2} dz dz'/(z-z')^2 + \text{analytic} = \frac{\kappa}{2} d_z d_{z'} \log(z-z') + \text{analytic}$ in any local coordinates. The difference of orders of integration yields the discontinuity of the log, namely $\pm \kappa\pi\ii$ depending on the orientation.
If we choose the behaviour \eqref{normalizedB}, we get $2\pi\ii \gamma\cap \gamma'$. 
If instead of \eqref{normalizedB}, we would choose $\kappa\neq 2\ \text{Id}$, this would define a generalized intersection theory for Cameral covers, that is considered in details in \cite{}.
\end{proof}

\subsubsection{Forms $\longleftrightarrow$ cycles }\label{sec:formcycleB}

The bilinear differential $B$ allows to define a "form--cycle duality":

\bd[Map $\hat B$: cycles $\longrightarrow$ 1-forms]
The integral --in the 1st projection-- of the 2nd kind differential $B$ along a  cycle is a meromorphic 1-form of the 2nd projection, of the same kind as the cycle: 1st kind $\to$ no poles, 3rd kind $\to$ simple poles, 2nd kind $\to$ poles of degree $\geq 2$.
This defines a linear map from the space of cycles to meromorphic 1-forms on $\curve$
\bea
\hat B : \mathfrak M_1(\curve) & {\to} & \mathfrak M^1(\curve)  \cr
\gamma & \mapsto & \oint_\gamma B.
\eea
\ed

We shall prove below that the map $\hat B$ is surjective, which shows that $\mathfrak M_1(\curve) = \hat B^{-1}(\mathfrak M^1(\curve))$ as mentioned in remark \ref{rem:otherdefofM1}. It is not injective, we shall see that it has a huge kernel.

Now we shall define a map $\hat C$: 1-forms $\to$ cycles, playing the role of a right inverse of $\hat B$.
If we would have a finite basis of cycles $\Gamma_i$, with intersection matrix $I_{i,j}=\Gamma_i\cap \Gamma_j$ we would define the map $\hat C$ as
\beq\label{eq:hatCsumij}
\hat C(\omega) =  \frac{1}{2\pi\ii}\sum_{i,j} \Gamma_i\ (I^{-1})_{i,j} \ \left(\oint_{\Gamma_j} \omega\right) \ 
\eeq
which is invariant under changes  of basis.
Unfortunately, expression \ref{eq:hatCsumij} is meaningless because the space $\mathfrak M_1(\curve)$ is infinite dimensional. 
But given a meromorphic $\omega$, it is possible (this is what the definition of $\hat C$ below does) to find a basis well adapted to $\omega$, such that only finitely many terms are not vanishing in the sum, or sums are absolutely convergent (in particular Fourier series at the boundaries), so in practice formula \ref{eq:hatCsumij} can be applied.
In a symplectic basis $I_{i,j}=\acycle_i\cap \bcycle_j=\delta_{i,j}$ we would have
\beq
\hat C(\omega) = \sum_i t_i(\omega) \bcycle_i
\qquad \text{with} \qquad 
t_i(\omega) = \frac{1}{2\pi\ii}\oint_{\acycle_i} \omega,
\eeq
it is such that 
\beq
\omega = \hat B(\hat C(\omega)) = \sum_i t_i(\omega) \hat B(\bcycle_i).
\eeq
Now we give the actual intrinsic definition of the map $\hat C$:

\bd[Map $\hat C$: 1-forms $\longrightarrow$ cycles]
If $\curve=\cup_i \curve_i$ is a union of connected components,
 choose a generic point $o_i\in\curve_i$, and a  fundamental domain of $\curve_i-\{o_i\}$ i.e. $\curve_i-\Gamma_i$ a 1-face cellular graph $\Gamma_i$ (for which $o_i$ is a 1-valent vertex).
The graph $\Gamma=\cup_i \Gamma_i$ has 2 kinds of edges: internal edges, and boundary edges that lie on $\partial\curve$.
On each internal edge $e$, choose a point $p_e$, corresponding to 2 points $p_{e\pm}$ in the fundamental domain, with $p_{e+}$ on the left of $e$ and $p_{e-}$ on the right. Let $e^\perp=\gamma_{p_{e-}\to p_{e+}}$ the unique (up to homotopy) arc from $p_{e-}$ to $p_{e+}$ in the fundamental domain.
We also identify $e^\perp$ with the homology class of a 1st kind cycle on $\curve$.

Then for a given choice of fundamental domain, define, for every meromorphic 1-form $\omega\in \mathfrak M^1(\curve)$:
\begin{itemize}
\item For every pole $p$ of $\omega$, of degree $d_p\geq 1$, and every $j \geq 0$: 
\beq\label{def:KPtimestpj}
t_{p,j} = \Res_p  (\xi_p)^j \ \omega =  \frac{1}{2\pi\ii}\oint_{\acycle_{p,j}} \omega
\eeq
the $t_{p,j}$s are called the "KP times" (see def.\ref{deftimes}).

We assumed the number of poles $p$ finite. Also all times with $j\geq \deg_p \omega$ are vanishing, only a finite number of times are non-zero.

\item define
\beq
\tilde\omega = \omega - \sum_i \sum_{p=\text{poles of} \ \omega\text{ in }\curve_i} 
\left(  t_{p,0} \hat B(\gamma_{o_i\to p}) 
 +   \sum_{j=1}^{\deg_p \omega-1} t_{p,j} \hat B(\bcycle_{p,j})  \right)
\eeq
$\tilde \omega$ is holomorphic on $\curve$, it has no poles.

\item In the fundamental domain of $\curve_i-\{o_i\}$, choose some generic points $o'_i$, and  define the holomorphic function
\beq
\tilde f(z) = \int_{o'_i}^z \tilde \omega .
\eeq

\item Then define
\begin{equation}
\begin{array}{llll}
\hat C(\omega)
& \overset{\text{def}}{=} & \sum_i \sum_{p=\text{poles of} \ \omega\text{ in }\curve_i} t_{p,0} \gamma_{o_i\to p} & \quad \to \quad\text{3rd kind}\cr
&& - \sum_{p=\text{poles of} \ \omega}  \sum_{j=1}^{\deg_p \omega-1} \frac{t_{p,j}}{2\pi\ii} \ \mathcal C_p. \frac{\xi_p^{-j}}{j} & \quad \to \quad\text{2nd kind}\cr
&&+ \frac{1}{2\pi\ii}\sum_{e=\text{internal edges}} (\oint_{e^\perp} \tilde{\omega}) e & \quad \to \quad\text{3rd + 1st kind}\cr
&& + \frac{1}{2\pi\ii}\sum_{e= \text{boundary edges}} e.\tilde f &\quad \to \quad \text{2nd kind}.
\end{array}
\end{equation}
Notice that the 3rd line may yield 1st kind cycles because a sum of open chains can be a closed cycle.

In the neighborhood of a boundary $b$ with its map $\xi_b: b\to S^1$, 
choose $p_0$ a boundary edge endpoint on $b$, then $\tilde f(z) - \frac{1}{2\pi\ii}\left(\oint_b \omega\right) \log (\xi_b(z)/\xi_b(p_0))$ -- with the log's cut on $\mathbb R_+$ --  is a periodic function on $b$, that can be decomposed into its Fourier modes:
\beq
\tilde f(z) = \frac{1}{2\pi\ii}\left(\oint_b \omega\right) \log (\xi_b(z)/\xi_b(p_0))
+ \sum_{k\in\mathbb Z} t_{b,k} \xi_b(z)^k.
\eeq
This shows that $\hat C(\omega)$ is a linear combination of all types of cycles introduced in section \ref{sec:cycles}, therefore
\beq
\hat C(\omega) \in \mathfrak M_1(\curve).
\eeq

\end{itemize}

\ed

\bl
\label{Lem:defChat}
The map:
\bea
\hat C: \mathfrak M^1(\curve) &\to & \mathfrak M_1(\curve) \cr
\omega & \mapsto & 
\hat C(\omega)
\eea
is linear and 
is independent of a choice of fundamental domain and of the choice of $o'_i$.

Moreover it satisfies for every $\omega$:
\beq\label{eq:BCId}
\omega = \hat B(\hat C(\omega)).
\eeq
\el

\proof
In appendix \ref{app:defChat}. The fact that $\omega=\hat B(\hat C(\omega))$ is in fact the Riemann bilinear identity.
\eproof

\bc
\beq
\Pi = \hat C \circ \hat B 
\qquad \text{is a projector}
\qquad
\Pi^2=\Pi.
\eeq

$\hat C$ is injective.
$\hat B$ is surjective.
We have the exact sequence
\bea
0 \to \Ker \hat B \to \mathfrak M_1(\curve) & \overset{\hat B}{\to} & \mathfrak M^1(\curve) \to 0 
\eea
We have  an isomorphism
 $\mathfrak M^1(\curve) \sim \mathfrak M_1(\curve)/\Ker\hat B$.

The map $\Pi=\hat C\circ\hat B: \mathfrak M_1(\curve)\to \mathfrak M_1(\curve)$ is a projector on $\Img \Pi \sim \mathfrak M_1(\curve)/\Ker\hat B$, parallel to $\Ker\hat B$. The map $\text{Id}-\Pi$ is the projector on $\Ker \hat B$ parallel to $\Img \Pi$.

We have
\beq
\mathfrak M_1(\curve) = \Img \Pi \oplus \Ker \hat B.
\eeq
\ec

\bp
\label{cor:LagrangeBPi}
Both 
$\Ker\hat B$ and 
$\operatorname{Im} \Pi$ are Lagrangian submanifolds of $\mathfrak M_1(\curve)$.

$\mathfrak M_1(\curve) = \operatorname{Im} \Pi \oplus \Ker \hat B$ is a Lagrangian decomposition.

\ep

\proof
In appendix \ref{app:LagrangeBPi}.
Notice that $\Ker\hat B$ Lagrangian is a trivial consequence of prop.\ref{prop:Binter}.
\eproof

Notice that all 2nd kind cycles $\acycle_{p,k}$ with $k\geq 0$ are in $\Ker\hat B$, whereas the 2nd kind cycles $\bcycle_{p,k}$  involving negative powers of $\xi_p^{-k}$  are  not in $\Ker\hat B$.

\bp[proved in \cite{EO07}]
If $\gamma\in \Ker\hat B$, we have for all $n\geq 1$ and $(g,n)\neq (0,1)$:
\beq
\int_{z_1\in\gamma} \omega_{g,n}(z_1,z_2,\dots,z_n) = 0.
\eeq
\ep
This holds by definition for $\omega_{0,2}=B$. 

The map $\gamma \mapsto \int_\gamma \omega_{0,1}$ has a kernel, which usually differs from $\Ker\hat B$.

\subsubsection{Positivity}

We define a complex structure on the space of cycles by complexifying integer cycles:
\beq
\mathfrak M_1(\curve,\mathbb C)=\mathfrak M_1(\curve,\mathbb Z)\otimes \mathbb C,
\eeq
we define the complex conjugate by acting only on the $\mathbb C$ factor.
In other words, if $\gamma=\sum_i t_i \gamma_i $, with $\gamma_i$ independent integer cycles and $t_i\in\mathbb C$, we define its complex conjugate as
\beq
\bar\gamma = \sum_i \bar{t_i} \gamma_i.
\eeq
This is independent of a choice of integer decomposition.

Then:
\bl\label{Lem:Qpositive}

The quadratic form $-2\pi\ii \mathcal Q$ defined in prop.\ref{prop:Binter} is a positive definite Hermitian form on $\Img\Pi$, for $\Pi(\gamma)\neq 0$:
\beq
-2\pi\ii\ \mathcal Q(\Pi(\gamma) ,\overline{\Pi (\gamma)}) >0 .
\eeq
\el

\proof
In appendix \ref{app:Qpositive}. This is a generalization of the Riemann bilinear inequality proving that $\Im\tau>0$, it is proved likewise using Riemann bilinear identity and Stokes theorem.
\eproof

\subsection{Lagrangian submanifolds and Darboux coordinates}

\framefig{
$$
\includegraphics[scale=0.65]{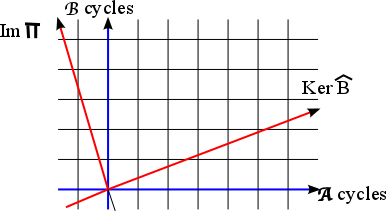}
\qquad \qquad
\includegraphics[scale=0.45]{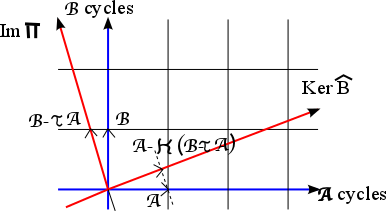}
$$
\caption{Consider $\acycle$-cycles and $\bcycle$-cycles forming a symplectic basis of the lattice of integer cycles. $\Ker \hat B$ and $\Img \Pi$ are orthogonal Lagrangian submanifolds, and $\Img \Pi$ is never parallel to the lattice.
A convenient basis for $\Img\Pi$ is to project $\bcycle$-cycles parallel to $\acycle$, and a basis for $\Ker\hat B$ is to project $\acycle$-cycles parallel to $\Img \Pi$.
}
}

\bd[Darboux basis]

A Darboux decomposition is  $\mathfrak M_1(\curve)= \acycle(\curve) \oplus \bcycle(\curve)$, with $\acycle(\curve)$ and $\bcycle(\curve)$ both Lagrangian.
An integer Darboux decomposition is
$\mathfrak M_1(\curve,\mathbb Z)= \acycle(\curve,\mathbb Z) \oplus \bcycle(\curve,\mathbb Z)$
with $\acycle(\curve,\mathbb Z)$ and $\bcycle(\curve,\mathbb Z)$ both integer Lagrangian sublattices.
Then  it is possible (not unique) to choose a Darboux Basis of cycles such that
\beq
\acycle_i \in \acycle(\curve)
\quad , \quad
\bcycle_i \in \bcycle(\curve)
\eeq
\beq
\acycle_i \cap \acycle_j = 0
\quad , \quad
\bcycle_i \cap \bcycle_j = 0
\quad , \quad
\acycle_i \cap \bcycle_j = \delta_{i,j}
\eeq
i.e. the intersection matrix takes the block--form 
$$
\begin{pmatrix}
0 & {\rm Id} \cr -{\rm Id} & 0
\end{pmatrix} .
$$
\ed

We have several usual decompositions:

\begin{enumerate}

\item The  decomposition
$\mathfrak M_1(\curve) = \Img \Pi \oplus \Ker\hat B$ is canonical, it is Darboux but not integer.
There is no canonical Darboux basis in it. 
Moreover, $\Ker \hat B$ and $\Img \Pi$ both get deformed under deformations of the spectral curve.
This Darboux decomposition, although canonical, is not very convenient for defining a connection on the bundle of cycles $\mathfrak M_1\to\mathcal M$.


\item Due to lemma \ref{Lem:Qpositive}, $\Im \Img \Pi$ and $\Re \Img \Pi$ are transverse and provide another canonical decomposition:
$$\mathfrak M_1(\curve) = \Im \Img \Pi \oplus \Re \Img \Pi .$$
This decomposition is real but is not integer, and there is no canonical Darboux basis in it, and it also gets deformed under deformations of the spectral curve.

\item There exists integer Darboux basis
$\mathfrak M_1(\curve,\mathbb Z)= \acycle(\curve,\mathbb Z) \oplus \bcycle(\curve,\mathbb Z)$ (we have constructed one in section \ref{sec:cycles}).
Any such decomposition is rigid but is not canonical. 
Going from one choice to another is called a modular transformation.

Due to lemma \ref{Lem:Qpositive},  $\Img \Pi$ is always transverse to that decomposition.
A basis of $\Img\Pi$ can be obtained by projecting $\bcycle(\curve,\mathbb Z)$ parallel to $\acycle(\curve,\mathbb Z)$, i.e. for each $\bcycle_i$, find a linear combination $\sum_{j} \tau_{i,j} \acycle_j$ such that
\beq\label{eq:BtauA}
\bcycle'_i = \bcycle_i-\sum_j \tau_{i,j} \acycle_j \in \Img \Pi.
\eeq
Lemma \ref{Lem:Qpositive} implies that the matrix $\tau_{i,j}$ is symmetric and  $\Im \tau>0$. In particular $\tau\neq 0$.
We can get a basis of $\Ker \hat B$ by projecting $\acycle(\curve,\mathbb Z)$ parallel to $\Img \Pi$.
Define 
\beq
\kappa_{i,j} = \acycle_i \cap \Pi(\acycle_j),
\eeq
then
\beq\label{eq:AkappaB}
\acycle'_i = \acycle_i-\sum_j \kappa_{i,j} \bcycle'_j \in \Ker \hat B.
\eeq

The basis
\beq\label{eq:defDarbouxi}
(\acycle'_i, \bcycle'_i)
\eeq
is Darboux, but not integer ($\tau$ can never be integer since $\Im\tau>0$), and it gets deformed under deformations of the spectral curve.
In this basis we have
\beq
\mathcal Q(\acycle'_i,\acycle'_j)=0
\qquad , \qquad
\mathcal Q(\acycle'_i,\bcycle'_j)=\frac12 \ \delta_{i,j}
\qquad , \qquad
\mathcal Q(\bcycle'_i,\bcycle'_j)=0 .
\eeq

\br 
The matrices $\tau$ and $\kappa$ are infinite dimensional.
However, it is possible to choose $\acycle(\curve,\mathbb Z)$ that differs from $\Ker \hat B$ by at most a finite dimension space.
For example, it suffices to choose $\acycle(\curve,\mathbb Z)$ (resp. $\bcycle(\curve,\mathbb Z)$) containing all but a finite number of positive (resp. negative) 2nd kind cycles, and completed with a finite number of negative (resp. positive) cycles. In this case the matrices $\tau$ and $\kappa$ have only a finite dimensional non-trivial part.
\er

\item
Define $\acycle''_i$ the projection of $\acycle_i$ onto $\Ker\hat B$, parallel to $\bcycle(\curve, \mathbb Z)$
\beq
\acycle''_i = \acycle_i-\sum_j X_{i,j} \bcycle_j \in \Ker \hat B.
\eeq
The symmetric matrix $X_{i,j}$ is related to the matrices $\tau$ and $\kappa$ above by
\beq
X = (1+\kappa \tau)^{-1}\kappa .
\eeq
Then, the basis
\beq\label{eq:defDarbouxii}
(\acycle''_i, \bcycle_i)
\eeq
is Darboux, but not necessarily integer.
However, we shall see below that it is rigid under "Rauch" deformations of the spectral curve. 
We define the projection on $\bcycle(\curve,\mathbb Z)$ parallel to $\Ker \hat B$:
\beq
\Pi_{\bcycle(\curve,\mathbb Z)}(\bcycle_i)=\bcycle_i
\qquad , \qquad
\Pi_{\bcycle(\curve,\mathbb Z)}(\acycle''_i)=0.
\eeq
In this basis we have
\beq
\mathcal Q(\acycle''_i,\acycle''_j)=0
\qquad , \qquad
\mathcal Q(\acycle''_i,\bcycle_j)=\frac12 \ \delta_{i,j}
\qquad , \qquad
\mathcal Q(\bcycle_i,\bcycle_j)=R_{i,j} 
\eeq
with the matrix $R$ defined below.

\item
Define the symmetric matrix
\beq
R =  \tau (1- X\tau)^{-1} ,
\eeq
and 
\beq
\bcycle''_i = \bcycle_i - \sum_j R_{i,j} \acycle''_j
= (1-\tau X)^{-1} \bcycle'_i  \in \Img \hat C.
\eeq
The basis
\beq\label{eq:defDarbouxiii}
(\acycle''_i, \bcycle''_i)
\eeq
is Darboux, but not integer.
It is a Lagrangian decomposition of $\Ker\hat B \oplus \Img \Pi$.
It is not rigid under Rauch deformations, because $\tau_{i,j}$ is not.

In this basis we have
\beq
\mathcal Q(\acycle''_i,\acycle''_j)=0
\qquad , \qquad
\mathcal Q(\acycle''_i,\bcycle''_j)=\frac12 \ \delta_{i,j}
\qquad , \qquad
\mathcal Q(\bcycle''_i,\bcycle''_j)=0 .
\eeq


\end{enumerate}

In each decomposition, choosing a basis, allows to introduce time coordinates parametrizing cycles and forms

\br
Observe that, since several of the decompositions introduced here depend on integers or on reals, then going from one decomposition to another, changes the coordinates (the times) in a possibly not analytic way.
This is related to the existence of several complex structures in our moduli space.
This is the origin of the HyperK\"ahler structure of the moduli space.

\er

\subsubsection{Times and periods of Darboux coordinates }

Having choosen a Darboux basis of cycles $(\acycle,\bcycle)$, 
\bd\label{deftimes}
we define the times as the periods of $\y$ over $\acycle$--cycles:
\beq
t_i = \frac{1}{2\pi\ii} \oint_{\acycle_i} \y.
\eeq

\ed

In many known examples, these times are indeed the "good" times of our system.
For example, if $\y$ is meromorphic with a pole at $p$, and with Laurent series expansion near $p$ as 
\beq
\y(z) \mathop{\sim}_{z\to p} \sum_{k=0}^{\deg_p \y} t_{p,k} \xi_p(z)^{-k-1} d\xi_p(z) + \text{analytic at } p,
\eeq
the coefficients $t_{p,k}$ are the $\acycle_{p,k}$ periods
\beq
t_{p,k} = \frac{1}{2\pi\ii} \oint_{\acycle_{p,k}} \y.
\eeq
These are called the KP times.

\subsubsection{Example with compact surfaces}

Consider a case where $\curve$ is compact of genus $\genus$, with a symplectic basis of 1st kind cycles $(\acycle_i,\bcycle_i)_{i=1,\dots,\genus}$, and define $\acycle''_i$ as in \eqref{eq:defDarbouxii}.
Define $\omega_i$ the holomorphic basis of 1-forms such that $\oint_{\acycle_i} \omega_j=\delta_{i,j}$ and define the period matrix $\tau_{i,j}=\oint_{\bcycle_i} \omega_j = \tau_{j,i}$, it actually coincides with the one in \eqref{eq:BtauA}.
Consider also the case where
\beq
B(z,z') = d_z d_{z'} \log \theta_c(\mathfrak a(z)-\mathfrak a(z');\tau)  + 2\pi\ii \sum_{i,j=1}^{\genus} \kappa_{i,j} \omega_i(z) \omega_j(z')
\eeq
with $\mathfrak a=(\mathfrak a_1, \dots, \mathfrak a_\genus)$ the Abel map (defined by $d\mathfrak a_i=\omega_i$)  and $c$ a regular odd characteristic, and $\theta$ the Siegel Theta function. 
The symmetric $\genus\times \genus$ matrix $\kappa_{i,j}$ coincides with the one introduced in  \eqref{eq:AkappaB}.

We have:
\bea
&& \hat B(\acycle'_i) = \hat B(\acycle''_i) = 0
\cr
&& \hat B(\bcycle'_i)  = 2\pi\ii  \omega_i
\qquad , \quad
\hat B(\bcycle_i) = \hat B(\bcycle''_i) = 2\pi\ii  \sum_j (1+\tau\kappa)_{i,j} \omega_j.
\eea
Writing the Taylor expansion of $B$ near a point $p\in\curve $ as
\beq
B(z,z') - \frac{d\xi_p(z)d\xi_p(z')}{(\xi_p(z)-\xi_p(z'))^2} \sim  \sum_{k,l=1}^\infty  k l \  \tau_{p,k;p,l} \xi_{p}(z)^{k-1} \xi_p(z')^{l-1}  \, d\xi_p(z) d\xi_p(z')
\eeq
we get
\beq
\hat B(\acycle_{p,k})=0
\qquad , \qquad
\hat B(\bcycle_{p,k})  =   \left( \xi_p^{-k-1} + \sum_{l=1}^\infty l \tau_{p,k;p,l} \ \xi_p^{l-1} \right)  d\xi_p \quad \text{near } p.
\eeq
\beq
\hat B(\gamma_{q\to p}) =  \int_q^p B =  \left( \xi_p^{-1} + \sum_{l=1}^\infty l \tau_{p,0;p,l} \ \xi_p^{l-1} \right)  d\xi_p \quad \text{near } p.
\eeq
In both cases near another point $p'\neq p$, these forms are analytic and have a regular Taylor expansion that we write
\beq
\hat B(\bcycle_{p,k})  =  \sum_{l=1}^\infty l \tau_{p,k;p',l} \ \xi_{p'}^{l-1}  d\xi_{p'} \quad \text{near } p',
\eeq
where the matrix $\tau$ coincides with that of \eqref{eq:BtauA}.

In the Darboux basis \eqref{eq:defDarbouxii} we have:
\beq
\mathcal Q(\acycle''_i,\acycle''_j) = 0
\quad , \quad
\mathcal Q(\acycle''_i,\bcycle_j) = \frac12 \ \delta_{i,j}
\quad , \quad
\mathcal Q(\bcycle_i,\bcycle_j) =  R_{i,j}
\eeq
For 3rd kind cycles, introducing the prime form (see \cite{Fay})
 we have:
\begin{eqnarray}
&& 2\pi\ii\mathcal Q(\gamma_{q\to p},\gamma'_{q'\to p'}) = \ln{\frac{E(p,p')E(q,q')}{E(q,p')E(p,q')}} \cr
&& 2\pi\ii\mathcal Q(\gamma_{q\to p},\gamma'_{q'\to p}) = \ln{\frac{E(q,q')}{E(q,p)E(p,q')d\xi_p(p)}}
\cr
&& 2\pi\ii\mathcal Q(\gamma_{q\to p},\gamma_{q\to p}) = \ln{\frac{-1}{E(q,p)^2 d\xi_p(p)d\xi_q(q)}}
\end{eqnarray}

\beq
2\pi\ii\mathcal Q(\sum_{i=1}^k \alpha_i \gamma_{q_i\to p_i},\sum_{i=1}^k \alpha_i \gamma_{q_i\to p_i})
=  \ln{\frac{\prod_{i<j} E(p_i,p_j)^{\alpha_i\alpha_j} \prod_{i<j} E(q_i,q_j)^{\alpha_i\alpha_j} }
{\prod_{i,j} E(p_i,q_j)^{\alpha_i\alpha_j}    \prod_i d\xi_{p_i}(p_i)^{\alpha_i^2}d\xi_{q_i}(q_i)^{\alpha_i^2} }  }
\eeq
We see here that indeed the result depends on an ordering $i<j$, echoing remark \ref{rem:chains}.
A different choice of ordering changes the result by $\pi\ii$ times an integer quadratic polynomials of the $\alpha_i$s.
In particular if the $\alpha_i$s are integer, a change of ordering just changes the result modulo $\pi\ii\mathbb Z$, i.e. a sign inside the log.

\section{Deformations of spectral curves}
\label{sec:deform}

We shall consider deformations of spectral curves at fixed $\curve$ (as a topological surface)  and fixed base $\curverond$.
We can therefore deform $\x$, $\y$ or $B$.
The deformation of $\x$ induces a deformation of the complex structure of 
$\curve$ as a Riemann surface whose complex structure is the pullback by $\x$ of $\curverond$.

The idea developed below, is that deformations --here we mean tangent vectors-- of $\y$ are 1-forms (resp. deformations of $B$ are $1 \boxtimes 1$--forms) and are dual to cycles (resp. dual to tensor products of cycles).
This will allow to identify  the tangent space  $T_{\spcurve}\modsp$ with a subspace of cycles $\mathfrak M_1(\curve,\mathbb Z)\otimes \mathbb C$ (resp. pairs of cycles).
Moreover, since at fixed base $\curverond$, integer cycles form a lattice, they are rigid and don't deform, this will induce a trivial connexion on the cycles bundle. 

\subsection{Meromorphic tangent space}

The space $\mathcal M$ of all spectral curves is too large and is not a manifold, in particular it doesn't always have a tangent space at a point. 
However, given a spectral curve, we shall consider a subspace, that is locally a manifold and has a tangent space. 
In this purpose we shall restrict to only a subset of possible deformations, those locally meromorphic.

\bd[Subspace of Meromorphic shifts of a spectral curve.]

Given a spectral curve $\spcurve$ represented by the data $(\curve,\x,\y,B)$, consider the subspace $\modspmero{\spcurve}$ whose objects are spectral curves $\tilde\spcurve=[(\curve,\tilde\x,\tilde\y,\tilde B)]$
with the same topological surface $\curve$, with the same base Riemann surface $\curverond$ for $\x$ and $\tilde \x$, and such that $\tilde\x-\x$, $\tilde\y -\y$, $\tilde B-B$ are $C^\infty$ plus at most a finite number of singularities, and such that at points where $\x$ and $\tilde \x$ are locally invertible, $\y\circ \x^{-1} - \tilde\y \circ \tilde \x^{-1}$ is locally meromorphic on $\curverond$, and similarly $\x_* B  - \tilde\x_*\tilde B$ is locally a bi-holomorphic differential on $\curverond\times \curverond$ 
(these are in general not holomorphic at branchpoints of $\x$ or $\tilde \x$).
\ed

\framefig{
\includegraphics[scale=0.5]{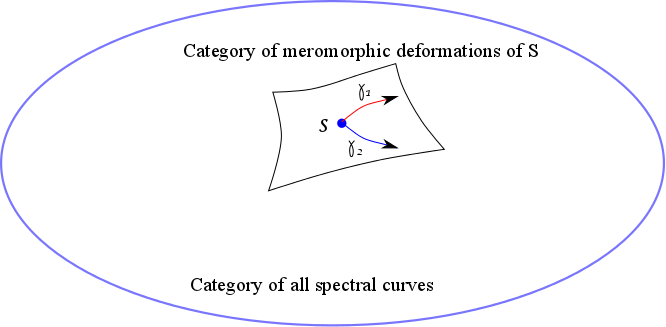}
\caption{The subspace of meromorphic deformations of a spectral curve.
Tangent vectors are cycles, and thus cycles generate flows.}
}

This subspace is a quotient (mod diffeomorphisms) of an affine bundle $\spcurve+ \mathfrak M^1(\curve)\oplus \mathfrak M^1(\curve)^{\boxtimes 2\,\text{sym}}\to C^\infty(\curve,\curverond)$, it is thus a differentiable manifold, having a tangent space at $\spcurve$.
tangent vectors in the total affine bundle are triplets $(\delta\x, \delta\y,\delta B)\in C^\infty(\curve,\curverond)\oplus  \mathfrak M^1(\curve)\oplus \mathfrak M^1(\curve)^{\boxtimes 2\,\text{sym}}$.

The quotient by diffeomorphisms, identifies 
$(\x,\y,B)\equiv (\phi^*\x,\phi^*\y,\phi^*B)$, and for  tangent vectors
\bea\label{defdeltaxyB}
(\delta\x, \delta\y,\delta B)
& \equiv &
(\delta\x+d\x.\delta\phi, \delta\y+ d\y.\delta\phi ,\delta B+ dB.\delta\phi \boxtimes \delta\phi ) \cr
& \equiv &
(0, \delta\y - \delta\x \, d(\y/d\x),\delta B-  d_1 (B/d\x_1) \delta\x_1  -  d_2 (B/d\x_2) \delta\x_2 ) 
\eea
indeed we may locally choose $\phi$ such that $d\x.\delta\phi=-\delta\x$, or equivalently we may choose $\phi$ so that $\delta\x=0$ locally.
In other words, away from branchpoints, we may consider deformations of $\y$ and $B$ at constant projection by $\x$.

By abuse of language we shall from now on call $(0,\delta\y,\delta B)$ the representative of the tangent vector for which $\delta \x=0$ locally.
The fact that we can choose such a representative only locally away from branchpoints, means that $\delta\y$ and $\delta B$ can have poles at branchpoints (indeed there is a ratio by $d\x$ in \eqref{defdeltaxyB}).

The tangent space at $\spcurve$ is thus
\beq
T_\spcurve \modspmero{\spcurve} \sim 
\mathfrak M^1(\curve) \oplus (\mathfrak M^1(\curve) \boxtimes  \mathfrak M^1(\curve))^{\rm sym} .
\eeq
where $\delta\y$ is a meromorphic 1-form and $\delta B$ a meromorphic symmetric bilinear differential, and both can have poles at the branchpoints of $\x$.

Deformations are thus made of 2 parts: one that deforms the 1-form $\y$ and one that deforms the bilinear form $B$.
We shall decompose the $B$-deformation into two pieces: a piece related to the deformation of $\y$, and a piece not related.

\bt[Cycles $\longrightarrow$ tangent vectors]
We have a surjective map from the space of cycles or pairs of cycles into the tangent space
\beq
 \mathfrak M_1(\curve) \oplus (\mathfrak M_1(\curve)\otimes \mathfrak M_1(\curve))^{\text{sym}}
 \to T_{\spcurve} \modspmero{\spcurve},
\eeq
that, to $\gamma\in \mathfrak M_1(\curve)$ associates the tangent vector $\partial_\gamma$ such that
\bea
\partial_\gamma \y &=& \int_\gamma B = \hat B(\gamma)  \cr
\partial_\gamma B &=& \int_{\gamma} \omega_{0,3}
\eea
and that, to a pair $(\gamma_1,\gamma_2) \in \mathfrak M_1(\curve)\times \mathfrak M_1(\curve)$ associates the tangent vector $\partial_{\gamma_1\otimes \gamma_2}$ such that
\bea
\partial_{\gamma_1\otimes \gamma_2} \y &=&  0 \cr
\partial_{\gamma_1\otimes \gamma_2} B 
&=& \frac12 \ \left( \hat B(\gamma_1)\boxtimes \hat B(\gamma_2)+\hat B(\gamma_2)\boxtimes \hat B(\gamma_1) \right).
\eea
\et
The 1-cycle part $\partial_\gamma$ is called the {\bf Rauch} part.
The 2-cycles part $\partial_{\gamma_1\otimes\gamma_2}$ can be called the \textbf{BCOV}-like part, for a reason that we shall see later on.


\proof
We need to prove that it is surjective.
Consider a tangent vector $(0,\delta\y,\delta B)$.
Define $\gamma=\hat C(\delta\y)$, and consider 
\beq
\tilde\delta B = \delta B - \int_\gamma \omega_{0,3}.
\eeq
By definition \cite{EO07} of $\omega_{0,3}$, $\tilde\delta B$ has no poles at branchpoints (this is Rauch variational formula), neither at coinciding points, therefore it is a tensor product of meromorphic forms.
Define $\tilde \Gamma = (\hat C\boxtimes \hat C)(\tilde \delta B)$.
Then we have
\beq
\delta = \partial_\gamma + \partial_{\tilde \Gamma}.
\eeq
\eproof

\br\label{remKerformcycle}
Notice that this map has an infinite dimensional kernel, namely $\Ker \hat B\oplus (\Ker\hat B \otimes T_{\spcurve}\modsp)^{\text sym}$. It is thus not injective.
\er

\br[Rauch variational formula]
Consider a family of spectral curves, $\spcurve=(\curve,\x,\y,B)$ where $\curve$ is a compact connected surface, and where $B$ is choosen to be the fundamental 2nd kind differential \cite{Fay, MumTata} on the curve $\curve$ equipped with a Torelli marking -- this $B$ is sometimes called the Bergman or Bergman-Schiffer kernel \cite{BergSchif} --. In other words, $B$ is determined by $(\curve_{\text{marked}},\x,\y)$, and thus gets deformed under deformations of $(\curve_{\text{marked}},\x,\y)$, at fixed marking.

Consider a deformation $\delta$ i.e. a tangent vector to that family. It induces a deformation $\delta\y=\Omega$ with $\Omega$ a meromorphic 1-form, and choosing $\gamma=\hat C(\Omega)$, we have
\beq
\delta\y = \Omega = \int_{\hat C(\Omega)} B.
\eeq
Then, Rauch's variational formula \cite{Rauchvar} implies that
\beq
\delta B = \int_{\hat C(\Omega)} \omega_{0,3}.
\eeq
In other words, $\partial_\gamma$ are tangent deformations that conserve the fundamental 2nd kind differential.
\er

\bp[Curvature]\label{prop:curvature1}

The Lie derivatives satisfy the following Lie algebra:
\beq
[\partial_{\gamma_1},\partial_{\gamma_2} ] =0.
\eeq
\bea
[\partial_{\gamma_1\otimes \gamma_2},\partial_{\gamma_3\otimes \gamma_4} ] &=&
\pi\ii \ \Big(  \gamma_3\cap\gamma_1  \ \partial_{\gamma_2\otimes \gamma_4} 
+\gamma_3\cap\gamma_2  \ \partial_{\gamma_1\otimes \gamma_4}  \cr
&& \qquad +\gamma_4\cap\gamma_1  \ \partial_{\gamma_2\otimes \gamma_3} 
+\gamma_4\cap\gamma_2  \ \partial_{\gamma_1\otimes \gamma_3}  \Big)
\eea
\beq\label{commutd2gd1g}
[\partial_{\gamma_2 \otimes \gamma_3} , \partial_{\gamma_1}] =
\pi\ii \left( (\gamma_1\cap \Pi(\gamma_2)) \ \partial_{\gamma_3} + (\gamma_1\cap \Pi(\gamma_3)) \ \partial_{\gamma_2} \right).
\eeq

\ep

\proof{easy computation.}
\eproof

The tangent space is isomorphic to the quotient of the space of cycles by $\Ker \hat B\oplus (\Ker\hat B \otimes T_{\spcurve}\modsp)^{\text sym}$, and thus the space of cycles is "twice larger" than the tangent space. Only a Lagragian submanifold of the space of cycles should actually be indentified to the tangent space.
And indeed, restricted to a Lagrangian space of cycles, the curvature \eqref{commutd2gd1g} vanishes.

\bt[proved in \cite{EO07}]
\label{th:deformomgn}
The invariants $\omega_{g,n}$ for $(g,n)\neq (0,0)$ are deformed as
\beq
\partial_\gamma \omega_{g,n} = \int_\gamma \omega_{g,n+1}.
\eeq
\begin{multline}
\partial_{\gamma_1\otimes \gamma_2} \omega_{g,n}(z_1,\dots,z_n) = \frac12 \int_{z\in\gamma_1}  \int_{z'\in\gamma_2} 
\Big( \omega_{g-1,n+2}(z,z',z_1,\dots,z_n) \\
+ \sum'_{\overset{h+h'=g}{I\uplus I'=\{z_1,\dots,z_n\}}} \omega_{h,1+\# I}(z,I) \omega_{h',1+\# I'}(z',I') \Big).
\end{multline}
where $\sum'$ in the last line means excluding the 2 terms $(h,I)=(0,\emptyset)$ and $(h',I')=(0,\emptyset)$.

\et
For the $F_g$s we have
\beq
g=1: \quad \partial_{\gamma_1\otimes \gamma_2} F_1 
= \frac12 \left( \oint_{\gamma_1} \oint_{\gamma_2} B + \oint_{\gamma_2} \oint_{\gamma_1} B\right) = 2\pi\ii\mathcal Q(\gamma_1,\gamma_2) ,
\eeq
\bea
g>1: \quad \partial_{\gamma_1\otimes \gamma_2} F_g 
&=& \partial_{\gamma_1}\partial_{\gamma_2} F_{g-1} + \sum_{h=1}^{g-1} \partial_{\gamma_1} F_h  \partial_{\gamma_2} F_{g-h} \cr
\eea

\br\label{rem:F0dgamma}
It would be tempting to define $F_0=\omega_{0,0}$ in such a way that $\partial_\gamma F_0 = \oint_\gamma \omega_{0,1}=\oint_\gamma \y$.
The problem is that 
\beq
\partial_{\gamma_i} \oint_{\gamma_j} \y  - \partial_{\gamma_j} \oint_{\gamma_i} \y
= \oint_{\gamma_j} \oint_{\gamma_i} B  - \oint_{\gamma_i} \oint_{\gamma_j} B
 = 2\pi\ii\ \gamma_j\cap \gamma_i
\eeq
which can be $\neq 0$. 
However, it vanishes on any Lagrangian sub-manifold.

Recall that the space of cycles is in some sense twice larger than the tangent space, only a Lagragian submanifold of the space of cycles should actually be indentified to the tangent space.
In other words, we could define $F_0$ such that $\partial_\gamma F_0 = \int_\gamma \omega_{0,1}$ only for $\gamma$ in a certain Lagrangian submanifold.
We shall do it in section \ref{sec:F0} below.

\er

\subsection{Hirota derivatives of spectral curves}

We define
\bd[Hirota derivative]
We define, for any generic smooth $z\in\curve$:
\beq
\Delta_z = d\x(z)\ \partial_{\bcycle_{z,1}}
\eeq
where $\bcycle_{z,1}\in \mathfrak M_1(\curve,\mathbb Z)$ is the 2nd kind cycle defined in \eqref{def:M12ndpm} with $p=z$ and $k=1$
\beq
\bcycle_{z,1} = \frac{1}{2\pi\ii}\mathcal C_z . \left( z'\mapsto (\x(z')-\x(z))^{-1}\right).
\eeq

\ed

\bt[\cite{EO07}]\label{prop:Deltaomgn}
We have
\beq\label{DeltayB}
\Delta_{z_2} \y(z_1) = B(z_1,z_2).
\eeq
And more generally,
the Hirota derivative acts on the invariants by shifting $n\to n+1$, i.e. we have, $\forall\,(g,n)\neq(0,0)$:
\beq
\Delta_z\ \omega_{g,n}(\spcurve;z_1,\dots,z_n) = \omega_{g,n+1}(\spcurve;z_1,\dots,z_n,z) .
\eeq
In other words
\beq
\partial_\gamma \omega_{g,n}  = <\gamma,\Delta_z\ \omega_{g,n}> = \int_\gamma \Delta_z\omega_{g,n}.
\eeq
We shall sometimes denote it:
\beq
\partial_\gamma = <\gamma,\Delta_z> = \int_\gamma \Delta_z.
\eeq
\et

\proof
\eqref{DeltayB} is immediate from the definition. For all invariants it is proved in \cite{EO07}.
\eproof

\br
In the random--matrix model literature $\Delta_z$ was named the "insertion operator".
\er

\br \textbf{Link to the usual Sato--Hirota notation}
\label{rem:Hirota}

Consider a meromorphic 1-form $\y$ written in a neighborhood of a pole $p$, in a local coordinate $\xi_p(z)=(\x(z)-\x(p))^{1/{{\rm order}_p(\x)}}$, as
\beq
\y = \sum_{k=0}^{d_p} t_{p,k} \xi_p^{-k-1} d\xi_p + {\rm analytic\,at}\,p
\eeq
whose coefficients $t_{p,k}=\frac{1}{2\pi\ii}\oint_{\acycle_{p,k}} \y $ are called the KP times.

If $z$ is near the pole $p$, the shift by $\epsilon \bcycle_{z,1}$ amounts to
\bea
\y(z') & \to & \y(z') + \epsilon d\xi_p(z) \hat B(\bcycle_{z,1})(z')  \cr
& = & \y(z') + \epsilon \frac{ 1}{(\xi_p(z)-\xi_p(z'))^2}\ d\xi_p(z) d\xi_p(z') 
+\text{analytic at}\,z'\to p \cr
 & \sim & \y(z') + \epsilon \sum_{k=1}^\infty  k \xi_p(z)^{k-1} \xi_p(z')^{-k-1} d\xi_p(z) d\xi_p(z')  \cr
\eea
and thus, locally near $p$, the Taylor expansion of the Hirota derivation operator is (locally, not globally) the familiar infinite sum of time derivatives
\beq
\Delta_z \sim  \sum_{k=1}^\infty k \xi_p(z)^{k-1}d\xi_p(z) \frac{\partial}{\partial t_{p,k}}  .
\eeq
However, we should keep in mind that this is only a Taylor expansion near $p$ and this has no meaning far away from $p$.

\er

\subsection{Loop equations and $\mathfrak W$ operators}

We denote $A\subset_k B$ if $A\subset B$ and $\#A=k$.
From now on we assume that  $\deg \x$ is finite.
We define

\bd[$\mathfrak W$ operators]\label{defWk}
For $x\in \curverond$ we define the operators $\mathfrak W_0={\rm Id}$ and for  $k\geq 1$:
\beq
\mathfrak W_k(x) = \x_*\sum_{I\subset_k \x^{-1}(x)} \prod_{z'\in I} \Delta_{z'} ,
\eeq
and for $(x,y)$ a point in the total cotangent bundle of $\curverond$, i.e. $x\in\curverond$ and $y\in T^*_x\curverond$, we define the operator
\beq
\mathfrak W(x,y) = \sum_{k=0}^{\deg \x} (-1)^k y^{\deg\x-k} \mathfrak W_k(x)
= \prod_{z\in \x^{-1}(x)} (y-\x_*\Delta_{z})
.
\eeq
\ed
Notice that $\mathfrak W_k=0$ if $k>\deg \x$.

Denote also:
\beq\label{defPxy}
P_k(\spcurve;x) = \x_*\sum_{\{z_1,\dots,z_k\}\subset_k \x^{-1}(x)} \y(z_1)\dots \y(z_k),
\eeq
it is a $k^{\rm th}$-order differential on the base $x\in\curverond$, it has singularities at the projections of the singularities of $\y$.

\br[Hitchin map]
Condider the case where the spectral curve is the eigenvalues locus of a Higgs pair $(\mathcal E,\Phi)$, i.e. the locus of solutions of $\det_\rho(y-\Phi(x))=0$ (with $\rho$ a faithful representation of a Lie group $G$ into $GL(r,\mathbb C)$, and thus $r=\dim\rho=\deg \x $). The map $\Phi\mapsto (P_1,P_2,\dots,P_{r}) \in \oplus_{k=1}^r H^0(\curverond,(K_{\curverond})^k) $ is the Hitchin map, it maps a Higgs field to its invariants.
\er

\bd
We define, for $y\in T^*_x\curverond$:
\beq
P(\spcurve;x,y) = \sum_{k=0}^{\deg x} (-1)^k y^{\deg\x-k} P_k(\spcurve,x) = \prod_{z\in \x^{-1}(x)}^{\deg x}(y-\x_*\y(z)).
\eeq
and
\beq
P_y(\spcurve;x,y) = \sum_{k=0}^{\deg x-1} (-1)^k (\deg\x-k)\,y^{\deg\x-k-1} P_k(\spcurve,x).
\eeq
\bea
P'(\spcurve;z)
&=& P_y(\spcurve;x,y)_{x=\x(z),\ y=\y(z)} \cr
& = &  \sum_{k=0}^{\deg x-1} (-1)^k (\deg\x-k)\,\y(z)^{\deg\x-k-1} P_k(\spcurve,\x(z)) = P_y(\spcurve;\x(z),\y(z)). \cr
\eea
\ed

Remark that $P'(\spcurve;z)$ vanishes whenever 2 or more branches meet, in particular it vanishes at all ramification points.
It vanishes also at double points, i.e. where 2 branches cross, these can also be seen as nodal points, or as cycles that have been pinched.

In order to take double points into account, we need to slightly enlarge $H_1(\curve,\mathbb Z)$ (the space of 1st kind cycles) to include all possible $\Omega$ that will arise in def.\ref{def:loopeq} below, that is:
\bd\label{defHprime1}
Let $H^{1'}(\curve,\mathbb C)$ be the space of 1-forms that have no poles at ramification points, and whose product with $P'(\spcurve;z)$ has no pole at the zeros of $P'(\spcurve;z)$.

Let $H_1'(\curve,\mathbb C)$ be the space generated by 1st kind cycles, and by 3rd kind cycles with boundary at self--intersection points, and 2nd kind cycles surrounding self-intersection points, with degree lower than that of $P'(\spcurve;z)$.

$H^{1'}(\curve,\mathbb C)$ is the image by $\hat B$ of $H_{1}'(\curve,\mathbb C)$. 
\ed

In the case where the spectral curve is a Lagrangian embedding
$\curve \to T^*\curverond \ , \    z \mapsto (\x(z),\y(z))$
into the cotangent space of $\curverond$, there is no double point, all zeros of $P'(\spcurve;z)$ are ramification points, and then
\beq
H'^1(\curve)=H^1(\curve)
\qquad , \qquad
H'_1(\curve)=H_1(\curve).
\eeq
Otherwise, the zeros of $P'(\spcurve;z)$ that are not ramification points, are self--intersections of the immersion of $\curve \hookrightarrow T^*\curverond$, they are in some sense pinched cycles, and it is natural to enlarge the space of holomorphic forms to include the forms duals to pinched cycles as well.
$H_1'(\curve,\mathbb Z)$ is thus the space of non-contractible cycles together with pinched cycles, in other words, $H_1'(\curve,\mathbb C)$ is the $H_1(\curve,\mathbb C)$ after desingularization. 
The dimension of $H^{1'}(\curve,\mathbb C)$ is the number of integer points of the Newton's polygon of $P(\spcurve;x,y)$, i.e. is the genus of the unpinched surface.

For example, if all zeros of $P'(\spcurve;z)$ are simple zeros, then $H_1'(\curve,\mathbb C)$ includes the  small circles $\acycle_{p,0}$ around these zeros, and the 3rd kind paths $\gamma_{p_-\to p_+}$ going from one side of the pinched cycle to the other.

Equipped with these notations we define the notion of "loop equations":

\bd[Loop equations]
\label{def:loopeq}
A local section $f$ of a line bundle over the subspace $\modspmero{\spcurve}$ is said to be solution of loop equations iff:
\begin{itemize}
\item
\beq\label{eq:Wkloopeq}
\forall\,1\leq k\leq \deg\x
\quad , \qquad
 \mathfrak W_k(x).  f(\spcurve)
\eeq
is a holomorphic $k^{\rm th}$-order differential on the base $\curverond$, that has no poles at branchpoints.

\item The 1-form
\beq\label{eq:Wxyloopeq}
\Omega(\spcurve;z)(f) = \left.
\frac{1}{P'(\spcurve;z)}  \ \mathfrak W(x,y).f(\spcurve) \right|_{x=\x(z),\,y=\x_*\y(z)}
\eeq
belongs to $H'^1(\curve)$.

\end{itemize}

\ed

Very often in the CFT literature the first condition, saying that the operator $\mathfrak W_k$ is holomorphic on the base $\curverond$, is written as
\beq
\bar\partial\ \mathfrak W_k=0,
\eeq
and is called "Ward identity".
This is in some sense equivalent to "Virasoro" constraints ($k=2$) and $\mathfrak W$--algebra constraints.

We have the obvious lemma
\bl
Loop equations are $\mathbb C$--linear, i.e. if 2 sections $f$ and $\tilde f$ satisfy loop equations, then so does $a f+b \tilde f$ (with $a$ and $b$ fixed complex numbers, not sections).
\el

Below, we shall define some "partition functions" and "Tau functions" that are solutions of loop equations.

\subsubsection{$\mathfrak W'$ operators}

It is also very useful to define the following operators (contrarily to the $\mathfrak W_k(x)$ that live on the base curve $x\in\curverond$, these live on the curve $\curve$):

\bd[$\mathfrak W'$ operators]\label{defWkprime}
For $z\in \curve$ we define the operators $\mathfrak W'_0(z)=0$ and when $k\geq 1$:
\beq
\mathfrak W'_k(z) = \sum_{I\subset_k \x^{-1}(\x(z))-\{z\}} \prod_{z'\in I} \Delta_{z'} ,
\eeq
and the operator
\beq
\mathfrak W'(z) = \sum_{k=0}^{\deg \x-1} (-1)^k \y(z)^{\deg\x-k} \mathfrak W'_k(z)
= \left(\prod_{z\in \x^{-1}(\x(z))-\{z\}} (y-\Delta_{z'})\right)_{y=\y(z)}
.
\eeq
\ed

\bp
We have
\beq
\mathfrak W_k(\x(z)) =  \mathfrak W'_k(z) +  \mathfrak W'_{k-1}(z) \Delta_z ,
\eeq
or equivalently
\beq
\mathfrak W(\x(z),\y(z)) = :(\y(z)-\Delta_z) \mathfrak W'(z) : ,
\eeq
where the ":( . ):" notation means that the $\Delta$  do not act on $\y(z)$, in fact this means $\mathfrak W(\x(z),\y(z)) = \y(z) \mathfrak W'(z) - \mathfrak W'(z)\Delta_z$.
\ep
The proof is obvious.

\subsection{Prepotential $\mathcal F_0$}
\label{sec:F0}

The idea would be to define $\mathcal F_0$ so that it satisfies prop. \ref{prop:Deltaomgn}.
There exists some $\mathcal F_0$ that satisfies $\Delta_z \mathcal F_0(\spcurve) = \omega_{0,1}(\spcurve;z)$ because $\Delta_{z'}\omega_{0,1}(\spcurve;z) = B(z,z') = \Delta_z \omega_{0,1}(\spcurve;z')$ is symmetric.
The problem is thus not the existence of $\mathcal F_0$, but uniqueness, because $\Delta$ might have a kernel.
This is also related to remark \ref{rem:F0dgamma}.
As we mentioned there, a definition of $\mathcal F_0$ needs to choose a Lagrangian submanifold.

Let us choose arbitrarily $\acycle(\curve,\mathbb Z) \oplus \bcycle(\curve,\mathbb Z)=\mathfrak M_1(\curve,\mathbb Z)$ an integer Darboux decomposition with an integer Darboux symplectic basis, and consider the basis $(\acycle''_i,\bcycle_i)$ of \eqref{eq:defDarbouxii}.
\beq
\acycle''_i = \acycle_i - \sum_{j} X_{i,j} \bcycle_j .
\eeq
From these, we define the times
\beq
t_i = \frac{1}{2\pi\ii} \oint_{\acycle''_i} \y .
\eeq

\bl
\label{lemma:deltaX}
We have
\beq
\Delta_z X = 0
\qquad , \qquad
\partial_\gamma X=0.
\eeq
\beq
\Delta_z t_i = 0 \quad \text{if}\ i\neq (z,1)
\qquad , \qquad
\Delta_z t_{z,1} =  d\x 
\qquad , \qquad
\partial_\gamma t_i = \acycle''_i \cap \gamma.
\eeq
\beq
\partial_{\gamma_1\otimes \gamma_2} X_{i,j}
= \frac12 \left( \acycle''_i \cap \gamma_1 \ . \ \acycle''_j \cap \gamma_2 + \acycle''_i \cap \gamma_2 \ . \ \acycle''_j \cap \gamma_1  \right) .
\eeq
\beq
\partial_{\gamma_1\otimes \gamma_2} \acycle''_i
= - \ \frac12 \left( \acycle''_i \cap \gamma_1 \ . \Pi_{\bcycle(\curve,\mathbb Z)}(\gamma_2) + \acycle''_i \cap \gamma_2 \ . \ \Pi_{\bcycle(\curve,\mathbb Z)}(\gamma_1)  \right) .
\eeq
\beq
\partial_{\gamma_1\otimes \gamma_2} \bcycle'_i = 0
\qquad , \qquad 
\partial_{\gamma_1\otimes \gamma_2} \tau_{i,j} = 0 .
\eeq
\beq
\partial_{\gamma_1\otimes \gamma_2} \acycle'_i = -  \frac12 \left( \acycle'_i \cap \gamma_1 \ \Pi(\gamma_2) + \acycle'_i \cap \gamma_2 \ \Pi(\gamma_1) \right) .
\eeq

\el

\proof
In appendix \ref{app:DeltaX}.
\eproof

The key point for us is that $X$ and the times $t_i$s are undeformed by Rauch deformations.

\bd
Define the prepotential, polarized along ${\bcycle(\curve,\mathbb Z)}$ as
\bea
{\mathcal F}_{0,[\bcycle(\curve,\mathbb Z)]}(\spcurve)
&=& \pi\ii \  (\Pi_{\bcycle(\curve,\mathbb Z)}(\hat C(\y))) \cap \hat C(\y) \cr
&=& \frac{1}{4\pi\ii} \sum_i \oint_{\acycle''_i} \y \oint_{\bcycle_i} \y  \cr
&=& \frac{1}{2} \sum_i t_i  \oint_{\bcycle_i} \y  \cr
&=& \pi\ii \mathcal Q(\Pi_{\bcycle(\curve,\mathbb Z)}(\hat C(\y))),\Pi_{\bcycle(\curve,\mathbb Z)}(\hat C(\y))))  .
\eea

\ed

\bp
\beq
\Delta_z {\mathcal F}_{0,[\bcycle(\curve,\mathbb Z)]}(\spcurve) = \y(z) ,
\eeq
\beq
\partial_\gamma {\mathcal F}_{0,[\bcycle(\curve,\mathbb Z)]}(\spcurve) = \oint_{\Pi_{\bcycle(\curve,\mathbb Z)}(\gamma)} \y,
\eeq
i.e.
\beq
\partial_{\acycle''_i} {\mathcal F}_{0,[\bcycle(\curve,\mathbb Z)]}(\spcurve) =0
\qquad , \qquad
\partial_{\bcycle_i} {\mathcal F}_{0,[\bcycle(\curve,\mathbb Z)]}(\spcurve) =  \int_{\bcycle_i} \y .
\eeq
and
\beq
\partial_{\gamma_1\otimes \gamma_2} {\mathcal F}_{0,[\bcycle(\curve,\mathbb Z)]}(\spcurve) =  \oint_{\Pi_{\bcycle(\curve,\mathbb Z)}(\gamma_1)} \y \oint_{\Pi_{\bcycle(\curve,\mathbb Z)}(\gamma_2)} \y
\eeq
\ep

\proof
Let $t_i = \frac{1}{2\pi\ii}\oint_{\acycle''_i} \y$ and $\eta_i = \oint_{\bcycle_i}\y$.
We have
\beq
2 \Delta_z \mathcal F_0 = \sum_i \eta_i \Delta_z t_i + \sum_i t_i \Delta_z \eta_i.
\eeq
If $i\neq (z,1)$ we have $\Delta_z t_i=0$, and if $i=(z,1)$ we have $\Delta_z t_{z,1} = d\x$, and $\eta_{z,1} =  \y(z)/d\x(z)$, therefore
\beq
\sum_i \eta_i \Delta_z t_i = \eta_{z,1} \Delta_z t_{z,1} =  \y(z) .
\eeq
Then compute
\beq
\sum_i t_i \Delta_z \eta_{i} = \oint_{\sum_i t_i \bcycle_i} B = \oint_{\sum_i t_i \bcycle_i- \eta_i \acycle''_i} B =  \oint_{\hat C(\y)} B =  \y ,
\eeq
which completes the proof that
\beq
\Delta_z \mathcal F_0 = \y.
\eeq
It is obvious that
\beq
\partial_{\acycle''_i} \mathcal F_0=0.
\eeq
Then we have
\beq
\partial_{\bcycle_i} t_j = \frac{1}{2\pi\ii} \oint_{\acycle''_j} \oint_{\bcycle_i} B = \acycle''_j \cap \bcycle_i =  \delta_{i,j},
\eeq
and $\bcycle_j\cap \bcycle_i=0$ implies that
\beq 
\partial_{\bcycle_i} \eta_j =  \oint_{\bcycle_j} \oint_{\bcycle_i} B = 2\pi\ii \bcycle_j \cap \bcycle_i + \oint_{\bcycle_i} \oint_{\bcycle_j} B  = \oint_{\bcycle_i} \oint_{\bcycle_j} B .
\eeq
This shows that
\beq
\sum_j t_j \partial_{\bcycle_i} \eta_j 
= \oint_{\bcycle_i} \oint_{\sum_j t_j \bcycle_j}   B
= \oint_{\bcycle_i} \oint_{\sum_j t_j \bcycle_j-\eta_j \acycle''_j}  B
= 2\pi\ii \oint_{\bcycle_i} \oint_{\hat C(\y)} B
= 2\pi \ii \oint_{\bcycle_i} \y
= 2\pi\ii \eta_i.  
\eeq
It follows that
\beq
\partial_{\bcycle_i} \mathcal F_0 = \oint_{\bcycle_i} \y.
\eeq

\eproof

\bp[Modular transformations]
Under a change of integer symplectic basis
\beq
\begin{pmatrix}
\tilde \bcycle \cr \tilde\acycle
\end{pmatrix}
=
\begin{pmatrix}
\alpha & \beta \cr \gamma & \delta
\end{pmatrix}
\ . \
\begin{pmatrix}
\bcycle \cr \acycle
\end{pmatrix}
\qquad \text{with}\quad 
\begin{pmatrix}
\delta^T & -\beta^T \cr -\gamma^T & \alpha^T
\end{pmatrix}
.
\begin{pmatrix}
\alpha & \beta \cr \gamma & \delta
\end{pmatrix}
= \text{Id} ,
\eeq
we have
\beq
\tau^{-1}  = - (\delta-\tilde \tau^{-1} \beta)^{-1} (\gamma - \tilde \tau^{-1} \alpha) , 
\eeq
\beq
X = - (\delta-\tilde X \beta)^{-1} (\gamma - \tilde X \alpha) , 
\eeq
\beq
R = (\delta-\tilde X \beta)^T \tilde R (\delta-\tilde X \beta) - (\delta-\tilde X \beta)^T \beta
\eeq
\beq
\tilde \acycle''_i
= \sum_j (\delta - \tilde X \beta)_{i,j} \acycle''_j.
\eeq
\beq
\tilde \bcycle''_i
= \sum_j (\alpha+\beta X)_{i,j} \bcycle''_j.
\qquad , \qquad
\bcycle''_i
= \sum_j (\delta-\tilde X \beta)_{j,i} \tilde \bcycle''_j.
\eeq

\beq\label{eq:F0changeL}
{\mathcal F}_{0,[\tilde \bcycle(\curve,\mathbb Z)]}  
= {\mathcal F}_{0,[\bcycle(\curve,\mathbb Z)]} + 
\pi \ii  \Pi_{\tilde{\mathcal L}}(\hat C(\y)) \cap \Pi_{{\mathcal L}}(\hat C(\y))
= {\mathcal F}_{0,[\bcycle(\curve,\mathbb Z)]} + \pi\ii \sum_{i,j} \tilde t_i \beta_{i,j} t_j.
\eeq

\ep
This implies
\bc
$\mathcal F_0$ depends only on the choice of the Lagrangian submanifold $\bcycle(\curve,\mathbb Z)$, it is independent of the choice of its symplectic complement $\acycle(\curve,\mathbb Z)$, and it is independent of a choice of basis of $\bcycle(\curve,\mathbb Z)$.
\ec
\proof
Indeed, changing $\acycle(\curve,\mathbb Z)$ with fixed $\bcycle(\curve,\mathbb Z)$ is done by $\alpha=\delta=\text{Id}$, $\beta=0$.
And changing the basis of $\bcycle(\curve,\mathbb Z)$ with fixed $\acycle(\curve,\mathbb Z)$ is done with $\beta=0$ and $\alpha$ arbitrary with $\delta=(\alpha^T)^{-1}$. Both have $\beta=0$ so that $\mathcal F_0$ is unchanged.
\eproof

\br The  $\mathcal F_0$ as defined in \cite{EO07} is only a special case of this construction, i.e. a particular choice of Darboux decomposition. Indeed in \cite{EO07}, the spectral curve was assumed algebraic, with $\curve$ compact and equipped with a Torelli marking: a choice of symplectic basis $\acycle_i,\bcycle_i$ of $H_1(\curve,\mathbb Z)$. $B$ was choosen as the "Bergman" kernel, the fundamental 2nd kind differential normalized on $\acycle$-cycles. 
The Lagrangian corresponding to the definition of \cite{EO07} is the one generated by all $\bcycle$-cycles of 1st, 2nd and 3rd kind.

\er

\subsection{Shifted spectral curve}

At each spectral curve we have defined a "tangent" space.
We now want to define tangent vector fields (and not only tangent vectors pointwise), we  need to compare tangent vectors of different spectral curves, we thus need a connexion on the tangent bundle.
In our case the tangent bundle is realized from the bundle of cycles $\mathfrak M_1(\curve,\mathbb Z)$, which is a rigid lattice, thus not deformable, with a trivial connexion. 
This uniquely defines how to transport an integer cycle, and extend a tangent vector to a tangent vector field in a neighborhood of $\spcurve$. This allows to integrate flows. 

Let us do it in details as the following proposition

\bp
Let $\spcurve=[(\curve,\x,\y,B)]$ be a spectral curve, and $\gamma\in \mathfrak M_1(\curve,\mathbb Z)$ an integer cycle.

There exists a radius $R>0$, such that, there exists a unique 1-parameter  family of spectral curves holomorphic on the disc $t\in\mathcal D_{0,R}=\{ |t|<R\}$, with constant base $\curverond_t=\curverond$, and such that
\bea
\spcurve_0 &=& \spcurve \cr
\spcurve_t &=& (\curve_t,\x_t,\y_t,B_t) \cr
\frac{d\y_t}{dt} &=& \oint_\gamma B_t \cr
\frac{d B_t}{dt} &=& \oint_\gamma \omega_{0,3}(\spcurve_t).
\eea
In other words $d/dt=\partial_\gamma$.

We denote it:
\beq
\spcurve_t= \spcurve+t\gamma,
\eeq
or also
\beq
\spcurve_t = e^{t\partial_\gamma}  \spcurve.
\eeq
\ep

\proof
As mentioned at the beginning of section \ref{sec:deform}, there is a subspace $\modspmero{\spcurve}$ of spectral curves with the same base as $\curverond$, which is a differentiable analytic manifold.

Let $\gamma$ denote a representative of the integer cycle $\gamma$ on $\curve$, and 
let $\x(\gamma)$ be its image by $\x$ in a universal cover of the base $\curverond$ with cuts at the branchpoints.

If $\spcurve$ and $\tilde\spcurve$ are two spectral curves in $\modspmero{\spcurve}$, whose branchpoints and poles are not on $\x(\gamma)$, then 
one uniquely defines a cycle $\in \mathfrak M_1(\tilde\spcurve)$, also denoted $\gamma$, by 
\beq
\gamma = {\tilde \x}^*(\x(\gamma)),
\eeq
by first pushing the representative of $\gamma$ to the universal cover of the base $\curverond$, and pull it back to $\tilde\curve$.
There is a neighborhood of $\spcurve$ such that the branchpoints and poles are never on $\x(\gamma)$.

In it, one can  define the tangent vector field $d/dt=\partial_\gamma$, such that  $d\y/dt = \hat B(\gamma)= \oint_\gamma B$.
The spectral curve $\spcurve_t$ is obtained by flowing $\spcurve$ along that tangent vector field.

A radius of convergence is reached typically when a branchpoint, moving with the flow, has to cross $\x(\gamma)$.
Explicit examples will be given in section \ref{sec:examples}.

\eproof

Also, remark that the flows commute ($\partial_\gamma \partial_{\gamma'} \spcurve-\partial_{\gamma'}\partial_{\gamma}\spcurve=0$) and thus
\beq
(\spcurve +t\gamma)+t' \gamma' = (\spcurve +t'\gamma')+t \gamma = \spcurve +(t\gamma+t' \gamma'),
\eeq
except for 3rd kind cycles because these don't commute, we shall study the case of 3rd kind cycles in greater details below in section \ref{sec:satoshift}

\bp[Shifted spectral curve]
For an integer cycle $\gamma\in \mathfrak M_1(\curve,\mathbb Z)$ we have
\beq
\omega_{g,n}(\spcurve+\alpha \gamma) = \omega_{g,n}(\spcurve) + \sum_{m=1}^\infty \frac{\alpha^m}{m!} \overbrace{\int_{\gamma}\dots\int_{\gamma}}^{m\,{\rm integrations}} \omega_{g,n+m}(\spcurve) \ .
\eeq
The series is absolutely convergent in a disc $|\alpha|<R$ with some $R>0$.

This can formally be written:
\beq
e^{\alpha \partial_\gamma} = e^{\alpha\int_{z\in\gamma} \Delta_z}
\eeq
(again a special care is needed for 3rd kind cycles, see section \ref{sec:satoshift}.)
\ep

\proof
This is just Taylor expansion.
\eproof

\subsubsection{Sato shifted spectral curve}
\label{sec:satoshift}

For 3rd kind cycles we define:

\bd[Sato shift]
Let $\gamma_{p_2\to p_1}$ be an integer 3rd kind cycle with boundary divisor $[p_1]-[p_2]$.
For $t$ small enough we have defined the Sato-shifted spectral curve
\beq
\spcurve+t \gamma_{p_2\to p_1}.
\eeq

More generally, let  $D\in {\rm Div}_0(\curve)$ be a divisor $D=\sum_{i=1}^k \alpha_i p_i$ on $\curve$, of degree $\deg D=\sum_i \alpha_i=0$.
We also introduce a norm $||D||^2 = \sum_i |\alpha_i|^2$.

Let us choose a (non-unique) complex linear combination (non commutative and non associative) of integer chains $\gamma_1,\dots,\gamma_k$ with boundary $D$:
\beq
\gamma_D = t_1\gamma_1+(t_2\gamma_2+(t_3 \gamma_3 + \dots+t_k\gamma_k 
) \dots )) \qquad \, \qquad
\partial \gamma_D=\sum_{i=1}^k t_i\partial\gamma_i = D.
\eeq

Thus one may construct, for $||D||$ small enough
\beq
\spcurve+\gamma_D =  ( \dots ((\spcurve+ t_1\gamma_1 )+ t_2\gamma_2 ) + \dots )  + t_k\gamma_k.
\eeq
One should keep in mind that the sum is not associative nor commutative and thus depends on the order of summation. 

\ed

\br
A typical way of choosing such a set of chains, is called a "channel" in CFT: an oriented trivalent tree ending at the $p_i$s and with vertices $v_i$s, the chains being the edges, ordered by following the tree from root to leaves.

\er

\br
As we shall illustrate below, the shifted spectral curve $\spcurve+t\gamma_{p_2\to p_1}$ is often not in the same subspace as $\spcurve$, typically the shift  breaks the symmetries. If $\spcurve$ is a Hitchin system's spectral curve for a group $G$, then most often $\spcurve+\gamma$ will be a Hitchin system's spectral curve for a larger group containing $G$. For example if $G=SL(2,\mathbb C)$, we have $\y^2 = \frac12\Tr \Phi(x)^2$ and $\Tr\Phi(x)=0$, there is a symmetry $\y\to-\y$, while $\td \y=e^{t\gamma_{p_2\to p_1}}\y$ will not have the same symmetry, it will satisfy an equation $\td\y^2+A(x)\td y+B(x)=0$ with $A(x)\neq 0$. In other words $\td y$ is the spectral curve of a $G=GL(2,\mathbb C)$ rather than $G=SL(2,\mathbb C)$ Hitchin system. 
And of course the shift adds new poles, that are simple poles, so generically it changes the cohomology class of $\y$.

In other words, trying to work in a restricted subspace of spectral curves (for instance Hitchin systems $\det(\y-\Phi(x))=0$ with $\Phi(x)\in H^0(\curverond,\text{End}\mathcal E\otimes K) $ with fixed group $G$ and fixed pole divisors for $\Phi$) prevents from performing Sato shifts, and is often damaging regarding the powerful Sato's formalism that we will see below.

\er

\br {\bf Link to the usual Sato notation:}\label{rem:Sato}
Like in remark \ref{rem:Hirota}, write $\y$ in a neighborhood of a pole $p$, in a local coordinate $\xi_p(z)=(\x(z)-\x(p))^{1/{{\rm order}_p(\x)}}$, as
\beq
\y = \sum_{k=0}^{d_p} t_{p,k} \xi_p^{-k-1} d\xi_p + {\rm analytic\,at}\,p
\eeq
with $t_{p,k}$ the KP times.
Then if one considers a 3rd kind cycle $\gamma_{p_2\to p_1}$ with boundary $\partial\gamma_{p_2\to p_1}=[p_1]-[p_2] = \sum_i \alpha_i [p_i]$ at $p_1,p_2$ with $\alpha_1=1=-\alpha_2$, and if some at $p_i$, $i=1,2$ is close to $p$, with local coordinates $z_i=\xi_p(p_i)$, we have, as power series of $z_i$:
\bea
\int_{\gamma_{p_2\to p_1}}  B 
&\sim & \sum_i \alpha_i \frac{d\xi_p}{\xi_p-z_i} + {\rm analytic\,at}\,p  \cr
&\sim & \sum_i \alpha_i \sum_{k=0}^\infty z_i^k \xi_p^{-k-1}  d\xi_p + {\rm analytic\,at}\,p  \cr
\eea
In other words the Sato shift locally (but in general not globally) amounts to
\beq
t_{p,k} \to t_{p,k} + \sum_i \alpha_i z_i^k.
\eeq
This is usually denoted
\beq
t_p \to t_p \pm  [z_i]
\qquad , \qquad
[z_i]=(1,z_1,z_1^2,z_1^3,z_1^4,\dots)
\qquad , \qquad
\pm = \alpha_i = \pm 1.
\eeq

\er

\bc[Sato shift]
Let $\gamma_D$ a third kind cycle with boundary divisor $D$, 
written as a non-commutative complex linear combination of integer chains
\beq
\gamma_D=\sum_i t_i \gamma_{q_i\to p_i}.
\eeq
we have
\beq
\omega_{g,n}(\spcurve+\alpha \gamma_D) = \omega_{g,n}(\spcurve) + \sum_{m=1}^\infty \frac{\alpha^m}{m!} \overbrace{\int_{\gamma_D}\dots\int_{\gamma_D}}^{m\,{\rm times}} \omega_{g,n+m}(\spcurve) \ .
\eeq
The series is absolutely convergent in a disc $|\alpha|<R$ for some $R>0$.

\ec

Notice that if $(g,n+m)\neq (0,2)$, the integrals are independent of the order of integration because $\omega_{g,n+m}$ has no pole at coinciding point, and thus independent of the order in which the relative homologies are defined in $\gamma_D$, i.e. we can ignore the non--associativity and non--commutativity of $\gamma_D$.

In the case $(0,2)$ however, we need to order the points of the divisor $D= \sum_{i=1}^k \alpha_i [p_i]$ and write, in this order
\beq
\gamma_D = \sum_{j=1}^{k-1} \alpha_j \gamma_{p_k\to p_j},
\eeq
and we have (we use the regularized integration along 3rd kind cycles, given  in appendix \ref{app:defint3rdkind})
\beq
e^{\frac12 \int_{\gamma_D}\int_{\gamma_D} \omega_{0,2}}
= \prod_{j=1}^{k-1} \prod_{i=j+1}^{k}  E(\gamma_{p_j\to p_k} - \gamma_{p_i\to p_k} )^{\alpha_i\alpha_j} \prod_{i=1}^k (\ii d\x(p_i))^{-\frac12 \alpha_i^2}
\eeq
where $E$ is the prime form on $\curve$ (see \cite{Fay}).
For example for a single chain $\alpha\gamma_{p_2\to p_1}$, with boundary $D=\alpha [p_1]-\alpha [p_2]$
\beq
e^{\frac12 \int_{\gamma_D}\int_{\gamma_D} \omega_{0,2}}
= \frac{1}{\left( E(\gamma_{p_1\to p_2}) \sqrt{- d\x(p_1)d\x(p_2)} \right)^{\alpha^2} } \ .
\eeq
If the divisor is integer, i.e. all $\alpha_i\in\mathbb Z$, since $\sum_i\alpha_i=0$ we see that $\sum_i \alpha_i^2 \in 2\mathbb Z$ is even, and we have
\beq\label{eqintom02spinor}
\prod_{i} d\x(p_i)^{\alpha_i^2/2} \ e^{\frac12 \int_{\gamma_D}\int_{\gamma_D} \omega_{0,2}}
=  \frac{(-1)^{\frac12 \sum_i \alpha_i^2}}{\prod_{i<j} E(\gamma_{p_i\to p_j})^{-\alpha_i\alpha_j} } ,
\eeq
which is a spinor form.

\subsection{Hirota equations}

\bd
A functional $f$ on the (local meromorphic subspace) space of spectral curves, is said to satisfy Hirota equations iff at all branchpoints $a$, and for every generic $p,p'q,q'$, one has
\beq
\Res_{z\to a} f(\spcurve+\gamma_{q\to z}+\gamma_{q'\to p})\ f(\spcurve+\gamma_{z\to p'}) = 0
\eeq
remember that with \eqref{eqintom02spinor}, the shift by a 3rd kind cycle is a spinor $1/2$-form, and thus the product is a 1-form.
\ed
If we write locally $\spcurve+\gamma_{q\to z}+\gamma_{q'\to p} =\tilde{\spcurve}+[z] + [u] $ and  
$\spcurve+\gamma_{z\to p'} =\tilde{\spcurve}-[z] - [u] $, this would read in the more familiar way:
\beq
\Res_{z\to a} f(\tilde{\spcurve}+[z]+[u])\ f(\tilde{\spcurve}-[z]-[u]) = 0
\eeq

Our goal from now on, will be to try to find a solution to these Hirota equations.

\section{Perturbative definition of the Tau function}

The idea is that we would like to define a "Tau function" -- for the moment we shall call it  "partition function" -- as:
\beq
\mathcal Z(\spcurve)  = e^{\sum_{g=0}^\infty \mathcal F_g(\spcurve)} ,
\eeq
but of course we have to give a meaning to the infinite sum.

Using the homogeneity of $\mathcal F_g$, we may rescale the spectral curve and write
\beq
\mathcal Z(\epsilon^{-1}\spcurve) =  e^{\sum_{g=0}^\infty \mathcal F_g(\epsilon^{-1}\spcurve)} =  e^{\sum_{g=0}^\infty \epsilon^{2g-2} \mathcal F_g(\spcurve)} .
\eeq
In other words, we shall consider the limit of "Large spectral curves" -- by rescaling with $\epsilon^{-1}$ for a small $\epsilon$ --, and write the Tau function and amplitudes as formal series in power of $\epsilon$.
In the context of integrable systems like KdV or KP,  $\epsilon$ is called the "dispersion" parameter, in the context of matrix model it is the inverse size of the matrix $\epsilon=1/N$, in the context of topological strings, it is the string coupling constant $\epsilon=g_s$. In the context of CFTs, the large spectral curve limit is called the "\textbf{heavy limit}".

Therefore we shall here define all our amplitudes as formal series of some formal variable $\epsilon$. We proposed in  \cite{BeEy19} a way to proceed for finite $\epsilon$.

\subsection{Definition of the perturbative partition function}

All definitions, theorems, propositions in this section are valid only in $\mathbb C[[\epsilon]]$, i.e. coefficientwise in the $\epsilon$ expansion (possibly multiplied by an exponential term in some expressions).

\bd[Perturbative partition function]
\beq
\mathcal Z_{[\mathcal L]}(\epsilon^{-1}\spcurve) 
=  \epsilon^{\frac1{12} \deg \x} \ \exp{ \left( \epsilon^{-2} \mathcal F_{0,\mathcal L}(\spcurve) + \mathcal F_1(\spcurve) +  \sum_{g=2}^\infty \epsilon^{2g-2} \mathcal F_g(\spcurve) \right)}.
\eeq
In other words $\epsilon^2\log \mathcal Z_{[\mathcal L]}$ is defined as a formal series.
Recall that $\mathcal F_{0,\mathcal L}$ is defined with a choice of integer Lagrangian submanifold $\mathcal L$ of $\mathfrak M_1(\curve,\mathbb Z)$.
From \eqref{eq:F0changeL}, under a change $\mathcal L\to \tilde{\mathcal L}$ we have
\beq
\frac{\mathcal Z_{[\tilde{\mathcal L}]}(\epsilon^{-1}\spcurve) }{\mathcal Z_{[\mathcal L]}(\epsilon^{-1}\spcurve) }
= e^{\pi \ii  \Pi_{\tilde{\mathcal L}}(\hat C(\y)) \cap \Pi_{{\mathcal L}}(\hat C(\y))}.
\eeq

\ed

As an immediate consequence of theorem \ref{th:deformomgn} we have
\bp \textbf{(Deformations)}
The partition function satisfies 
\beq
\epsilon\ \Delta_z \log \mathcal Z_{[\mathcal L]}(\epsilon^{-1}\spcurve)    =  \sum_{g=0}^\infty \epsilon^{2g-1} \omega_{g,1}(\epsilon^{-1}\spcurve,z) ,
\eeq
and
\beq
\epsilon\ \partial_{\gamma} \log \mathcal Z_{[\mathcal L]}(\epsilon^{-1}\spcurve)   
 = \int_{\Pi_{\mathcal L}(\gamma)} \sum_{g=0}^\infty \epsilon^{2g-1} \omega_{g,1}(\epsilon^{-1}\spcurve,z) .
\eeq
\ep
This coincides with Seiberg-Witten relations for 1st kind deformations, with Miwa-Jimbo for 2nd kind deformations (see section \ref{secMiJi}), and with the perturbative version of Malgrange-Bertola \cite{Bertola} for general deformations. For example, 2nd kind deformations of poles in a Fuchsian $\mathfrak{sl}_n(\mathbb C)$ system coincide with Schlessinger equations.

\bp[Heat kernel equation, proved in \cite{EO07}]
If $\gamma_1$ and $\gamma_2$ are in $\mathcal L$, 
the partition function satisfies
\beq
\left( \partial_{\gamma_1\otimes\gamma_2} \mathcal Z_{[\mathcal L]}(\epsilon^{-1}\spcurve) 
=    \left.
(\epsilon\partial_{\gamma_1}-\eta_1) (\epsilon\partial_{\gamma_2}-\eta_2) \right) \mathcal Z_{[\mathcal L]}(\epsilon^{-1}\spcurve) \right|_{\eta_i = \oint_{\Pi_{\mathcal L}(\gamma_i)}\y } \  .
\eeq
\ep

\proof
Done in \cite{EO07}
\eproof

These are also equivalent to BCOV equations \cite{BCOV}.

\bp[Dilaton equation, proved in \cite{EO07}]
The partition function satisfies
\beq
\epsilon\frac{d Z_{[\mathcal L]}(\epsilon^{-1}\spcurve)}{d\epsilon}
=  \sum_a \Res_{z\to a} F_{0,1}(z)\,\Delta_z \mathcal Z_{[\mathcal L]}(\epsilon^{-1}\spcurve)
\eeq
where $dF_{0,1}=\omega_{0,1}=\y$.

\ep

\bp
For any $\gamma\in \mathfrak M_1(\curve,\mathbb Z)$, 
\beq
\tilde \psi(\epsilon^{-1}\spcurve,\gamma) \overset{\text{def}}{=}
\frac{\mathcal Z_{[\mathcal L]}(\epsilon^{-1}\spcurve + \gamma)}{\mathcal Z_{[\mathcal L]}(\epsilon^{-1}\spcurve)}
= e^{\sum_{n\geq 1}\sum_{g=0}^\infty \frac{\epsilon^{2g-2+n}}{n!}\int_\gamma \dots \int_\gamma \omega_{g,n}(\spcurve)}
\eeq
is independent of $\mathcal L$.
\ep


\subsubsection{Monodromies}

Let $\gamma_1,\gamma_2$ be 2 integer cycles.
Since
\beq
(\partial_{\gamma_1} \partial_{\gamma_2}-\partial_{\gamma_2} \partial_{\gamma_1}) \log  \mathcal Z_{[\mathcal L]}(\epsilon^{-1}\spcurve) = 2\pi\ii\ \gamma_2\cap \gamma_1
\eeq
the parallel transport along $\gamma_1$ and along $\gamma_2$ don't commute, there is a curvature. Since the curvature is constant, the monodromy of $\log \mathcal Z$ is the integral of the curvature
\beq
\frac{ \mathcal Z_{[\mathcal L]}(\epsilon^{-1}\spcurve+t_1\gamma_1+t_2\gamma_2-t_1\gamma_1-t_2\gamma_2)}{ \mathcal Z_{[\mathcal L]}(\epsilon^{-1}\spcurve)} = e^{2\pi\ii \ t_1 t_2 \ \gamma_2\cap \gamma_1}
\eeq
If $t_1=t_2=1$, since intersections of integer cycles are integers, then there is no monodromy.
This shows that for integer cycles
\beq
 \mathcal Z_{[\mathcal L]}(\epsilon^{-1}\spcurve+\gamma_1+\gamma_2)
=  \mathcal Z_{[\mathcal L]}(\epsilon^{-1}\spcurve+\gamma_2+\gamma_1),
\eeq
except for 3rd kind cycles, whose intersections can be half--integer, and for 3rd kind integer cycles one may have a sign $\pm$, in agreement with the fact that it is a spinor (see \eqref{eqintom02spinor}):
\beq
 \mathcal Z_{[\mathcal L]}(\epsilon^{-1}\spcurve+\gamma_1+\gamma_2)
= \pm \mathcal Z_{[\mathcal L]}(\epsilon^{-1}\spcurve+\gamma_2+\gamma_1).
\eeq

\subsection{Loop equations}

\bt[Loop equations]\label{th:Zloopeq} 
(proved in \cite{EO07} and \cite{BEM15})\\
$\mathcal Z_{[\mathcal L]}(\epsilon^{-1}\spcurve)$
is solution of loop equations:
\beq
\mathfrak W_k(\x(z)).\mathcal Z_{[\mathcal L]}(\epsilon^{-1}\spcurve)
\eeq
has no pole at branchpoints, and
\beq
\Omega(z)= \frac{1}{P'(\spcurve;z)} \left(\mathfrak W(\x(z),y).\mathcal Z_{[\mathcal L]}(\epsilon^{-1}\spcurve)\right)_{y=\y(z)} \quad \in H'^1(\curve) .
\eeq

\et

\bp\label{prop:loopeqH1}
If $\gamma\in H'_1(\curve,\mathbb Z)$ is a 1st kind cycle (with possibly pinched cycles) and $t\in \mathbb C$, then
$f_{t\gamma}(\epsilon^{-1}\spcurve) = \mathcal Z_{[\mathcal L]}(\epsilon^{-1}(\spcurve+t\gamma))$ is solution of loop equations.
More generally, if $\gamma\in H'_1(\curve,\mathbb Z)\otimes \mathbb C[[\epsilon]]$, then
$f_{\gamma}(\epsilon^{-1}\spcurve) = \mathcal Z_{[\mathcal L]}(\epsilon^{-1}\spcurve+\gamma)$ is solution of loop equations.

\ep

\proof
\beq
\left. \mathfrak W(x,\epsilon^{-1}y).f(\epsilon^{-1}\spcurve +t\gamma) \right|_{x=\x(z),\,y=\y(z)}
\eeq
can't have poles at branchpoints. 
It could have poles at the poles of $\y$ or at poles of $\hat B(\gamma)$.
If we assume $\gamma$ to be of 1st kind, then $\hat B(\gamma)$ has no poles.
The only remaining possible poles are those of $\y$, and are the same as those without shifting by $\gamma$.

\eproof

\subsection{Definition of the Tau function}

\subsubsection{Motivation for the definition}

However, $\mathcal Z_{[\mathcal L]}(\epsilon^{-1}\spcurve)$ is not yet our Tau function, for 3 reasons:
\begin{enumerate}
\item it has bad modular properties,

\item the would-be Baker-Akhiezer function $\mathcal Z_{[\mathcal L]}(\epsilon^{-1}\spcurve + \gamma_{z_2\to z_1})$
 depends on the homotopy class of path $\gamma_{z_2\to z_1}$, and not just on the 2 boundary points, the divisor  $\partial \gamma_{z_2\to z_1} = [z_1]-[z_2]$.

\item for many spectral curves, it does not satisfy Hirota equations.
Hirota equations are in fact equivalent to the existence of a quantum curve as we shall see below.

\end{enumerate}

To cure the second point, notice that 
the space of integer chains with boundary $[z_1]-[z_2]$ is an affine lattice
\beq
\gamma_{z_2\to z_1} + H_1(\curve,\mathbb Z).
\eeq
The idea would be to consider
\beq
\hat{\mathcal Z}(\epsilon^{-1}\spcurve) = 
\sum_{\mathbf n\in H_1(\curve,\mathbb Z)} \mathcal Z_{[\mathcal L]}(\epsilon^{-1}\spcurve + \mathbf n)
\eeq
but remember that cycles modulo $\Ker\hat B$ are redundant and the sum would diverge.
Therefore we want to sum only over a sublattice $\Lambda \in H_1(\curve,\mathbb Z)$ that is a representent of $H_1(\curve,\mathbb Z)$ modulo $\Ker\hat B$.
So instead we would like to define
\beq
\hat{\mathcal Z}(\epsilon^{-1}\spcurve) = 
\sum_{\mathbf n\in \Lambda } \mathcal Z_{[\mathcal L]}(\epsilon^{-1}\spcurve + \mathbf n).
\eeq
This partition function would be quasi--periodic both modulo $\Lambda$ and $\Ker\hat B$, and its Sato shifted function would depend only on the boundary divisors of the 3rd kind cycle.
Moreover we shall see that under a good choice, it would also solve Hirota equations, and it would have nice modular properties.

\subsubsection{Choice of a Lagrangian decomposition}

We have 2 situations:
\begin{itemize}

\item $\Lambda^\perp = H'_1(\curve,\mathbb Z)\cap \Ker\hat B$
is a lattice, we say that $B$ is rational.

\item $H'_1(\curve,\mathbb Z)\cap \Ker\hat B$ is not a lattice, we say that $B$ is irrational.
By shifting along tangent vectors of type $\partial_{\gamma_1\otimes \gamma_2}$, we can shift $\Ker\hat B$, and eventually, we may assume that we have choosen a spectral curve whose $B$ is rational.

\end{itemize}

The property of being rational or not is unaffected by Rauch deformations.
On the contrary it is changed under $\partial_{\gamma_1\otimes \gamma_2}$ deformations.

From now on, assume that $\spcurve$ is choosen so that  $B$  is rational.

This means that 
\beq
H'_1(\curve,\mathbb Z) = \Lambda \oplus \Lambda^\perp
\qquad , \qquad
\Lambda^\perp \subset \Ker\hat B,
\eeq
where $\Lambda$ and $\Lambda^\perp$ are Lagrangian integer sublattices.

\bd
A characteristic is a pair $\chi=(\Lambda,\nu)$ where $\Lambda$ is an integer  Lagrangian sublattice with a complementary $\Lambda^\perp\subset \Ker\hat B$, and $\nu\in \Lambda^\perp$.
\ed

\subsubsection{Definition of the Tau function}

Having choosen a spectral curve with a rational $B$ and having choosen an integer Lagrangian decomposition, we define the Tau function as

\bd[Tau function]
Given a characteristic $\chi=(\Lambda,\nu)$,
we define the Tau function
\beq
\Tau_{\chi}(\epsilon^{-1}\spcurve)=
\sum_{\mathbf n\in \Lambda}  e^{\pi\ii \ \mathbf n \cap \nu} \ \mathcal Z_{\Lambda}(\epsilon^{-1}\spcurve + \mathbf n ).
\eeq

\ed
We obviously have
\bp\label{propTauperiodic}
It is $H'_1(\curve,\mathbb Z)$ periodic -- up to a sign:
\beq
\forall \, \gamma\in H'_1(\curve,\mathbb Z) \, , \qquad
\Tau_{\chi}(\epsilon^{-1}\spcurve+\gamma)
= e^{\pi\ii \ \gamma \cap \nu} \ \Tau_{\chi}(\epsilon^{-1}\spcurve)
= \pm\  \Tau_{\chi}(\epsilon^{-1}\spcurve)
.
\eeq
\ep
The Tau function can be written as a formal $\epsilon$ power series, whose coefficients are Theta functions, so let us define
\bd[Theta function]
Let $\chi=(\Lambda,\nu) $ a characteristic, 
we  define the Theta function
\beq
\Theta_{\chi}(u,Q)=
\sum_{\mathbf n\in \Lambda} e^{\pi\ii \ \mathbf n \cap \nu} \  e^{ \int_{\mathbf n} u } \ e^{\pi \ii Q(\mathbf n,\mathbf n)}
= 
\sum_{\mathbf n\in \Lambda} e^{\pi\ii \ \mathbf n \cap \nu} \  e^{ \int_{\mathbf n} u } \ e^{\frac12 \int_{\mathbf n} \int_{\mathbf n} B} ,
\eeq
where $u\in \mathfrak M^1(\curve)$ is a meromorphic 1-form, and where $Q$ is a quadratic form, whose imaginary part is positive definite, so that the series is absolutely convergent for any $u\in \mathfrak M^1(\curve)$.
We also define the Theta derivatives
\beq
\Theta_{\chi}^{(k)}(u,Q)=
\sum_{\mathbf n\in \Lambda} e^{\pi\ii \ \mathbf n \cap \nu} \ e^{ \int_{\mathbf n} u } \ e^{\pi \ii Q(\mathbf n,\mathbf n)}\,  \ \overbrace{\mathbf n \otimes \mathbf n \otimes \dots \otimes \mathbf n}^{k}
\quad 
\in \mathfrak M_1(\curve)^{\otimes k}.
\eeq

\ed
An order $k$ Theta derivative can be contracted --with the Poincarr\'e integration pairing-- with a $k$-form on $\curve^k$, typically with some $\omega_{g,k}$.
For short we denote $\Theta^{(1)}=\Theta'$, $\Theta^{(2)}=\Theta''$, $\Theta^{(3)}=\Theta''' \dots $

\bp

Expanding into powers of $\epsilon$, we have, as a formal series of $\epsilon$ whose coefficients are Theta derivatives:
\bea
\Tau_{\chi}(\epsilon^{-1}\spcurve)
&=& \mathcal Z_{\Lambda}(\epsilon^{-1}\spcurve)
\, \Big( \Theta_{\chi}(\epsilon^{-1}\y, \mathcal Q) 
  + \sum_{k=1}^\infty \sum_{2g_i-2+n_i>0, n_i>0} \frac{\epsilon^{\sum_i (2g_i-2+n_i)}}{k!\,\prod_i n_i!} \cr
  && \qquad \left<  \Theta_{\chi}^{(\sum_i n_i)}(\epsilon^{-1}\y, \mathcal Q) , \prod_{i=1}^k \omega_{g_i,n_i}(\epsilon^{-1}\spcurve) \right> \ \Big) .
\eea
\ep

To the first few orders:
\bea
\Tau_{\chi}(\epsilon^{-1}\spcurve)
&=& \mathcal Z_{\Lambda}(\epsilon^{-1}\spcurve)
\, \Big( \Theta_{\chi}(\epsilon^{-1}\y, \mathcal Q) \cr
&& + \epsilon \Big(  \left< \Theta_{\chi}', \omega_{1,1}  \right> + \frac16 \left<  \Theta'''_{\chi} , \omega_{0,3}  \right> \ \Big) \cr
&& + \epsilon^2 \Big(  F_2 \Theta_{\chi} 
+ \frac12 \left< \Theta_{\chi}'', \omega_{1,1} \omega_{1,1} \right> + \frac1{72} \left<  \Theta^{(6)}_{\chi} , \omega_{0,3}\omega_{0,3}  \right> 
+ \frac{1}{6} \left<  \Theta^{(4)}_{\chi} , \omega_{1,1}\omega_{0,3}  \right> \cr
&& + \frac12 \left<  \Theta^{''}_{\chi} , \omega_{1,2}  \right>
+ \frac1{24} \left<  \Theta^{''''}_{\chi} , \omega_{0,4}  \right>
 \Big) \cr
&& + O(\epsilon^3) \Big) .
\eea
This expression was first introduced in \cite{Eyntheta, EMtheta}.
Nice diagrammatic representations of the terms in this series are provided in \cite{EMtheta}, as well as \cite{BE12}.

\subsection{Some properties}

\subsubsection{Loop equations}

Since $\Tau$ is a linear combination of partition functions, as a corollary of theorem \ref{th:Zloopeq} and prop \ref{prop:loopeqH1}
\bt
$\Tau(\epsilon^{-1} \spcurve) $ is solution of loop equations, and so is $\Tau(\epsilon^{-1} \spcurve + \Gamma ) $ for every $\Gamma\in H'_1(\curve)$.
\et

\subsubsection{Modular transformations}

Consider a change of Lagrangian decomposition that doesn't change $\Ker\hat B$:
\bea
\Lambda\to \Lambda'=\alpha \Lambda+\beta \Lambda^\perp , \cr
\Lambda^\perp \to  \Lambda^{\perp'} = \gamma \Lambda^\perp ,
\eea
where $\alpha,\beta,\gamma$ are integer linear maps such that $\alpha^T\gamma=\rm{Id}$ and $\alpha\beta^T=\beta\alpha^T $ is symmetric.

\bp[Proved in \cite{EMtheta}]

\bea
\Tau_{(\Lambda',\nu')}(\epsilon^{-1}\spcurve) = \rho \ \Tau_{(\Lambda,\nu)}(\epsilon^{-1}\spcurve)
\ e^{-\pi\ii t^T \beta t'}
\eea
where, in a basis
\beq
\nu' = \nu + \sum_i (\alpha\beta^T)_{ii} \Lambda^{\perp'}_i
\quad \text{mod} \ 2\Lambda^\perp
\eeq
and $\rho$ is an 8th root of unity.

\ep

\subsubsection{The Baker-Akhiezer function}

\bd
Given $D$ an integer divisor of degree 0, we define the Baker-Akhiezer function, as the Sato shifted Tau-function
\beq
\check\psi_{\chi}(\epsilon^{-1}\spcurve;D) = \frac{\Tau_{\chi}(\epsilon^{-1}\spcurve+\gamma_D)}{\Tau_{\chi}(\epsilon^{-1}\spcurve)} . 
\eeq
It depends only on $D=\partial \gamma_D$, not on the homology class of the 3rd kind cycle $\gamma_D$, thanks to prop. \ref{propTauperiodic}.
It is a spinor form, if $D=\sum_{i=1}^k \alpha_i [z_i]$, we have
\beq
\check\psi_{\chi}(\epsilon^{-1}\spcurve;\sum_i \alpha_i [z_i])
\in H^0(\curve^k, \boxtimes_i K_\curve^{\alpha_i^2/2}).
\eeq
It has poles at $z_i=z_j$ of degree $-\alpha_i\alpha_j$.

\ed
In the formal $\epsilon$ expansion, the first orders coincide with the twisted Szeg\"o kernel:
\bea
\check\psi_{\chi}(\epsilon^{-1}\spcurve;D)
&=& \frac{e^{\frac{1}{\hbar} \int_{\gamma_D} \y } \ e^{- \gamma_D \cap \Pi_\Lambda(\hat C(\y))) }}{E(D)} \ \frac{\Theta_{\chi}(\epsilon^{-1}\y + \hat B(\gamma_D),\mathcal Q) }{\Theta_{\chi}(\epsilon^{-1}\y ,\mathcal Q) } \ \Big( 1 + 
O(\epsilon) \Big)
\eea
where we recall that
\beq
E(\sum_i \alpha_i [z_i]) = \prod_{i<j} E(z_i,z_j)^{-\alpha_i \alpha_j} = e^{-\frac12 \int_{\gamma_D} \int_{\gamma_D} B}.
\eeq

\section{Quantum curve, KZ and Hirota}

Our goal is to show that, appropriately shifted, the Tau function satisfies the Hirota equation.
We do it by first deriving a "quantum curve", a quantization of the spectral curve into a differential operator, annihilating the Sato shifted Tau function (the Baker-Akhiezer function).

\subsection{Preliminaries: Loop equations for $\mathcal Z$}

We notice that a 1st kind deformation $\partial_\Gamma $ of the spectral curve, deforms $P(x,y)$ (\eqref{defPxy} in def \ref{defWk}) only through terms inside its Newton's polygon, i.e. the 1-form $\left. \partial_\Gamma P(x,y)/P_y(x,y) \right|_{x=\x(z), y=\y(z)}$ is a 1st kind form $\in H'_1(\curve)$.
This defines a (non--linear) bijection between 1st kind forms and 1st kind cycles. 
We state it as the lemma

\bl\label{lemmahatOmega}
The map $\hat \Omega$:
\bea
\hat \Omega: \qquad  H'_1(\curve,\mathbb Z)\otimes \mathbb C / \Ker\hat B & \to &  H'^1(\curve,\mathbb C)\cr
 \Gamma & \mapsto & \left. \frac{P(\spcurve-\Gamma;x,y)}{P_y(\spcurve;x,y)}\right|_{x=\x(z), \ y=\y(z)}  = \hat\Omega(\Gamma;z) \cr
\eea
is well defined and invertible in a neighborhood of zero.

\el

\proof
First observe that adding an element of $\Ker\hat B$ to $\Gamma$ doesn't change the right hand side, so the map indeed descends to the quotient by $\Ker\hat B$.

We have
\beq
\hat\Omega(\Gamma+t \gamma) = \hat\Omega(\Gamma) + t \hat B(\spcurve-\Gamma,\gamma) + O(t^2)
\eeq
so the differential is $d\hat\Omega= \hat B_{\spcurve-\Gamma} $, which is invertible with inverse $d\hat\Omega^{-1}  = \hat C_{\spcurve-\Gamma}$.
Therefore, there is a non-empty neighborhood of zero in which the map is invertible.

Loosely speaking, it is solution of the ODE $d\hat\Omega/d\Gamma = \hat B$ or $d\Gamma/d\hat\Omega = \hat C$.
\eproof

From theorem \ref{th:Zloopeq}, we know that the polynomial $\mathfrak W(x,y).\mathcal Z_\Lambda(\epsilon^{-1}\spcurve)$ differs also from the classical spectral curve by a 1st kind deformation in $H'_1(\curve)$, this leads to
\bc For every spectral curve $\spcurve$, there exists 
$\Gamma_{\spcurve}\in H'_1(\curve)$ such that
\beq
\frac{1}{\mathcal Z_{\Lambda}(\epsilon^{-1}\spcurve)} \mathfrak W(x,y) .\mathcal Z_{\Lambda}(\epsilon^{-1}\spcurve)
 = P(\epsilon^{-1}\spcurve-\Gamma_{\epsilon^{-1}\spcurve};x,y)
\eeq

\ec

To leading order in $\epsilon$ we have
\bea
\Gamma_\spcurve = \epsilon \ \hat C\left( \omega_{1,1}(\spcurve;z)-  \sum_{z'\in \x^{-1}(\x(z))-\{z\}} \frac{\omega_{0,2}(\spcurve;z,z')}{\y(z)-\y(z')}  \right) + O(\epsilon^3)
\eea
One can easily check that indeed the 1-form in the brackets has no pole at branchpoints and belongs to $H'^1(\curve,\mathbb C)$.

What we have seen is that in general $\Gamma_{\spcurve}\neq 0$, and thus
\beq
\frac{1}{\mathcal Z_{\Lambda}(\epsilon^{-1}\spcurve)} \mathfrak W(x,y) .\mathcal Z_{\Lambda}(\epsilon^{-1}\spcurve)
\neq 
P(\epsilon^{-1}\spcurve;x,y).
\eeq
However, in almost all cases where quantum curves could be derived in the litterature \cite{BE09,Quantum,MS12,Iwaki,BouchardEynard2016,BEM15,N15}, the 2 polynomials were in fact equal.

Therefore we shall look for a small (of order $O(\epsilon)$) shift that makes them equal.

\subsection{Shift of the spectral curve}

Having choosen an integer Lagrangian decomposition
\beq
H'_1(\curve,\mathbb Z) = \Lambda \oplus \Lambda^\perp 
\qquad , \quad \Lambda^\perp \subset \Ker\hat B,
\eeq
notice that
\beq
H'_1(\curve,\mathbb C) /( \Lambda \oplus \Lambda^\perp)
\eeq
is compact, it is a torus.

\bl For every spectral curve $\spcurve$, there exists $\tilde\Gamma_{\spcurve}\in H'_1(\curve,\mathbb C) /\Lambda^\perp   $ such that
\beq
\frac{1}{\Tau_{\chi}(\epsilon^{-1}\spcurve) }
\mathfrak W(x,y). \Tau_{\chi}(\epsilon^{-1}\spcurve) 
= 
P(\epsilon^{-1}\spcurve-\tilde\Gamma_{\epsilon^{-1}\spcurve};x,y).
\eeq
and
\beq
\tilde\Gamma_{\epsilon^{-1}\spcurve+\mathbf n} = \tilde\Gamma_{\epsilon^{-1}\spcurve}+\mathbf n
\qquad \forall \ \mathbf n \in \Lambda.
\eeq
To leading order we have
\beq
\tilde\Gamma_{\epsilon^{-1}\spcurve} =  \frac{\Theta'_{\chi}(\epsilon^{-1}\y,\mathcal Q)}{\Theta_{\chi}(\epsilon^{-1}\y,\mathcal Q)}   + O(\epsilon)
\eeq

\el

\proof
We write
\bea
&& \frac{1}{\Tau_{\chi}(\epsilon^{-1}\spcurve) }
\mathfrak W(x,y). \Tau_{\chi}(\epsilon^{-1}\spcurve)  \cr
&=& \frac{1}{\Tau_{\chi}(\epsilon^{-1}\spcurve) }\sum_{\mathbf n\in \Lambda} \mathfrak W(x,y). \mathcal Z_{\Lambda}(\epsilon^{-1}\spcurve + \mathbf n) \cr
&=& \frac{1}{\Tau_{\chi}(\epsilon^{-1}\spcurve) }\sum_{\mathbf n\in \Lambda}  \mathcal Z_{\Lambda}(\epsilon^{-1}\spcurve + \mathbf n) P(\epsilon^{-1} \spcurve+\mathbf n-\Gamma_{\epsilon^{-1}\spcurve+\mathbf n};x,y) \cr
&=&   \sum_{\mathbf n\in \Lambda}  \frac{\mathcal Z_{\Lambda}(\epsilon^{-1}\spcurve + \mathbf n)}{\Tau_{\chi}(\epsilon^{-1}\spcurve) } \ P(\epsilon^{-1} \spcurve+\mathbf n-\Gamma_{\epsilon^{-1}\spcurve+\mathbf n};x,y) \cr
&=&  Q(x,y) \cr
\eea
where $Q$ is such that
\beq
\frac{Q(\x(z),\y(z))}{P_y(\epsilon^{-1}\spcurve;\x(z),\y(z))} 
\in H'^1(\curve,\mathbb C)
\eeq
so that we can use lemma \ref{lemmahatOmega}, and deduce the existence of $\tilde\Gamma_{\spcurve}$.
$Q$ is invariant under a shift $\epsilon^{-1}\spcurve \to \epsilon^{-1}\spcurve+\mathbf n$ for $\mathbf n\in \Lambda$, therefore 
\beq
\epsilon^{-1}\spcurve-\tilde \Gamma_{\epsilon^{-1}\spcurve} = 
\epsilon^{-1}\spcurve+\mathbf n-\tilde\Gamma_{\epsilon^{-1}\spcurve+\mathbf n} ,
\eeq
which implies that
\beq
\tilde\Gamma_{\epsilon^{-1}\spcurve+\mathbf n}
= \tilde\Gamma_{\epsilon^{-1}\spcurve} + \mathbf n.
\eeq
\eproof

\bl
The map
$\gamma \mapsto \gamma+\tilde\Gamma_{\epsilon^{-1}\spcurve-\gamma} \mod \Lambda$ 
must have a fixed point that we shall denote $<\epsilon^{-1} \spcurve >$, such that
\beq
\tilde\Gamma_{\epsilon^{-1}\spcurve-<\epsilon^{-1} \spcurve >} \in\Lambda.
\eeq

To leading orders we have
\bea
<\epsilon^{-1} \spcurve >
&=& \epsilon\hat C\Big( -\omega_{1,1}(z)+\frac12 \sum_{z'\in \x^{-1}(\x(z))-\{z\}} \frac{B(z,z')}{\y(z)-\y(z')}\Big) \cr
&& + \frac\epsilon 2 \hat C \left(  \sum_{z'\in \x^{-1}(\x(z))-\{z\}} \frac{1}{\y(z)-\y(z')} 
\ \hat B\boxtimes \hat B( \frac{\Theta''}{\Theta}- \frac{\Theta'}{\Theta}\otimes \frac{\Theta'}{\Theta},z,z') \right) \cr
&& + O(\epsilon^2).
\eea
\el

\proof
Starting from $\gamma_1=0$, define recursively 
\beq
\gamma_{n+1} = \gamma_n+\tilde\Gamma_{\epsilon^{-1}\spcurve-\gamma_n} \mod \Lambda .
\eeq
This defines an infinite sequence in the compact torus $H'_1(\curve,\mathbb C) /( \Lambda \oplus \Lambda^\perp)$, and compactness implies that this sequence must have at least one accumulation point.
The accumulation point must be a fixed point.
\eproof

As an immediate corollary we get (we call it theorem rather than corollary because it is the main result)
\bt\label{th:PloopequalP}
We have
\beq
\frac{1}{\Tau_{\chi}(\epsilon^{-1}\spcurve-<\epsilon^{-1}\spcurve>) }
\mathfrak W(x,y). \Tau_{\chi}(\epsilon^{-1}\spcurve-<\epsilon^{-1}\spcurve>)) 
= 
P(\epsilon^{-1}\spcurve-<\epsilon^{-1}\spcurve>;x,y).
\eeq

\et

In other words, we have shifted the spectral curve $\spcurve \to \spcurve -\epsilon <\epsilon^{-1}\spcurve>$ in such a way that the loop equations polynomial coincides with the classical spectral curve.
The set of spectral curves for which $<\epsilon^{-1}\spcurve>=0 \mod H_1(\curve,\mathbb Z)$ is a discrete subset. 
We call these the "integral" spectral curves.
We shall denote
\beq
\lfloor \epsilon^{-1}\spcurve \rfloor = \epsilon^{-1}\spcurve - <\epsilon^{-1}\spcurve>
\eeq
and call it the "integer part" of $\epsilon^{-1}\spcurve$.
$<\epsilon^{-1}\spcurve>$ is called the "fractional part" of $\epsilon^{-1}\spcurve$.
To summarize, any spectral curve is at distance $O(\epsilon)$ from an integral spectral curve, and we are going to find a quantum curve, for integral spectral curves.

\subsubsection{The Baker-Akhiezer function}

\bd
Given $D$ an integer divisor of degree 0, we define the Baker-Akhiezer function, as the Sato shifted Tau-function
\beq
\psi_\chi(\epsilon^{-1}\spcurve;D) = \check\psi_\chi(\lfloor\epsilon^{-1}\spcurve \rfloor;D) = \frac{\Tau_\chi(\lfloor \epsilon^{-1}\spcurve \rfloor+\gamma_D)}{\Tau_\chi(\lfloor \epsilon^{-1}\spcurve \rfloor)} . 
\eeq
It depends only on $D=\partial \gamma_D$, not on the homology class of the 3rd kind cycle $\gamma_D$.

\ed

Since we shall now work at fixed spectral curve and fixed characteristic, we shall write for short
\beq
\psi(D) = \psi_\chi(\epsilon^{-1}\spcurve;D).
\eeq

\subsection{Quantum curve and KZ equations}

Let $D=\sum_{i=1}^{\ell} \alpha_i [z_i]$ be a divisor of degree 0.
Here we assume that $\curverond=\bar{\mathbb C}$ the Riemann sphere.

\bd
For $i=1,\dots,\ell$, we define
\beq
\psi_{k,\alpha_i}(z_i;D) = \frac{1}{\Tau_{\chi}(\lfloor \epsilon^{-1}\spcurve\rfloor+\gamma_0)}
\mathfrak W'_k(z_i) . \Tau_{\chi}(\lfloor \epsilon^{-1}\spcurve\rfloor+\gamma_0+\gamma_D)
\eeq
and when $k=0$ we denote 
\beq
\psi_{0,\alpha_i}(z_i;D)=\psi(D).
\eeq
$\frac{\psi_{k,\alpha_i}(z_i;D)}{\psi(D)}$
is a $k^{\rm th}$order form in the variable $z_i$ and scalar in the other variables.
\ed

\subsubsection{KZ equations}

\bt[KZ equation]
If $\alpha_i=\pm 1$ we have
\bea
\epsilon \alpha_i  \ d_{z_i} \psi_{k,\alpha_i}(z_i;D)
 = P_{k+1}(\x(z_i)) \psi(D) - \psi_{k+1,\alpha_i}(z_i;D) \cr
 - \epsilon\ \sum_{j\neq i} \alpha_j \frac{\psi_{k,\alpha_i}(z_i;D)-\psi_{k,\alpha_j}(z_j;D)}{\x(z_i)-\x(z_j)}
\eea
In particular
\beq
\epsilon  \alpha_i d_{z_i} \psi(D) =  P_1(\x(z_i)) \psi(D) - \psi_{1,\alpha_i}(z_i;D)
\eeq

\et

\proof
This is the same proof presented in \cite{BouchardEynard2016}. 
It relies on theorem \ref{th:PloopequalP}.



\eproof

\br[KZ equations]
We call it KZ (Knizhnik-Zamlodchikov) equations for reasons that will be apparent below in corollary \ref{corAk}, where it will indeed take the familiar form of KZ equations.

\er

\br[Link with CFT]
The proof in \cite{BouchardEynard2016} needs that
\beq
\alpha_i = \frac{1}{\alpha_i}
\eeq
i.e. $\alpha_i=\pm 1$.
This is reminiscent of CFT, indeed in a more general CFT context with backgroud charge $Q$, one can get closed differential equations only for a vertex operator $\mathcal V_{\alpha_i}(z_i)$ with a "degenerate charge", i.e. $\alpha_i  - \frac{1}{\alpha_i} =\ii  Q$. Here we have $Q=0$.

\er

\bc\label{corpsiDzz}
If we choose $D=[z]-[z'] $, we have
\bea\label{eq:KZ}
\epsilon d_z \psi_{k,+}(z;[z]-[z']) 
&=& P_{k+1}(\x(z)) \psi([z]-[z']) - \psi_{k+1,+}(z;[z]-[z']) \cr
&& + \epsilon\ \frac{\psi_{k,+}(z;[z]-[z'])-\psi_{k,-}(z';[z]-[z'])}{\x(z)-\x(z')} \cr
-\epsilon d_{z'} \psi_{k,-}(z';[z]-[z']) 
&=& P_{k+1}(\x(z')) \psi([z]-[z']) - \psi_{k+1,-}(z';[z]-[z']) \cr
&& + \epsilon\ \frac{\psi_{k,+}(z;[z]-[z'])-\psi_{k,-}(z';[z]-[z'])}{\x(z)-\x(z')} \ .
\eea
\ec

\bc\label{corAk}
There are linear operators, polynomials of $\hat y$ and $\hat y'$
\beq
\hat A_{k,\pm}(x,\hat y; x',\hat y')
\eeq
such that
\bea
\psi_{k,+}(z;[z]-[z']) = \hat A_{k,+}(\x(z),\epsilon d_z; \x(z'),-\epsilon d_{z'}).\psi([z]-[z'])
\cr
\psi_{k,-}(z';[z]-[z']) = \hat A_{k,-}(\x(z),\epsilon d_z; \x(z'),-\epsilon d_{z'}).\psi([z]-[z']).
\eea
They satisfy the recursion
\bea
\hat A_{0,\pm} &=& \text{Id} \cr
\hat A_{k+1,+}(x,\hat y; x',\hat y') 
&=& P_{x+1}(x)  - \hat y \hat A_{k,+}(x,\hat y; x',\hat y') \cr
&& +  \frac{\epsilon}{x-x'}  \left( \hat A_{k,+}(x,\hat y; x',\hat y')
- \hat A_{k,-}(x,\hat y; x',\hat y')  \right) \cr
\hat A_{k+1,-}(x,\hat y; x',\hat y') 
&=& P_{x+1}(x')  -  \hat y' \hat A_{k,-}(x,\hat y; x',\hat y')   \cr
&& +  \frac{\epsilon}{x-x'}  \left( \hat A_{k,+}(x,\hat y; x',\hat y')
- \hat A_{k,-}(x,\hat y; x',\hat y')  \right) \ .
\eea
They are such that
\beq
\hat A_{k,-}(x,\hat y; x',\hat y') = \hat A_{k,+}(x',\hat y'; x,\hat y) .
\eeq

The operators $\hat A_{d,\pm}$ annihilate $\psi([z]-[z'])$:
\beq
\hat A_{d,\pm}(z,\epsilon d_z; z',-\epsilon d_{z'}).\psi([z]-[z'])=0.
\eeq

We have
\beq
\deg_{\hat y} \hat A_{k,+}=k
\qquad , \qquad
\deg_{\hat y'} \hat A_{k,+}\leq k-1.
\eeq

\ec

\proof{Simple computation.}\eproof

Examples:
\beq
\hat A_{0,\pm}=1
\quad , \quad
\hat A_{1,+}=P_1(x)-\hat y ,
\eeq
\beq
\hat A_{2,+}=P_2(x)-\hat y P_1(x) + \hat y^2 + \epsilon \ \frac{P_1(x)-P_1(x')}{x-x'}  -  \frac{\epsilon}{x-x'}(\hat y -\hat y')
\eeq
\bea
\hat A_{3,+}
&=& P_3(x)-\hat y P_2(x)  + \hat y^2 P_1(x) - \hat y^3 + \epsilon \ \frac{P_2(x)-P_2(x')}{x-x'}  \cr
&&  - \epsilon \ \hat y \ \frac{P_1(x)-P_1(x')}{x-x'}  - \ \frac{\epsilon}{x-x'} \ (\hat y P_1(x) - \hat y' P_1(x') ) \cr
&&  + \epsilon \ \hat y \ \frac{\epsilon}{x-x'}(\hat y -\hat y')+ \ \frac{\epsilon}{x-x'} \ ( \hat y^2-\hat y'^2) 
\eea
and so on...

\subsubsection{Quantum curve}

The KZ equations are PDE, they involve both $d/dx$ and $d/dx'$.
By elimination, we shall find an ODE, involving only $d/dx$.

Let us decompose into powers of $\hat y'$:
\beq
\hat A_{d,+} = \sum_{l=0}^{k-1} \hat A_{l}(x,\hat y;x') \hat y'^l
\eeq
\beq
\hat A_{d,-} = \sum_{l=0}^{k} \hat B_{l}(x,\hat y;x') \hat y'^l.
\eeq
and define
\beq
\hat y'^l \psi = \psi^{(l)}(x,x').
\eeq
and the vector 
\beq
\vec \psi = (\psi^{(0)},\dots,\psi^{(d-1)}).
\eeq

\bt
The following $d\times d$ matrix of operators annihilates the vector $\vec \psi(x,x')$:
\beq
\forall\ k=0,\dots,d-1
\quad : \quad
\sum_{l=0}^{d-1} \hat L_{k,l}(x,\hat y;x') \psi^{(l)} = 0.
\eeq
The operators $\hat L_{k,l}$ are defined by the recursion
\beq
\hat L_{0,l} = \hat A_l
\eeq
\beq
\hat L_{k+1,l} = \hat L_{k,l-1} - \epsilon \frac{d\hat L_{k,l}}{dx'}  - \hat L_{k,d-1} \hat B_{l}
.
\eeq
Each $\hat L_{k,l}$ is polynomial of $\hat y$ of degree $\leq d$, with
\beq
\hat L_{k,l}(x,\hat y;x') = \delta_{k,l} \hat y^d  + \sum_{j=0}^{d-1} L_{k,l,j}(x,x') \hat y^j.
\eeq

\et

\proof
Simple computation
\eproof

\bc
The vector of dimension $d^2$, with coordinates  $(\hat y^j \hat y'^k \psi )_{0\leq \leq d-1} $, satisfies a 1st order matricial ODE
\bea
\forall\ j=0,\dots,d-2
\quad : \quad \epsilon \frac{d}{dx} \psi^{(k,j)} &=& \psi^{(k,j+1)} \cr
\epsilon \frac{d}{dx} \psi^{(k,d-1)} &=& -\sum_{0\leq l,j\leq d-1} \hat L_{k,l,j}(x,x') \psi^{(l,j)} .
\eea

\ec

As a consequence, $\psi([z]-[z'])$ satisfies an ODE of order $d^2$, whose coefficients are rational functions of $x$ and $x'$.
The only possible poles, are either at $x=x'$ or at poles of some $P_k(x)$ or $P_k(x')$.
In particular there can be no pole at branchpoints.

\bt[Quantum curve]\label{thqcurve1}

\beq
\dim \Ker \hat A_{d,+} \cap \Ker \hat A_{d,-}  = d^2.
\eeq
A  basis of solutions is given by:
\beq
\psi([z^i(x)]-[z^j(x')])
\eeq
with $z^i(x)$ (resp. $z^j(x')$) the $d$ solutions of $\x(z^i(x))=x$ (resp. $\x(z^j(x'))=x'$).

Moreover, there exists some ordinary differential operators, of order $d^2$ with coefficients rational in $x$ and $x'$, and regular at branchpoints, that annihilate $\psi$,
\beq
\hat R_{d^2,+}(x,\hat y;x').\psi=0
\quad ,\quad
\hat R_{d^2,-}(x;x',\hat y').\psi=0 .
\eeq

\et

\proof
$\hat A_{d,+}$ is a polynomial of $\hat y'$ of degree $d-1$, and $\hat A_{d,-}$ is a polynomial of $\hat y'$ of degree $d$, whose coefficients are polynomials of $\hat  y$ and functions of $x$ and $x'$.
We can eliminate $\hat y'$ and arrive to a polynomial of $\hat y$ only, annihilating $\psi$.
It is a non-commutative resultant of $(x-x')\hat A_{d,+}$ and $\hat A_{d,-}$:
\beq
\hat R_{d^2,+} = \text{Resultant}_{\hat y'}((x-x')\hat A_{d,+},\hat A_{d,-}).
\eeq
It is a polynomial of the coefficients of $\hat A_{d,\pm}$ and thus a polynomial of $\y$. 
The highest power of $\hat y$ is found by keeping only the terms with highest power of $\hat y$ at each step, and since the highest power of $\y$'s coefficient is constant, the highest power of $\hat y$ is found with the commutative resultant and is easily seen to be $\hat y^{d^2}$. 

Therefore $\psi$ satisfies an ODE of order $d^2$ in $x$, and thus the dimension of the space of solutions is at most $d^2$.

Moreover, since the coefficients of $\hat A_{d,\pm}$ are rational functions of $x$ and $x'$, then $\psi([z^i(x)]-[z^j(x')])$ is solution for every pair $(i,j)$.
Moreover, by looking at their leading order in powers of $\epsilon$, these are clearly linearly independent solutions, so the dimension of the space of solutions is at least $d^2$.
Therefore it is $d^2$.

\eproof

Explicit examples of operators $\hat R_{d^2,\pm}$ will be given in section \ref{sec:examples}.

\subsubsection{Abelianization}

\bd[Hurwitz cover]\label{defHurwitzcover}
Within any simply--connected and connected local chart $U\subset\curverond$ that doesn't contain branchpoints nor singularities of $\x$ or $\y$, it is possible to define an ordering of preimages of $\x$, analytic in $U$:
\beq
\x^{-1}(x) = \{z^1(x),z^2(x),\dots,z_d(x)\}.
\eeq
The transition maps from chart $U_1$ to chart $U_2$ are permutations $\sigma_{1,2}\in\mathfrak S_d$.

\ed

\bd
We define the matrix $\mathbf \Psi(x',x)\in GL_d(\mathbb C) \otimes K_{\curverond}^{\frac12} \boxtimes K_{\curverond}^{\frac12} $:
\beq\label{eqdefPsi}
\left( \mathbf \Psi(x',x) \right)_{j,i} = \psi([z^i(x)]-[z^j(x')]).
\eeq
For each given $x'$, it defines a section of a meromorphic spinor $GL_d(\mathbb C)$ bundle over $\curverond=\bar{\mathbb C}$.
It is invertible as a formal series of $\epsilon$.
\ed
Notice that:
\beq
\mathbf \Psi(\epsilon^{-1}\spcurve;
x',x) = - \ \mathbf \Psi(-\epsilon^{-1}\spcurve;x',x)^T
\eeq
in other words, changing the sign of $\epsilon$ changes $x\leftrightarrow x'$ and transposes the matrix.

Near $x\to x'$ we have a simple pole
\beq\label{BPsixdiag}
\mathbf \Psi(x',x) \sim \frac{\sqrt{dx dx'}}{x-x'} \ \text{Id} \  (1+O(x-x')).
\eeq

\bl[Isomonodromy]

After going around a cycle surrounding a singularity of the operators $\hat A_{d,\pm}$ (i.e. poles of $P_k(x)$) define
\beq
S_{\gamma,+}=  \mathbf \Psi(x',x)^{-1} . \mathbf \Psi(x',x+\gamma) 
\eeq
\beq
S_{\gamma,-}^{-1}=  \mathbf \Psi(x',x)^{-1} . \mathbf \Psi(x'+\gamma,x) 
\eeq
Although $\hat A_{d,\pm}$ has a pole at $x=x'$, there is no monodromy around the pole at $x=x'$ due to \eqref{BPsixdiag}.

The monodromy matrices $S_{\gamma,\pm}$ are independent of $x$ and $x'$, they are constant, and they are permutation matrices.
\el

\proof
This can be read on the definition \eqref{eqdefPsi}. A monodromy just permutes the preimages of $x$ (reps. $x'$).
\eproof

\bc
The matrix (defined as a formal power series of $\epsilon$)
\beq
\mathcal D_+(x',x) = \epsilon   d_{x} \mathbf \Psi(x',x) . \mathbf \Psi(x',x)^{-1}
\eeq
\beq
\text{resp.} \qquad \mathcal D_-(x',x) = -\epsilon  \mathbf \Psi(x',x)^{-1} . d_{x'} \mathbf \Psi(x',x)   
,
\eeq
is meromorphic in $x$ (resp. $x'$), and has poles only at poles of $P_k$ and at $x=x'$. They have a simple pole at $x=x'$.
In particular they have no poles at the branchpoints.
They satisfy
\bea
\det(y -\mathcal D_+(x',x)) = P(x,y) + O(\epsilon) \cr
\det(y -\mathcal D_-(x',x))= P(x,y) + O(\epsilon) .
\eea
They have a simple pole at coinciding point with residue $\pm \epsilon \text{Id}$:
\beq
\mathcal D_+(x',x) = - \epsilon \ \text{Id}\ \frac{ dx }{x-x'}  \ (1+O(x-x')) 
\eeq
\beq
{\mathcal D}_-(x',x) = - \epsilon \ \text{Id}\ \frac{ dx' }{x-x'} \ (1+O(x-x')) .
\eeq

\ec

\subsubsection{Quantum curve}

Usually, the quantum curve is obtained at $x'=\infty$.
Let $\infty_1,\dots\infty_l=\x^{-1}(\infty)$, each with multiplicity $d_1,\dots,d_l $.
Let $V_{\infty_j}(z) = \sum_{k=1}^{d_j} \frac{t_{\infty_j,k}}{k} \xi_{\infty_j}^{-k}$ be such that $\y-dV_{\infty_j}$ is analytic at $\infty_j$.
Notice that the following limits exist
\beq
\psi_{j,k}(z) = \lim_{z'\to \infty_j} \left(\frac{\epsilon d}{d\xi_{\infty_j}(z')}\right)^k
 e^{\epsilon^{-1}V_{\infty_j}(z')} \ \psi([z]-[z']) \ .
\eeq
For the same reasons as above, there is a quantum curve, i.e. an ODE satisfied by each $\psi_{j,k}(z)$, whose coefficients are rational functions of $\x(z)$, with poles only at the poles of $\x$ and $\y$, and in particular no pole at branchpoints.

In many cases, this quantum curve equation has degree $d$ instead of $d^2$, because in the limit $x'\to\infty$, the term $1/(x-x')$ tends to vanish and the PDE becomes an ODE.
This was proved for many spectral curves in \cite{BouchardEynard2016}, and it had been proved case by case many times as in \cite{DM14, Quantum, N15, MS15, Iwaki}

\subsection{Miwa-Jimbo equation}\label{secMiJi}

Let $p$ a pole of $\y$, and $t_{p,k} = \frac{1}{2\pi\ii}\oint_{\acycle_{p,k}}\y$ the corresponding 2nd kind times, so that $\y\sim \sum_k t_{p,k} \xi_p(z)^{-k-1}d\xi_p(z)$ near $p$.

\bt[Miwa-Jimbo]
For every 2nd kind time $t_{p,k}$ with $k\geq 1$, we have 
\beq
\epsilon \frac{\partial}{t_{p,k}} \log\Tau 
= \Res_{x\to \x(p)}  \Tr  \frac{\partial T(x)}{\partial t_{p,k}} \, \, \left(\lim_{x'\to x} \frac{1}{x-x'} \Psi(x',x)^{-1} d_x\left((x-x') \Psi(x',x)\right) \right)   .
\eeq
where the diagonal matrix $T(x)$ is a primitive of $\operatorname{diag}(\y(z^i(x)))$:
\beq
dT(x) = \operatorname{diag}(\y(z^i(x))).
\eeq
This implies that our Tau function coincides with Miwa-Jimbo Tau function, possibly up to multiplicative factors that are independent of the 2nd kind times.
These could possibly depend on 1st and 3rd kind times.

\et

\proof

When $x\to x'$ we have on the diagonal 
\beq
\frac{x-x'}{\sqrt{dx dx'}} \ \left(\mathbf \Psi(x',x)\right)_{i,i} \sim 1 + \frac{\epsilon(x-x')}{dx} \Delta_{z^i(x)} \log\Tau + O((x-x')^2)
\eeq
and off diagonal $i\neq j$
\beq
\frac{x-x'}{\sqrt{dx dx'}} \ \left(\mathbf \Psi(x',x)\right)_{i,j} \sim   O(x-x') .
\eeq
Therefore, on the diagonal we have
\beq
\left(\frac{1}{x-x'} \Psi(x',x)^{-1} d_x(x-x') \Psi(x',x) \right)_{i,i} \sim  \frac{\epsilon}{dx} \Delta_{z^i(x)} \log\Tau + O(x-x').
\eeq

We defined an antiderivative of $\y$, such that $dF_{0,1}=\y$. 
Near a pole $p$ it behaves like
\beq
F_{0,1}(z) = - \sum_{k>0} \frac{t_{p,k}}{k} \xi_p(z)^{-k} + t_{p,0} \log{\xi_p(z)} + O(1)
\eeq
Define the diagonal matrix
\beq
T(x) = \text{diag}(F_{0,1}(z^i(x))).
\eeq
By definition we have
\beq
\frac{\partial T(x)}{\partial t_{p,k}} \sim - \text{diag}\left(\frac{\xi_p(z^i(x))^k}{k}\right) + \text{analytic at } p.
\eeq

We have
\beq
\epsilon \frac{\partial}{t_{p,k}} \log\Tau 
= \oint_{\bcycle_{p,k}} \epsilon\Delta_z \log\Tau 
= - \Res_{z\to p} \frac{\xi_p(z)^{-k}}{k} \, \epsilon\Delta_z \log\Tau .
\eeq
Subtracting $F_0$ we have
\beq
\epsilon \frac{\partial}{t_{p,k}} \left(\log\Tau -\epsilon^{-2}F_{0,\Lambda}(\spcurve)\right)
= - \Res_{z\to p} \frac{\xi_p(z)^{-k}}{k} \, \left( \epsilon\Delta_z \log\Tau - \epsilon^{-1}\y(z)\right)
\eeq
where the bracket in the RHS is analytic at $p$, so that we may replace 
\beq
\epsilon \frac{\partial}{t_{p,k}} \left(\log\Tau -\epsilon^{-2}F_{0,\Lambda}(\spcurve)\right)
=  \Res_{z\to p} \frac{\partial F_{0,1}(z)}{\partial t_{p,k}} \, \left( \epsilon\Delta_z \log\Tau - \epsilon^{-1}\y(z)\right).
\eeq
Projecting the $z$ integration contour to the $x$-plane, we get
\bea
&& \epsilon \frac{\partial}{t_{p,k}} \left(\log\Tau -\epsilon^{-2}F_{0,\Lambda}(\spcurve)\right) \cr
&=& \Res_{x\to \x(p)} \sum_i \frac{\partial F_{0,1}(z^i(x))}{\partial t_{p,k}} \,  \left( \epsilon\Delta_{z^i(x)} \log\Tau - \epsilon^{-1}\y(z^i(x))\right) \cr
&=& \Res_{x\to \x(p)}  \Tr  \frac{\partial T(x)}{\partial t_{p,k}} \, \, \left(\lim_{x'\to x} \mathbf\Psi(x',x) - \frac{\sqrt{dx dx'}}{x-x'} - \epsilon^{-1} dT(x) \right)   \cr
&=& \Res_{x\to \x(p)}  \Tr  \frac{\partial T(x)}{\partial t_{p,k}} \, \, \left(\lim_{x'\to x} \frac{1}{x-x'} \Psi(x',x)^{-1} d_x(x-x') \Psi(x',x) - \epsilon^{-1} dT(x) \right)   \cr
\eea
re-adding $F_0$ we get the result.

\eproof

\subsection{Hirota equations}

We consider $\spcurve$ algebraic, with $\curverond=\overline{\mathbb C}$, with $\curve$ compact and $\x,\y$ satisfying a polynomial equation $P(\x,\y)=0$.

\bl\label{lemmaHirota1}
given 4 distinct generic smooth points $p,q,p',q'$ of $\curve$, the following
\beq
\omega(z)= \Tau(\epsilon^{-1}\spcurve+[p']-[z]) \Tau(\epsilon^{-1}\spcurve+[z]-[q]+[p]-[q']) 
 \in K_\curve
\eeq
is a meromorphic 1-form on $\curve$, with simple poles at $p,q,p',q'$, and with poles, order by order in powers of $\epsilon$, at the ramification points, and no other poles
The meromorphic function $f(z)=\omega(z)/d\x(z)$ satisfies an ODE, with rational coefficients $\in\mathbb C(\x(z))[d/d\x(z)]$, having poles at $\x(p),\x(q),\x(p'),\x(q')$, at the poles of $\x$ and $\y$, but no poles at branchpoints.
It satisfies
\beq\label{eq:res0bpHirota2}
\forall\ a\in \mathcal R \ : \qquad \Res_a \omega(z) = 0.
\eeq

\el

\proof
each of the 2 factors of $\omega$ is a 1/2 forms $\in K_\curve^{\frac12}$, so their product is a 1-form.
Moreover the exponential singularities coming from $e^{\frac{1}{\epsilon} \int_z^p \y} \ e^{\frac{1}{\epsilon} \int_q^z \y}$ cancel each other, so the only possible singularities are poles, and thus the product of the 2 Tau is a meromorphic 1-form.
There are obviously simple poles at $z=p,p',q,q'$.
The fact that there can be --order by order in $\epsilon$-- poles at ramification points is because each $\omega_{g,n}$ has such poles.

From theorem \ref{thqcurve1}, we know that each factor satisfies a rational finite order ODE, with coefficients in $\mathbb C(\x(z))$, with poles at the poles of $\x,\y$ and at $\x(p),\x(p'),\x(q),\x(q')$ and no pole at ramification points.
When 2 functions $f(x),g(x)$ obey 2 ODEs of orders $d_1,d_2$, then their product $h(x)=f(x)g(x)$ obeys an ODE of order $d_1 d_2$.
This is because among the derivatives $f^{(i)}g^{(j)}$, at most $d_1 d_2$ are linearly independent over the field $\mathbb C(\x(z))$, and therefore at most $d_1 d_2$ of the $h^{(k)}$ can be linearly independent.
The coefficients are found as linear combinations of the coefficients of the equation for $f$ and $g$, and thus have at most the same poles.

Now, a solution of a linear ODE is singular at most where the coefficients of the ODE are singular, therefore a (true) solution $h(x)$ can have no pole at branchpoints, and thus any contour integral with $\mathcal C_a$  a contour that surrounds the branchpoint $\x(a)$ on $\curverond$ and no other special point, vanishes:
\beq\label{eq:intCahdx0}
\oint_{\mathcal C_a} h(x)dx=0
\eeq
A formal solution, series in powers of $\epsilon$, satisfying the same ODE, can have poles at branchpoints --order by order in powers of $\epsilon$--, because branchpoints are at the boundaries of Stokes sectors.
However, \eqref{eq:intCahdx0} holds to all orders in powers of $\epsilon$, and thus the residue vanishes for formal solutions.
This implies the result.

In other words, Hirota equation is a consequence of the existence of a quantum curve, as is well known in the theory of integrable systems \cite{BBT}.

\eproof

\br
Here we used that the Baker-Akhiezer function is a function of $z$ and not of a homotopy class $\gamma_{q\to z}$, in other words it is meromorphic on $\curve$ rather than on a universal cover.
Lemma.\ref{lemmaHirota1} would fail for $\mathcal Z$ instead of $\Tau$.
\er

As an immediate corollary:
\bt[Hirota equation]\label{thHirota}
The Tau function satisfies Hirota equations
\beq
0 = \sum_a \Res_{z\to a}  \Tau(\epsilon^{-1}\spcurve+[p']-[z]) \Tau(\epsilon^{-1}\spcurve+[z]-[q]+[p]-[q']) .
\eeq
\et

\bc
The Tau function satisfies the determinantal formula
\bea\label{eqHirota2}
\Tau(\epsilon^{-1}\spcurve) \Tau(\epsilon^{-1}\spcurve+[p']-[q]+[p]-[q']) \cr
=\Tau(\epsilon^{-1}\spcurve+[p']-[q]) \Tau(\epsilon^{-1}\spcurve+[p]-[q'])
- \Tau(\epsilon^{-1}\spcurve+[p']-[q']) \Tau(\epsilon^{-1}\spcurve+[p]-[q]) \cr
\eea
and
\beq\label{eqHirota1}
\Delta_p \left(\frac{ \Tau(\epsilon^{-1}\spcurve+[p']-[q]) }{\Tau(\epsilon^{-1}\spcurve)} \right)
=-\ \frac{\Tau(\epsilon^{-1}\spcurve+[p']-[p])}{\Tau(\epsilon^{-1}\spcurve)} \ \frac{\Tau(\epsilon^{-1}\spcurve+[p]-[q])}{\Tau(\epsilon^{-1}\spcurve)}.
\eeq
\ec

This last equation is illustrated schematically by fig.\ref{fighirota} saying that the $\Delta_p$ operator (insertion of a double pole 2nd kind cycle) is in some sense like inserting 2 coalescing simple poles:
\framefig{
\includegraphics[scale=0.13]{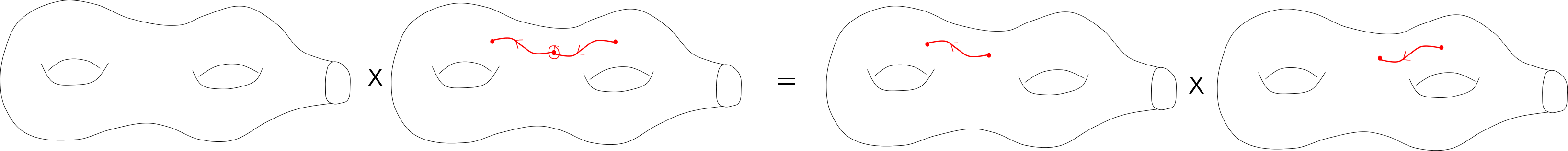}
\caption{Geometric Hirota equation.\label{fighirota}}
}

\proof
The equivalence between Hirota equations and \eqref{eqHirota2} and/or \eqref{eqHirota1} was proved in \cite{BE10}. 
To obtain \eqref{eqHirota2} from Hirota equation, just say that the sum of all residues vanishes, and since the residues at ramification points vanish, it only remains the sum of residues at $z=p,p',q,q'$, which give \eqref{eqHirota2}.
To get \eqref{eqHirota1} from \eqref{eqHirota2}, take the limit $q'\to p$.

\eproof

\section{CFT}

The link between integrable systems and 2D CFT (2-dimensional Conformal Field Theory) has been observed since \cite{JMS-1,JMS-2,JMS-3,JMS-4,JMS-5,DoreyTateo}, and has recently gained a lot of interest with \cite{Lis12,Lis14}.
Here we shall show that the geometric description of integrable systems indeed leads to a model obeying the axioms of CFT \cite{ER14,CER12,BEM15,ABO15}.

\subsection{CFT notations}

The Hirota operator $\Delta_z$, and all operators built from it act on functions of a spectral curve, i.e. on local sections over the local meromorphic space of spectral curves.
We introduce the Sugawara notation \cite{sugawara} borrowed from CFT (Conformal Field Theory).
Any operator $\mathcal O$ acting on $\mathcal Z(\epsilon^{-1}\spcurve)$, will be denoted:
\beq
\mathcal O.\mathcal Z(\epsilon^{-1}\spcurve)
\overset{\text{notation}}{=} \left< \mathcal O \ \mathcal V(\epsilon^{-1}\spcurve) \right>
\eeq
where the symbol $\mathcal V(\epsilon^{-1}\spcurve)$ is called the "generalized vertex operator" associated to the spectral curve $\spcurve$, it just serves to say that we are acting with $\mathcal O$ on the function $\mathcal Z(\epsilon^{-1}\spcurve)$.
The Hirota operator $\Delta_z$ is called the "\textbf{current}" and denoted $J(z)$.
For example we have
\beq
\hat W_1(\epsilon^{-1}\spcurve; z) = \epsilon\Delta_z \mathcal Z(\epsilon^{-1}\spcurve) = \epsilon \left< \mathcal V(\epsilon^{-1}\spcurve) J(z) \right>.
\eeq

\bd[CFT notations]
We denote 
\beq
\Delta_z = J(z)
\eeq
For a cycle $\Gamma\in \mathfrak M_1(\curve)$, we denote
\beq
\mathcal V(\Gamma)  = e^{\int_\Gamma J}.
\eeq
Denoting $\Phi(z)$ a (in general multivalued) primitive of $J(z)$, such that
\beq
d\Phi(z) = J(z),
\eeq
then for a 3rd kind cycle $\gamma_D$ with boundary divisor $D=\sum_i \alpha_i [z_i]$, we have locally
\beq
\mathcal V(\gamma_D) \propto \prod_i e^{ \alpha_i \Phi(z_i)}  d\x(z_i)^{\frac{\alpha_i^2}{2}}
\eeq
(where the proportionality constant depends on the homotopy class of $\gamma_D$ and in particular can contain a $\pm$ sign).
The (multivalued) operator
\beq
\mathcal V_{\alpha_i}(z_i) = 
e^{ \alpha_i \Phi(z_i)} 
\eeq
is called a vertex operator of charge $\alpha_i$ at $z_i$.

We have already introduced the following operators, called $\mathfrak W$-algebra generators
\beq
\mathfrak W_k(x) = \sum_{i_1<\dots<i_k} J(z^{i_1}(x)) \dots J(z^{i_k}(x)),
\eeq
and
\beq
\mathfrak W(x,y) = \sum_k (-1)^k y^{r-k} \mathfrak W_k(x)
= \prod_{i=1}^r (y-J(z^i(x))).
\eeq
In particular
\beq
T(x) = \mathfrak W_2(x)
\eeq
is called the stress energy operator.

Then, taking the summation over $\Lambda \subset H'_1(\curve,\mathbb Z)$, we define
\bea
\hat{\mathcal V}(\epsilon^{-1}\spcurve) 
&=& \sum_{\bcycle\in \Lambda} \mathcal V(\epsilon^{-1}\spcurve+\bcycle)  \cr
&=& \mathcal V(\epsilon^{-1}\spcurve)  \sum_{\bcycle\in \Lambda} \mathcal V(\bcycle) \cr
&=& \mathcal V(\epsilon^{-1}\spcurve)  \sum_{\bcycle\in \Lambda} e^{\oint_{\bcycle} J}
\eea

\ed

We then have
\beq
\mathcal Z(\epsilon^{-1} \spcurve) = \left< \mathcal V(\epsilon^{-1}\spcurve) \right>
\eeq
\beq
\hat W_n(\epsilon^{-1}\spcurve;z_1,\dots,z_n)
 = \left< \mathcal V(\epsilon^{-1}\spcurve) J(z_1) \dots J(z_n) \right>
\eeq

\beq
\Tau(\epsilon^{-1} \spcurve) = \left< \hat{\mathcal V}(\epsilon^{-1}\spcurve) \right>
\eeq

The notations $\mathcal V(\Gamma)$ and $\mathcal V(\epsilon^{-1}\spcurve)$ are consistant, due to 
\beq
\mathcal V(\epsilon^{-1}\spcurve+\Gamma) = \mathcal V(\epsilon^{-1}(\spcurve+\epsilon\Gamma))
= \mathcal V(\epsilon^{-1}\spcurve) \mathcal V(\Gamma)
= e^{\epsilon \oint_\Gamma J} \mathcal V(\epsilon^{-1}\spcurve).
\eeq

\bd[Chiral amplitudes]

The "chiral amplitude map" is a linear map on the algebra of operators generated by the currents $J(z)$ and vertex operators $\mathcal V(\gamma)$, defined by acting on $\mathcal Z(\epsilon^{-1}\spcurve)$.
Chiral amplitudes are denoted with a bracket:
\beq
\left< \prod_i \mathcal O_i \quad \mathcal V(\epsilon^{-1} \spcurve)  \right>
:= \prod_{i} \mathcal O_i . \Tau(\epsilon^{-1}\spcurve).
\eeq

\ed

\subsubsection{OPE}

A CFT is defined by a set of axioms, named OPE and Ward identities.
In this sense, any chiral amplitude (linear form on the algebra of operators) that satisfies the OPEs and Ward identities defines a CFT.

The amplitudes that we have defined do indeed satisfy the axioms of a CFT:

\bt[OPE]

At short distances we have
\beq
\hat{\mathcal V}(\alpha [z]+\alpha' [z'] + D') \mathop{\sim}_{z'\to z} 
(\x(z)-\x(z'))^{\alpha \alpha'}  \hat{\mathcal V}(D')
\eeq
\beq 
J(z) J(z') \mathop{\sim}_{z'\to \sigma(z)} \frac{2 \kappa_{\sigma} d\x(z) d\x(z')}{(\x(z)-\x(z'))^2} + O(1)
\eeq
\beq
J(z) \hat{\mathcal V}(\epsilon^{-1}\spcurve)
\mathop{\sim }_{z\to \text{poles of} \ \y} \y(z) \  (1+o(1))
\eeq

where these equations are just a symbolic notation to say that they hold within any brackets.

\et

Moreover they satisfy 
\bt[Ward identities]
\beq
\bar{\partial} \ \mathfrak W_k(x) = 0
\eeq
meaning that all amplitudes containing a $\mathfrak W_k(x)$ is meromorphic in $x$.
 
\et

The problem in defining a CFT is to prove the existence of an amplitude linear form that satisfies the axioms.
Once the existence is acquired, then uniqueness follows, and the so-called Bootstrap method \cite{coursRibault} allows an explicit computation of nearly any amplitude.

Here we have proved the existence, as a formal $\epsilon$ power series.

\proof
We have
\bea
\left< J(z)J(z') \mathcal V(\Gamma) \mathcal V(\epsilon^{-1}\spcurve) \right>
&=&  \left< J(z)J(z')  \mathcal V(\epsilon^{-1}\spcurve+\Gamma) \right> \cr
&=& \sum_{g\geq 0} \epsilon^{2g} \omega_{g,2}(\epsilon^{-1}\spcurve+\Gamma; z,z') \cr
&=& B(z,z') + \sum_{g\geq 1} \epsilon^{2g} \omega_{g,2}(\epsilon^{-1}\spcurve+\Gamma; z,z') \cr
\eea
where the first term has a double pole at coinciding point, and the other terms are regular at $z=z'$, and the pole behaves as \eqref{normalizedBkappa}.
The other limits are similar.

The ward identities are  the loop equations.

\eproof

\subsection{Real amplitudes}

Chiral amplitudes depend meromorphically on the moduli of the spectral curve, and on the location of vertex operators.
Real amplitudes are bilinear combinations of chiral and anti-chiral amplitudes, which must be real and have no monodromies.

Let us build a possible real amplitude with our formalism, consider a bilinear combination as follows:
\beq
\left<\left< \widehat{\mathcal V}(\epsilon^{-1}\spcurve) \right>\right> 
= \sum_{n,m \in H'_1(\curve,\mathbb Z) } c_{n,m} \left< \mathcal V(\epsilon^{-1}\spcurve+n) \right> \ \overline{\left< \mathcal V(\epsilon^{-1}\spcurve+m) \right>}.
\eeq
If we require that the result is real, we see that the matrix $c$ must be chosen Hermitian:
\beq
c_{m,n} = \overline{c_{n,m}}.
\eeq
If we require that the result has no monodromy we must have $\forall \ k\in H'_1(\curve,\mathbb Z)$
\beq
c_{n+k,m+k} = c_{n,m} ,
\eeq
and thus
\beq
c_{n,m} = c_{n-m,0} = \overline{c_{m-n,0}}.
\eeq
We may thus write
\beq
\left<\left< \widehat{\mathcal V}(\epsilon^{-1}\spcurve) \right>\right> 
= \sum_{k \in H'_1(\curve,\mathbb Z) } c_{k}   \sum_{n \in H'_1(\curve,\mathbb Z), \ n+k=\text{even} } \left< \mathcal V(\epsilon^{-1}\spcurve+\frac12 n+\frac12 k) \right> \ \overline{\left< \mathcal V(\epsilon^{-1}\spcurve+\frac12 n-\frac12 k) \right>}
\eeq
where
\beq
c_{-k} = \overline{c_k}.
\eeq
In particular a diagonal theory is obtained by choosing $c_k=\delta_{k,0}$, and thus
\beq
\left<\left< \widehat{\mathcal V}(\epsilon^{-1}\spcurve) \right>\right> 
=  \sum_{n \in H'_1(\curve,\mathbb Z) } \left| \left< \mathcal V(\epsilon^{-1}\spcurve+ n) \right> \right|^2
\eeq
is a valid real CFT amplitude.

\section{Examples}\label{sec:examples}

Very often it will be convenient to define $\check y=\y/d\x$ so that  $\check y$ is a scalar meromorphic function rather than a 1-form,  and we have
\beq
\y = \check y d\x.
\eeq
Also, in all examples below, $B$ is choosen to be the fundamental 2nd kind form on $\curve$, thus satisfying Rauch variational formula \cite{Rauchvar}.

\subsection{Example KdV}

Consider  the spectral curve -- often called the "\textbf{Airy curve}" :
\beq
\spcurve
= \left\{\begin{array}{l}
\curve=\bar{\mathbb C} , \curverond=\bar{\mathbb C}, B(z_1,z_2) = \frac{dz_1 dz_2}{(z_1-z_2)^2} \cr
\x: z\mapsto z^2 \cr
\check y : z\mapsto z \cr
\y=\check y d\x = 2z^2dz .
\end{array}\right.
\eeq
It has equation
\beq
\check y^2 - \x =0 .
\eeq

\subsubsection{Cycles}
$\x$ and $\y$ have poles only at $z=\infty$, and these are 2nd kind poles, we thus consider  the set of 2nd kind cycles
\beq
\acycle_{\infty,k} = \mathcal C_{\infty}.x^{-k/2}
\quad , \quad
\bcycle_{\infty,k} = \frac{1}{2\pi\ii}\mathcal C_{\infty}. \frac{x^{k/2}}{k} .
\eeq
\beq
\hat B(\acycle_{\infty,k}) = 0
\quad , \quad
\hat B(\bcycle_{\infty,k}) = -z^{k-1} dz
\quad , \quad
\acycle_{\infty,k}\cap \bcycle_{\infty,j} = -\Res_\infty z^{j-1-k}dz = \delta_{k,j} .
\eeq
The only non-vanishing time is
\beq
t_{\infty,3} = \frac{1}{2\pi\ii} \oint_{\acycle_{\infty,3}} \y = -2.
\eeq

Notice that all $\acycle_{\infty,k}$-cycles belong to $\Ker\hat B$, and chosing the integer Lagrangian submanifold generated by all $\bcycle_{\infty,k}$-cycles, we have
$X=0$ and $R=0$, and  $\acycle''_{\infty,k}=\acycle_{\infty,k}$.

\subsubsection{Invariants}

One easily computes from \cite{EO07}
\beq
\omega_{0,3}(\spcurve;z_1,z_2,z_3) = \frac{dz_1 dz_2 dz_3 }{2\ z_1^2 z_2^2 z_3^2}.
\eeq
\beq
\omega_{0,4}(\spcurve;z_1,z_2,z_3,z_4) = \frac34 \prod_{i=1}^4 \frac{dz_i}{z_i^2} \sum_{i=1}^4 \frac{1}{z_i^2}
\eeq
and so on ...
For $n=1$ we have \cite{EO07}
\beq
\omega_{1,1}(\spcurve;z) = \frac{dz }{16\ z^4}
\qquad , \qquad
\omega_{g,1}(z) = \frac{(6g-3)!!\ dz}{2^{5g-1}\ 3^{g}\ g! \ z^{6g-2}} .
\eeq
With our choice of Lagrangian submanifold, we have
\beq
\forall \ g\geq 0 \quad , \quad F_g=0.
\eeq

Since the only branchpoint is at $z=0$, the stable invariants $\omega_{g,n}$ are rational fractions with poles only at $0$, and antisymmetric under $z_i\to -z_i$,  they can therefore be uniquely written as
\beq
\omega_{g,n}(z_1,\dots,z_n) = \sum_{d_1,\dots,d_n} C_{g,n}(d_1,\dots,d_n)  \prod_{i=1}^n \frac{(2d_i-1)!! dz_i}{2^{d_i} z_i^{2d_i+2}} .
\eeq
It was proved in \cite{EO07,zhou13} that the coefficient  $C_{g,n}(d_1,\dots,d_n)$ is actually equal to the Witten--Kontsevich intersection number:
\beq\label{AirywgnWK}
C_{g,n}(d_1,\dots,d_n) = \left<\tau_{d_1}\dots \tau_{d_n} \right>_g  
= \int_{\overline{\mathcal M}_{g,n}} \prod_{i=1}^n c_1(\mathcal L_i)^{d_i} . 
\eeq

\subsubsection{Deformations:}

\begin{itemize}

\item 
Deformations along $\bcycle_{\infty,k}$ for $k\geq 2$ are very easy because $\int_{\bcycle_{\infty,k}} \omega_{0,3} = 0  $ so $B$ gets undeformed, and this allows to keep $\x$ undeformed at fixed $z$, we then get:
\beq
\spcurve+\sum_{k\geq 2} t_{\infty,k} \bcycle_{\infty,k}
= \left\{\begin{array}{l}
\curve=\bar{\mathbb C} , \curverond=\bar{\mathbb C}, B(z_1,z_2) = \frac{dz_1 dz_2}{(z_1-z_2)^2} \cr
\x: z\mapsto z^2  \cr
\y: z\mapsto \left( z-\frac{1}{2}\sum_k  t_{\infty,k} z^{k-2} \right) 2z dz \cr
\end{array}\right.
\eeq
We have
\beq
\partial_{\bcycle_{\infty,k}} = \partial_{t_{\infty,k}}
\quad , \quad
t_{\infty,k} -2\delta_{k,3} = \frac{1}{2\pi\ii} \oint_{\acycle_{\infty,k}} \y.
\eeq
Indeed one easily checks that $\partial_{\bcycle_{\infty,k}} \y = \int_{\bcycle_{\infty,k}} B$ and $\partial_{\bcycle_{\infty,k}} B = \int_{\bcycle_{\infty,k}} \omega_{0,3} = 0  $.
For $k\geq 2$
\beq
\partial_{ t_{\infty,k}} \omega_{g,n} = \frac{1}{k}\Res_{\infty} z^k \omega_{g,n+1}
\eeq

\item
Deformations along $\bcycle_{\infty,1}$ are more subtle because  $\int_{\bcycle_{\infty,1}} \omega_{0,3}(z,z_1,z_2) = -\frac12 \ \frac{dz_1 dz_2}{z_1^2 z_2^2}  $ so $B$ gets deformed, and $\x$ can't be constant at fixed $z$.  
We then get:
\beq
\spcurve+ t_{\infty,1} \bcycle_{\infty,1}
= \left\{\begin{array}{l}
\curve=\bar{\mathbb C} , \curverond=\bar{\mathbb C}, B(z_1,z_2) = \frac{dz_1 dz_2}{(z_1-z_2)^2} \cr
\x: z\mapsto z^2 +t_{\infty,1} \cr
\y: z\mapsto 2z^2 dz  \cr
\end{array}\right.
\eeq
We leave the reader to check that 
$\frac{1}{2\pi\ii} \oint_{\acycle_{\infty,k}} \y =  t_{\infty,1} $ if $k=1$ and $-2$ if $k=3$ and $0$ otherwise, and that $\partial_{t_{\infty,1}} \y = \int_{\bcycle_{\infty,1}} B$ and $\partial_{t_{\infty,1}} B = \int_{\bcycle_{\infty,1}} \omega_{0,3}  $.

\item
Combining all $k$ we have
\beq
\spcurve+ \sum_{k\geq 1} t_{\infty,k} \bcycle_{\infty,k} 
= \left\{\begin{array}{l}
\curve=\bar{\mathbb C} , \curverond=\bar{\mathbb C}, B(z_1,z_2) = \frac{dz_1 dz_2}{(z_1-z_2)^2} \cr
\x: z\mapsto z^2 +\check t_{1} \cr
\y: z\mapsto 2z^2 dz (1-\sum_k \check t_k z^{k-3}) \cr
\end{array}\right.
\eeq
where the coefficients $\check t_1, \check t_2,\check t_3,\dots$ are  convergent series of $t_{\infty,1},t_{\infty,2},t_{\infty,3},\dots,$ in a small disc, $\check t_k = t_{\infty,k} ( 1+\frac{k}2 t_{\infty,1} t_{\infty,k+2} + \dots )$, solutions of
\beq
t_{\infty,k} 
= \sum_{l=0}^\infty  \frac{(-1)^l (k+2l-2)!!}{2^l \ l! \ (k-2)!! }\  \check t_{k+2l} \ (\check t_1)^l 
.
\eeq
We leave to the reader to  check that $\frac{1}{2\pi\ii} \oint_{\acycle_{\infty,k}} \y =  t_{\infty,k} $ for all $k>0$, and that 
$\partial_{t_{\infty,k}} \y = \int_{\bcycle_{\infty,k}} B$ and $\partial_{t_{\infty,k}} B = \int_{\bcycle_{\infty,k}} \omega_{0,3}  $ for all $k>0$.

\item

For 3rd kind cycles we have
\beq
\spcurve+\alpha \gamma_{q\to p}
= \left\{\begin{array}{l}
\curve=\bar{\mathbb C} , \curverond=\bar{\mathbb C}, B(z_1,z_2) = \frac{dz_1 dz_2}{(z_1-z_2)^2} \cr
\x: z\mapsto z^2 +\alpha\left(\frac{1}{\tilde p}-\frac{1}{\tilde q}\right) \cr
\y: z\mapsto \left( z+\frac{\alpha}{2\tilde p(z-\tilde p)} - \frac{\alpha}{2\tilde q(z-\tilde q)} \right) 2z dz \cr
\end{array}\right.
\eeq
where $\tilde p=p+O(\alpha)$ and $\tilde q=q+O(\alpha)$ are series of $\alpha$, convergent in a small disc, solutions of the algebraic equations
\beq
\x(\tilde p) = p^2 \qquad , \qquad \x(\tilde q) = q^2.
\eeq
The radius of convergence is reached at
\beq
\alpha  = \frac{2 \tilde p^3 \tilde q^3}{\tilde q^3-\tilde p^3}.
\eeq

\end{itemize}

\br
Notice that at $\alpha=0$ the spectral curve has a $\mathbb Z_2$ symmetry $\y\to -\y$.
Remark that the symmetry is broken as soon as $\alpha\neq 0$.
This illustrates the fact that the Sato shift generically breaks the symmetries of $\spcurve$.

Typically, the symmetry $\y\to -\y$ corresponds to $\spcurve$ being the spectral curve of a $\mathfrak sl_2(\mathbb C)$ Hitchin system, and thus when $\alpha\neq 0$, $\spcurve_\alpha$ is no more a $\mathfrak sl_2(\mathbb C)$ Hitchin system, but a $\mathfrak gl_2(\mathbb C)$ Hitchin system.

\er

\subsubsection{Tau function}

Since $\curve$ has genus 0, and no double points, we have $H_1(\curve)=H'_1(\curve)=\{0\}$, and thus
\beq
\Tau(\spcurve)=\mathcal Z(\spcurve) = e^{\sum_g F_g}=1.
\eeq
When times $t_{\infty,k}$ are turned on, it can be proved \cite{EO07,zhou13} that we get the KdV tau function (with odd times only)
\beq
\mathcal Z(\spcurve + \sum_k t_{\infty,k}) = \Tau_{\text{KdV}}(t_{\infty,1},t_{\infty,3}, t_{\infty,5},\dots).
\eeq
Combined with \eqref{AirywgnWK}, this is nothing but the Witten--Kontsevich theorem.

\subsubsection{Loop equations}
\beq
\mathfrak W_1(x).\mathcal Z(\spcurve) = 0
\quad , \quad
\mathfrak W_2(x).\mathcal Z(\spcurve) = -x dx^2,
\eeq
\beq
\mathfrak W(x,y).\mathcal Z(\spcurve) = y^2 - x d x^2. 
\eeq
The 1-form
\beq
\Omega(\spcurve;z)(\mathcal Z(\spcurve)) = \left.\frac{y^2-x dx^2}{2 y}\right|_{x=\x(z), \ y=\y(z)} =0,
\eeq
is indeed holomorphic with no poles,  showing that $\Gamma_{\spcurve}=0$.

\subsubsection{Quantum curve and KZ equation}

The KZ equations are then (these were first derived in \cite{BE09}):
\beq
\epsilon d_z\psi(z,z') = - \psi_{1,+}(z,z') 
\eeq
\beq
\epsilon d_z\psi_{1,+}(z,z') = \x(z) \psi(z,z') + \epsilon \frac{\psi_{1,+}(z,z')-\psi_{1,-}(z,z')}{\x(z)-\x(z')}
\eeq
\beq
\epsilon d_{z'}\psi(z,z') = \psi_{1,-}(z,z') 
\eeq
\beq
\epsilon d_{z'}\psi_{1,-}(z,z') = \x(z') \psi(z,z') + \epsilon \frac{\psi_{1,+}(z,z')-\psi_{1,-}(z,z')}{\x(z)-\x(z')}
\eeq
The KZ and quantum curve equations can be solved with the Airy functions and the Airy kernel as done in \cite{BE09}.
Let $A$ and $B$ be solutions of:
\beq
\epsilon^2 \frac{d^2}{dx^2} \ f(x) = x\ f(x),
\eeq
with the asymptotic behaviors at $x\to\infty$
\bea
A(x) \sim \frac{e^{\frac2{3\epsilon} x^{\frac32}}}{\sqrt 2 \ x^{\frac14}} \ \left( 1 + O(x^{\frac{-3}{2}})\right) 
\qquad , \qquad
A'(x) \sim \frac{e^{\frac2{3\epsilon} x^{\frac32}}}{\epsilon \sqrt 2 \ x^{-\frac14}} \ \left( 1 + O(x^{\frac{-3}{2}})\right) \cr
B(x) \sim \frac{- \ e^{-\frac2{3\epsilon} x^{\frac32}}}{\sqrt 2 \ x^{\frac14}} \ \left( 1 + O(x^{\frac{-3}{2}})\right) 
\qquad , \qquad
B'(x) \sim \frac{ e^{-\frac2{3\epsilon} x^{\frac32}}}{\epsilon \sqrt 2 \ x^{-\frac14}} \ \left( 1 + O(x^{\frac{-3}{2}})\right)  .
\eea
These are the Airy and Bairy function respectively, with $x$ rescaled by $\epsilon^{2/3}$.

One can check that
\beq
\psi(z,z')
= \epsilon
 \frac{A(\x(z)) B'(\x(z')) - A'(\x(z)) B(\x(z))}{\x(z)-\x(z')}\ \sqrt{d\x(z) \ d\x(z')}
\eeq
\bea
\psi_{1,+}(z,z') &=&  \Big( \frac{\x(z) A(\x(z)) B(\x(z')) - \epsilon A'(\x(z)) B'(\x(z))}{\x(z)-\x(z')}  \cr
&& - \epsilon\ \frac{A(\x(z)) B'(\x(z')) - A'(\x(z)) B(\x(z))}{(\x(z)-\x(z'))^2} \Big) \sqrt{d\x(z) \ d\x(z')} \cr
\psi_{1,-}(z,z') &=& \Big( \frac{\x(z') A(\x(z)) B(\x(z')) - \epsilon A'(\x(z)) B'(\x(z))}{\x(z)-\x(z')}  \cr
&& + \epsilon\ \frac{A(\x(z)) B'(\x(z')) - A'(\x(z)) B(\x(z))}{(\x(z)-\x(z'))^2} \Big) \sqrt{d\x(z) \ d\x(z')}
\eea
are solutions of the KZ equation \eqref{eq:KZ} of corollary \ref{corpsiDzz}.
The operators of corollary \ref{corpsiDzz} are
\beq
\hat A_0=\text{Id}
\quad , \quad
\hat A_1 = -\hat y 
\eeq
\beq
\hat A_2 = -x +\hat y^2  - \frac{\epsilon}{x-x'}(\hat y -\hat y').
\eeq
We leave to the reader to check that $\hat A_2$ annihilates the Airy kernel $\psi([z]-[z'])$, i.e.
\bea
\left( \epsilon^2 \frac{d^2}{dx^2} -x \right) \psi([z]-[z']) 
=
\left( \epsilon^2 \frac{d^2}{dx'^2} -x' \right) \psi([z]-[z'])  \cr
= \frac{\epsilon}{x-x'} \ \left(\epsilon \frac{d}{dx} +  \epsilon \frac{d}{dx'} \right) \psi([z]-[z'])
\eea

By eliminating the $d/d x'$ derivative, the quantum curve is found to be a 4th order differential operator
\beq
\hat R_{4,+} = (\hat y^2-x)^2 + \frac{2\epsilon}{x-x'} \left((\hat y^2-x)\hat y - \epsilon\right) .
\eeq
In the limit $x'\to \infty$ it reduces to 
\beq
R_{4,+}(x,\hat y;\infty) = (\hat y-x)^2. 
\eeq
Also in the limit $\epsilon\to 0$ it reduces to
\beq
\lim_{\epsilon\to 0}
R_{4,+}(x,\hat y;x') = (\hat y-x)^2. 
\eeq

\subsection{Matrix models and enumeration of maps}

Consider the spectral curve with $\curverond = \bar{\mathbb C}$
\beq
\left\{
\begin{array}{l}
\x(z) = \alpha+\gamma(z+1/z) \cr
\y(z) = \left(\sum_{k=1}^d u_k z^{-k}\right) \x'(z) dz \cr
B(z_1,z_2) = \frac{z_1 dz_2}{(z_1-z_2)^2}
\end{array}
\right.
\eeq
It is topologically the Riemann sphere $\curve = \bar{\mathbb C}$, it has genus 0.
Each $\x(z)$ has 2 preimages $z$ and $1/z$, the degree is 2.
There are 2 ramification points at $z=1$ and $z=-1$, with the corresponding branchpoints $\x(\pm 1) = \alpha\pm 2\gamma$.

$\x$ and $\y$ satisfy a second degree algebraic equation
\beq
P(\x,\y)= \y^2 - \y dV(\x)  + P(\x) d\x^2 = 0
\eeq
where $V$ and $P$ are polynomials of respective degrees $d+1$ and $d-1$, in particular the polynomial $V'(x)$ is given by
\beq
V'(x) = \sum_{k=1}^d u_k T_k\left(\frac{x-\alpha}{\gamma}\right)
\eeq
where $T_k$ is the $k^{\text{th}}$ Chebychev polynomial $T_k(z+z^{-1})=z^k+z^{-k}$.

This spectral curve appears in matrix models and in the enumeration of maps \cite{Eynbook}.

\subsubsection{Cycles and times}

The only non-vanishing 2nd kind times are $t_{p,1},\dots,t_{p,d+1}$ at the 2 poles $p=0$ and $p=\infty$
\beq
t_k =t_{0,k} = -t_{\infty,k}= \frac{1}{2\pi\ii} \oint_{\acycle_{0,k}}\y = \Res_{z\to 0} \x(z)^k \y(z) = - \Res_{z\to \infty} \x(z)^k \y(z).
\eeq 
There is a 3rd kind time
\beq
t = t_{0,0}=-t_{\infty,0} = \Res_{z\to 0} \y(z) = - \Res_{z\to \infty} \y(z) =  \gamma u_1.
\eeq 
These times are such that
\beq
V(x) = \sum_{k=1}^{d+1} \frac{t_{k}}{k} \ x^k,
\eeq
and
\beq
P(x) = t \ t_{d+1}\ x^{d-1} + O(x^{d-2}).
\eeq

Since $\curve$ has genus 0, there is no 1st kind time, we have $H_1(\curve)=0$.
$\curve$ has $d-1$ nodal points and $\dim H'^1(\curve)=d-1$.
The nodal points are the pairs $(p_{i},1/p_{i})$ solutions --different from $(1,1)$ and $(-1,-1)$-- of
\beq
\sum_{k=1}^{d-1} u_k (p_i^k-p_i^{-k}) = 0.
\eeq
The cycles $\acycle_{i} = \acycle_{p_i,0}$ and $\bcycle_i = \gamma_{1/p_{i}\to p_{i}}$ make a basis of $H'_1(\curve,\mathbb Z)$.
Their corresponding $\acycle$-times vanish:
\beq
0 = \oint_{\acycle_{i}} \y .
\eeq
The corresponding $\bcycle$-periods are the Kazakov--Kostov instantons \cite{kkinstantons}:
\beq
\oint_{\bcycle_i} \y = 2 t \log{p_i} + \gamma \sum_k \frac{u_{k+1}-u_{k-1}}{k} \ (p_i^{k}-p_i^{-k}).
\eeq
We have
\beq
\oint_{\bcycle_i} \oint_{\bcycle_j} B = \ln\frac{(p_{i}-p_{j})(1/p_{i}-1/p_{j})}{(p_{i}-1/p_{j}) (1/p_{i}-p_{j})} 
= 2 \ln\frac{p_{i}-p_{j}}{ 1-p_{j}p_j}.
\eeq

We have
\beq
\partial_{t_k} \omega_{g,n} = \Res_{0} x^k \omega_{g,n+1} ,
\eeq
\beq
\partial_{t} \omega_{g,n} = \int_\infty^0 \omega_{g,n+1} . 
\eeq

\subsubsection{Quantum curve}

It is well known that matrix integrals satisfy Hirota equations \cite{kostovhirota, DFGZ,eynbookRMT}, and a quantum curve, however its expression is not so simple and we shall not write it here.
We just mention that the Baker-Akhiezer function is known as the "orthogonal polynomial" and we refer the readers to random matrices lecture books \cite{DFGZ,eynbookRMT}.

\subsubsection{Quadratic case $d=1$}

This case is particularly simple, it corresponds to the Gaussian matrix model.
choose $d=1$, $\alpha=0$, $\gamma=1$, $u_1=1$.
\beq
\left\{
\begin{array}{l}
\x(z) = z+1/z \cr
\y(z) = (z-1/z) \frac{dz}{z^2} \cr
B(z_1,z_2) = \frac{z_1 dz_2}{(z_1-z_2)^2}
\end{array}
\right.
\eeq
It corresponds to $V(x) = \frac{x^2}{2}$ and $P(x)=1$.
$\check y$ satisfies the quadratic equation
\beq
\check y^2 - x \check y +1=0 
\qquad , \qquad
P(x,\y) = \y^2-\x d\x \ \y + d\x^2=0.
\eeq
The times are $t_2=1$, $t=1$ and all other times vanishing.

There is no double point and $H'_1(\curve)=0$.
This is a case where $\Tau=\mathcal Z$ and theorem \ref{th:PloopequalP} is satisfied.
We have
\beq
\frac{1}{\mathcal Z} \mathfrak W(x,y) \mathcal Z = y^2-\x d\x \  y+d\x^2 = P(x,y).
\eeq
The KZ equations are solved \cite{Mul10} with the "Hermite kernel":
\beq
\mathbf \Psi(x',x) = \frac{1}{x-x'} \mathbf \Psi(x')^{-1} \mathbf \Psi(x)
\qquad \text{with} \qquad
\mathbf \Psi(x) = \begin{pmatrix}
H_{\epsilon^{-1}}( x/\sqrt{\epsilon})  & \tilde H_{\epsilon^{-1}}( x /\sqrt{\epsilon}) \cr
H'_{\epsilon^{-1}}( x /\sqrt{\epsilon})  & \tilde H'_{\epsilon^{-1}}( x /\sqrt{\epsilon}) \cr
\end{pmatrix}
 ,
\eeq
where $H_{m}(x)$ is the $m^{\text{th}}$ Hermite polynomial if $m=\epsilon^{-1}$ is integer, or the parabolic cyclindric function if $m$ is not integer, and $\tilde H_m$ is the associated function.
Both are solutions of 2nd order ODE $f''-xf+mf=0$, yielding  the quantum curve equation
\beq
\hat y^2 -x \hat y  + 1.
\eeq

The operator that annihilates $\psi([z]-[z'])$ is 
\beq
\hat R_{4,+} = (\hat y^2-x\hat y) (\hat y^2-x\hat y+1)+\frac{\epsilon}{x-x'} \left( 4\hat y^3 - 6 x \hat y^2 + 2 x^2 \hat y -2\epsilon \hat y \right) .
\eeq

\subsection{Example: Fuchsian sphere with 3 points}

Let $\alpha_0,\alpha_1,\alpha_\infty$ be 3 complex numbers, called the "charges".
Let
\beq
A = \sqrt{\alpha_0^4+\alpha_1^4+\alpha_\infty^4 -2 \alpha_0^2 \alpha_1^2-2 \alpha_0^2 \alpha_\infty^2-2 \alpha_1^2 \alpha_\infty^2 }
\eeq

Consider the spectral curve with $\curverond=\bar{\mathbb C}$, $\curve=\bar{\mathbb C}$ and
\beq
\left\{\begin{array}{lll}
\x(z)
 &=& \frac{\alpha_\infty^2 + \alpha_0^2-\alpha_1^2}{2\alpha_\infty^2} + \frac{A}{4\alpha_\infty^2}\, (z+z^{-1}) \cr
 \y(z) 
 &=& \alpha_0\left(\frac{dz}{z-\ii }-\frac{dz}{z+\ii }\right)
 + \alpha_1\left(\frac{dz}{z-z_1}-\frac{dz}{z-z_1^{-1}}\right)
+ \alpha_\infty \ \frac{dz}{z}
 \cr
B(z,z') &=& \frac{dz dz'}{(z-z')^2}
\end{array}\right.
\eeq
where $z_1$ is such that $\x(z_1)=1$, i.e.
\beq
z_1 =  \frac{(\alpha_1 \pm \alpha_\infty)^2 - \alpha_0^2  }{A}
\eeq
It satisfies the quadratic equation
\beq\label{eqyeqhypergeom}
\y^2 = \frac{\alpha_\infty^2 x^2 - (\alpha_\infty^2 + \alpha_0^2-\alpha_1^2) x + \alpha_0^2}{x^2 (x-1)^2}\ dx^2.
\eeq

\subsubsection{Cycles}

$\y$ has 6 simple poles, and thus we consider the cycles
\beq
\acycle_{0,\pm} = \mathcal C_{\pm\ii}
\quad , \quad
\acycle_{1,\pm} = \mathcal C_{z_1^{\pm 1}}
\quad , \quad
\acycle_{\infty,+} = \mathcal C_{\infty}
\quad , \quad
\acycle_{\infty,-} = \mathcal C_{0}
\eeq
The non-vanishing times are 3rd kind times, they are the charges:
\beq
t_{p,\pm} = \frac{1}{2\pi\ii} \oint_{\acycle_{p,\pm}} \y = \pm \alpha_p.
\eeq
The deformations are the Seiberg-Witten equations
\beq
\partial_{\alpha_p} F_g = \int_{p_-}^{p_+} \omega_{g,1}
\eeq

\subsubsection{Quantum curve}

The hypergeometric functions
\beq
A(x) = x^{\frac{\alpha_0}{\epsilon}} (x-1)^{\frac{\alpha_1}\epsilon} {}_2F_1(\frac12+\frac{\alpha_0+\alpha_1+\alpha_\infty}{\epsilon},\frac12+\frac{\alpha_0+\alpha_1-\alpha_\infty}{\epsilon},1+\frac{2\alpha_0}\epsilon ;x)
\eeq
\beq
B(x) = x^{\frac{-\alpha_0}{\epsilon}} (x-1)^{\frac{-\alpha_1}\epsilon} {}_2F_1(\frac12-\frac{\alpha_0+\alpha_1+\alpha_\infty}{\epsilon},\frac12-\frac{\alpha_0+\alpha_1-\alpha_\infty}{\epsilon},1-\frac{2\alpha_0}\epsilon ;x)
\eeq
both satisfy the differential equation
\beq
\epsilon^2 f'' = \frac{\alpha_\infty^2 x^2 - (\alpha_\infty^2 + \alpha_0^2-\alpha_1^2) x + \alpha_0^2}{x^2 (x-1)^2} \ f 
\eeq
which is the quantum curve associated to \eqref{eqyeqhypergeom}.

One can check that
\beq
\frac{1}{\mathcal Z} \mathfrak W(x,y) \mathcal Z = 
P(x,y) = y^2 -\frac{\alpha_\infty^2 x^2 - (\alpha_\infty^2 + \alpha_0^2-\alpha_1^2) x + \alpha_0^2}{x^2 (x-1)^2}\ dx^2.
\eeq
The KZ equations are solved with the "Hypergeometric kernel":
\beq
\mathbf \Psi(x',x) = \frac{1}{x-x'} \mathbf \Psi(x')^{-1} \mathbf \Psi(x)
\qquad \text{with} \qquad
\mathbf \Psi(x) = \begin{pmatrix}
A(x) & B(x) \cr
A'(x)  & B'(x) \cr
\end{pmatrix}
 ,
\eeq

The Tau function is  the $\mathfrak sl_2(\mathbb C)$, $c=1$ Liouville theory 3 point amplitude on the sphere
\beq
\Tau = \mathcal Z= \left<\mathcal V_{\alpha_0}(0) \ \mathcal V_{\alpha_1}(1) \  \mathcal V_{\alpha_\infty}(\infty) \right> .
\eeq
This was verified in \cite{CER12,ER14}.

\subsection{Example: Fuchsian sphere with 4 points}

Let $z_1,z_2,z_3,z_4$, 4 points on the Riemann sphere, and $\alpha_{z_1},\alpha_{z_2},\alpha_{z_3},\alpha_{z_4}$ 4 complex numbers called the charges.
We denote their biratio
\beq
q=\frac{(z_1-z_2)(z_3-z_4)}{(z_1-z_4)(z_3-z_2)}.
\eeq

Consider the spectral curve with $\curve$ a torus of modulus $\tau$, of equation
\beq
\y^2 = \frac{d\x^2}{\prod_{i=1}^4 (\x-z_i)} \ \left( c+\sum_{i=1}^4 \frac{\alpha_{z_i}^2 \prod_{j\neq i}(z_i-z_j)}{\x-z_i} \right)
\eeq
where $c$ is a constant, called the auxiliary parameter, that we may choose to our will.
$c$ is a function of $\tau$ or vice--versa.

It is often convenient to choose $z_1=0,z_2=q,z_3=\infty,z_4=1$, and then the spectral curve has the equation
\beq
\y^2 = \frac{dx^2}{x(x-1)(x-q)} \left(\alpha_\infty^2 x  +  \frac{q \alpha_0^2}{x} +\frac{(1-q) \alpha_1^2}{x-1}+\frac{q(q-1)\alpha_q^2}{x-q} + c \right).
\eeq

\subsubsection{Cycles and times}

$\y$ has 8 3rd kind poles, that we call $p_{\pm}$ for $p\in\{z_1,z_2,z_3,z_4\}$,  and therefore there are 8 cycles $\acycle_{p_\pm,0}$.
The corresponding times are the charges
\beq
t_{p_\pm} = \frac{1}{2\pi\ii} \oint_{\acycle_{p_\pm,0}} \y = \pm \alpha_p.
\eeq
The deformation with respect to $p\in\{z_1,z_2,z_3,z_4\}$ is a 2nd kind cycle:
\beq
\frac{\partial}{\partial p} = \bcycle_{p_+,1}-\bcycle_{p_-,1}
\qquad \Rightarrow \qquad
dp\ \frac{\partial}{\partial p} = \Delta_{p_+}-\Delta_{p_-}.
\eeq
Moreover, since $\curve$ is a torus, there are 2 1st kind cycles, let us name $\acycle,\bcycle$ a symplectic basis of $H_1(\curve,\mathbb Z)$. 
The corresponding time
\beq
\eta = \frac{1}{2\pi\ii} \oint_{\acycle} \y 
\eeq
is a way of parametrizing the auxiliary parameter $c$, it is called the intermediate charge.

We choose $B$ as the fundamental 2nd kind differential on the torus, normalized on the $\acycle$-cycle, it is then rational, and its variation obeys the Rauch formula.

Deformation equations with respect to charges are the Seiberg-Witten equations
\beq
\partial_{\alpha_p} F_g = \int_{p_-}^{p_+} \omega_{g,1},
\eeq
\beq
\partial_{\eta} F_g = \oint_{\bcycle} \omega_{g,1} ,
\eeq
The deformation with respect to $p\in\{z_1,z_2,z_3,z_4\}$ is
\beq
dp\ \partial_p F_g = \omega_{g,1}(p_+)-\omega_{g,1}(p_-).
\eeq

The deformation of $F_0$ is 
\beq
\partial_{z_i} \partial_{z_j} F_0 = \log \frac{({z_i}_+-{z_j}_+)({z_i}_--{z_j}_-)}{({z_i}_+-{z_j}_-)({z_i}_--{z_j}_+)}
\eeq

This equation, in its Miwa-Jimbo form, is the Schlesinger equation, leading to the Painlev\'e 6 equation.

\subsubsection{Tau function and conformal block}

There is generically no double points, and we have $H'_1(\curve)=H_1(\curve)$ both of dimension 2, and $\Ker\hat B \cap H'_1(\curve)$ is generated by $\acycle$, we choose $\Lambda$ to be generated by $\bcycle$:
\beq
\Lambda = \mathbb Z.\bcycle.
\eeq

For each choosen $\eta$, we write
\beq
\mathcal F_{\Lambda}(\eta) = \mathcal Z_{\Lambda}(\epsilon^{-1}\spcurve)  = \left<\mathcal V_{\alpha_0}(0) \ \mathcal V_{\alpha_1}(1) \  \mathcal V_{\alpha_\infty}(\infty)\ \mathcal V_{\alpha_q}(q) \right>_{\eta}.
\eeq
For each $\eta$ it satisfies the OPE and Ward identities, therefore it is a chiral amplitude for the $\mathfrak sl_2(\mathbb C)$ Liouville theory with 4 points on the sphere.
However it is not modular invariant, it depends on the choice of $\Lambda$, and thus on a choice of channel, in other words it does not satisfy the crossing symmetry.

The Tau function obtained by summing over the lattice $\Lambda$
\beq
\Tau = \sum_{n\in\mathbb Z} \left<\mathcal V_{\alpha_0}(0) \ \mathcal V_{\alpha_1}(1) \  \mathcal V_{\alpha_\infty}(\infty)\ \mathcal V_{\alpha_q}(q) \mathcal V(n\bcycle)  \right>
\eeq
is now modular, and thus satisfies the crossing symmetry. $|\Tau|^2$ is the Liouville theory 4-points amplitude.

%
%

\subsection{Example: Weighted Hurwitz numbers}

This example, coming from the combinatorics of Hurwitz covers, or the combinatorics of certain maps called "constellations" \cite{ACEH}, is interesting because this is an example where the "natural times" coming from the enumerative geometry, are not periods of integer cycles. They are periods of non--integer cycles and that has consequences.

Let $G(z)$ a polynomial of degree $M$, such that $G(0)=1$, written either as a sum or product 
\beq
G(z) = \prod_{k=1}^M (1+c_k z) = 1+\sum_{k=1}^M g_k z^k
\eeq  
and  let $S(z)$ a polynomial of degree $L$, such that $S(0)=0$, written
\beq
S(z)=\sum_{k=1}^L k s_k z^k.
\eeq 
The case $S(z)=z$ with $L=1$ and $s_k=\delta_{k,1}$ is particularly interesting.

Consider the spectral curve computed in \cite{ACEH} (where it was proved that its invariants $\omega_{g,n}$ are generating functions of the weighted Hurwitz numbers of genus $g$ and with a ramification profile of length $n$):
\beq
\left\{
\begin{array}{l}
\curverond=\bar{\mathbb C} \cr
\curve=S^2 \cr
\x(z) = \frac{z}{G(S(z))} \cr
\check y(z) = \frac{S(z)}{z}\ G(S(z))  \cr
B(z_1,z_2) = \frac{dz_1 dz_2}{(z_1-z_2)^2}
\end{array}\right.
\eeq
It satisfies the algebraic equation of degree $LM$:
\beq
\x \check y = S(\x G(\x \check y)).
\eeq

We have
\beq
\y(z)  = \check y(z) d\x(z) = S(z)\ \frac{G(S(z))-z S'(z) G'(S(z))}{G(S(z))} \ \frac{dz}{z} .
\eeq
The ramification points are the $LM$ solutions of $G(S(z)-z S'(z) G'(S(z))=0$.
$\x$ and $\y$ have poles at $\infty$, at $0$, and at all the $LM$ points $p_{k,l}$ such that 
\beq
S(p_{k,l})=\frac{-1}{c_k} .
\eeq

\subsubsection{Cycles and times}

$\y$ has a pole of degree $L+1$ at $z=\infty$, and all the other poles are generically simple poles.
$\x$ has a zero of order $LM-1$ at $z=\infty$, so that the local parameter is $\xi_\infty(z)=\x(z)^{\frac{1}{LM-1}}$, and generically simple poles at the $p_{k,l}$s, where $\xi_{p_{k,l}}(z)=\x(z)^{-1}$ is the local parameter.

The times with respect to the corresponding $\acycle$--cycles are:
\beq
t_{p_{k,l},0} = \Res_{p_{k,l}} \y = \Res_{p_{k,l}} S(z) \frac{d\x(z)}{\x(z)} = \frac{1}{c_k}.
\eeq
At $\infty$ we have
\beq
t_{\infty,k} = \Res_{\infty} \y(z) \x(z)^{\frac{k}{LM-1}}
= \Res_{\infty} S(z) \ \x(z)^{\frac{k}{LM-1}} \frac{d\x(z)}{\x(z)}
\eeq
they are the Laurent expansion coefficients of $S(z)$:
\beq
S(z) \sim \frac{1}{LM-1}\ \sum_{k=0}^{LM-1} t_{\infty,k} \xi_\infty(z)^{-k} + \text{analytic at }z\to\infty .
\eeq

In the case $S(z)=z$ we have for example
\beq
t_{\infty,1}=\left(\prod_{i=1}^M c_i\right)^{\frac{1}{1-M}}.
\eeq

\subsubsection{Deformations}

Under a deformation $\delta$ of the $c_k$s (or the $g_k$s) and the $s_k$s we have
\bea
\delta \y 
&=& \delta S(z) \frac{dz}{z} + d \left( \left( \delta \log G(u)\right)_{u=S(z)} \right) \cr
&=& \sum_{k=1}^L \delta s_k \ d(z^{k} )
+ \sum_{k=1}^M \delta c_k \ d\left(\frac{S(z)}{1+c_k S(z)}\right) \cr
&=& -\sum_{k=1}^L \delta s_k \ \Res_{z'=\infty} B(z,z') z'^k
- \sum_{k=1}^M \delta c_k \sum_{l=1}^L \Res_{z'=p_{k,l}} B(z,z') \frac{S(z')}{1+c_k S(z')} \cr
\eea

It follows that the deformations with respect to the $s_k$s and $c_k$s are generated by the 2nd kind cycles
\beq
\bcycle_{s_k} = \frac{-1}{2\pi\ii} \mathcal C_{\infty}.z^k
\eeq
\beq
\bcycle_{c_k} = \frac{-1}{2\pi\ii} \sum_{l=1}^L \mathcal C_{p_{k,l}}.\frac{S(z)}{1+c_k S(z)} .
\eeq
These generate a Lagrangian submanifold $\mathcal L$, isomorphic to the tangent space.
But they are not integer cycles.

For example, if $S(z)=z$, we have
\beq
\bcycle_{c_k} = \frac{G'(-1/c_k)}{c_k}\ \bcycle_{\frac{-1}{c_k},1} \qquad \text{mod}\ \Ker\hat B,
\eeq
which gets deformed under deformations of the $c_k$s.
There is a connection
\bea
j\neq k: \qquad && \partial_{c_j} \bcycle_{c_k} = \frac{c_k}{c_k-c_j} \bcycle_{c_k} \qquad \text{mod}\ \Ker\hat B, \cr
j=k: \qquad && \partial_{c_k} \bcycle_{c_k} = \frac{-1}{c_k} \left( 1- \frac{G''(-1/c_k)}{c_k G'(-1/c_k)}\right) \bcycle_{c_k} \qquad \text{mod}\ \Ker\hat B, \cr
\eea

As a consequence we have as usual that the derivative is an integral with a cycle:
\beq
\partial_{c_k} \omega_{g,n} = \oint_{\bcycle_{c_k}} \omega_{g,n+1}
\eeq
but a second derivative takes an extra term
\beq
\partial_{c_j}\partial_{c_k} \omega_{g,n} = \oint_{\bcycle_{c_k}} \oint_{\bcycle_{c_j}} \omega_{g,n+2} + \oint_{\partial_{c_j}\bcycle_{c_k}}  \omega_{g,n+1}.
\eeq

\subsubsection{Tau function and quantum curve}

Notice that the curve has genus zero, and no double points, and thus 
\beq
H_1(\curve)=H'_1(\curve)=\{0\}.
\eeq
Therefore
\beq
\mathcal Z = \Tau.
\eeq
It is proved in \cite{ACEH} that the Baker--Akhiezer function
\beq
\psi(z) = \frac{e^{\epsilon^{-1} \int^z \y} }{\sqrt{\x'(z)}} \ e^{\sum^{\text{stable}}_{g,n} \frac{\epsilon^{2g-2+n}}{n!} \int_\infty^z \dots \int_\infty^z \omega_{g,n}}
\eeq
satisfies the quantum curve ODE:
\beq
\left(x\epsilon \frac{d}{dx} - S(x G(x\epsilon\frac{d}{dx})) \right).\psi=0 .
\eeq

\subsubsection{Simple Hurwitz numbers}

Chosing $S(z)=z$, and $c_k=\frac{1}{M}$ and letting $M\to\infty$, corresponds to $G(z)=e^z$,  the spectral curve is then
\beq
\left\{
\begin{array}{l}
\curverond=\bar{\mathbb C} \cr
\curve=\mathbb C \cr
\x(z) = z e^{-z} \cr
\check y(z) = e^z \qquad \longrightarrow \qquad \y(z)=(1-z)dz \cr
B(z_1,z_2) = \frac{dz_1 dz_2}{(z_1-z_2)^2}
\end{array}\right.
\eeq
It satisfies:
\beq
\x = \x\check y \ e^{-\x\check y}
\eeq
or equivalently
\beq
\x\check y = L(\x)
\eeq
where $L$ is the Lambert function.

It is proved in \cite{MS12} that $\psi$ satisfies the quantum curve equation
\beq
\frac{d}{dx} \psi(qx) = x\psi(x).
\eeq
with $q=e^{-\epsilon}$.

The unique ramification point is at $z=a=1$, with $\x(a)=e^{-1}$.

It is well known \cite{BM07,EMS09,DKOSS13} that the invariants are the generating functions of Hurwitz numbers and satisfy the ELSV formula
\bea
\omega_{g,n}(z_1,\dots,z_n)
&=& \sum_\mu H_g(\mu) \prod_{i=1}^n \mu_i\x(z_i)^{\mu_i-1}d\x(z_i) \cr
&=& \sum_{d_1,\dots,d_n} \left< \Lambda(-1) \tau_{d_1}\dots \tau_{d_n} \right>_g \prod_{i=1}^n d\zeta_{d_i}(z_i) 
\eea
where 
\beq
\zeta_d(z) = (\x(z) d/d\x(z))^{d+1} z = \sum_{\mu=1}^\infty \frac{(-1)^{\mu-1} \mu^{d+\mu}}{\mu!} \x(z)^{\mu}
\eeq
and $\Lambda(\alpha)=\sum_k (-\alpha)^k c_k(\mathbb E)$ is the Hodge class: the sum of Chern classes of the Hodge bundle $\mathbb E\to \overline{\mathcal M}_{g,n}$.

\section*{Conclusion}

We have presented a geometric formulation of integrable systems, based on homology cycles.
Here, we have defined the Tau function and all amplitudes as formal power series of $\epsilon$.
In \cite{BeEy19} we present the analogous formulation for finite $\epsilon$.
The notion of spectral curve as a branched cover gets lost for finite $\epsilon$, however, the notions of homology and cohomology survive, almost unaffected, and most of the cycle-formalism survives for finite $\epsilon$.

Beyond, we shall in some upcoming article, also define a notion of non-commutative spectral curve, and quantum Tau-function and amplitudes, for which most of the formalism continues to hold.

\section*{Acknowledgments}

This work is supported by the ERC synergy grant ERC-2018-SyG  810573, ``ReNewQuantum''.
The begining of this work was also supported by the ERC Starting Grant no. 335739 ``Quantum fields and knot homologies'' funded by the European Research Council under the European Union's Seventh Framework Programme. 
It is also partly supported by the ANR grant Quantact : ANR-16-CE40-0017.
I wish to thank IHES for its hospitality.
I wish to thank the Centre de Recherches Math\'ematiques de Montr\'eal, the FQRNT grant from the Qu\'ebec government.
I wish to thank M. Mulase, because the first time I thought about this geometric description was when I wrote a lecture on topological recursion given at IPMU Tokyo in 2008, and I discussed it with M. Mulase in the Shinkansen between Tokyo and Kyoto.
I also thank G. Borot with whom I developed several of these ideas, many of which are presented in a common work with him in \cite{BE10}.
I also thank O. Babelon, R. Belliard, L. Chekhov and S. Ribault for numerous discussions.

\section*{Appendix}

\appendix

\section{Integration pairing for 3rd kind cycles}
\label{app:defint3rdkind}

Let $\gamma_{q\to p}$ a 3rd kind cycle, or in fact a Jordan arc representative, starting at $q$ and ending at $p$.

let $o$ and $o'$ some smooth points along the path $\gamma_{q\to p}$ such that
\beq
\gamma_{q\to p} = \gamma_{q\to o'}+\gamma_{o'\to o}+\gamma_{o\to p}
\eeq
and such that $o$ (resp. $o'$) is in a vicinity of $p$ (resp. $q$) where $\xi_p=(\x-\x(p))^{1/{\rm order}_\x(p)}$ (resp. $\xi_q=(\x-\x(q))^{1/{\rm order}_\x(q)}$) is a well defined local coordinate. Then we define
\beq
t_p(\omega) = \Res_p \omega 
\quad , \quad 
V_p(\omega)(z) = \Res_{z'\to p} \omega(z') \, \ln{\left(1-\frac{\xi_p(z')}{\xi_p(z)}\right)}
\eeq
Our definition of the pairing is then:
\bea
\oint_{\gamma_{q\to p}} \omega = <\gamma,\omega> 
&\overset{{\rm def}}{=}& \int_{\gamma_{o'\to o}} \omega \cr
&& + \int_{o\to p} \left( \omega - t_p(\omega) \frac{d\xi_p}{\xi_p}  - dV_p(\omega) \right) \cr
&& - V_p(\omega)(o) - t_p(\omega) \ln{\xi_p(o)} \cr
&& - \int_{o'\to q} \left( \omega - t_q(\omega) \frac{d\xi_q}{\xi_q}  - dV_q(\omega) \right) \cr
&& + V_q(\omega)(o') + t_q(\omega) \ln{\xi_q(o')} \ .
\eea
This definition of the pairing is independent of the choice of $o$ and $o'$.

If $\omega$ has no pole at $q$ and $p$ we simply have the usual integration along the path $\gamma_{q\to p}$.
\beq
<\gamma_{q\to p} ,\omega > = \int_{\gamma_{q\to p}} \omega.
\eeq

\section{Proof of lemma \ref{Lem:defChat}}
\label{app:defChat}

\bl
The map:
\bea
\hat C: \mathfrak M^1(\curve) &\to & \mathfrak M_1(\curve) \cr
\omega & \mapsto & 
\hat C(\omega)
\eea
is linear and 
is independent of a choice of fundamental domain and of the choice of $o'_i$.

Moreover it satisfies for every $\omega$:
\beq\label{eqapp:BCId}
\omega = \hat B(\hat C(\omega)).
\eeq
\el

\proof
First, observe that in the last line $\tilde f$ depends on the choice of $o'_i$. Changing $o'_i$ changes $\tilde f\to \tilde f+c_i $ by a constant $c_i$ on $\curve_i$, and the last term changes by $\sum_i \frac{c_i}{2\pi\ii}\sum_{e\in \partial\curve_i } \mathcal C_e=0$.
This proves also that $\hat C$ is linear.

The let us prove \eqref{eq:BCId}, this is Riemann bilinear identity:
Write that
\beq
2\pi\ii \ \tilde\omega(z) = 2\pi\ii\ d\tilde f(z) =  \oint_{z'\in \mathcal C_z} \tilde f(z')\ B(z,z'),
\eeq
and deform the integration contour homotopically from a small circle around $z$ to the boundary of the fundamental domain 
$\sum_{e=\text{internal edges}} (e_+ - e_-)  + \sum_{e\in \partial\curve} e$.
Notice that for internal edges, $B$ takes the same value on $e_+$ and $e_-$, and $\tilde f|_{e_+}-\tilde f|_{e_-}=\int_{e^\perp} \tilde \omega$ is constant along $e$, this implies that
\bea
2\pi\ii \ \tilde \omega
&=& \sum_{e=\text{internal edges}} (\oint_{e^\perp} \tilde{\omega}) \int_e B 
 + \sum_{e\in \partial\curve} \oint_{\mathcal C_e} \tilde f B,
\eea
and thus implies \eqref{eq:BCId} that $\omega=\hat B(\hat C(\omega))$.

Let $v$ be a vertex of $\Gamma$ (possibly some $o_i$).
Appart form the last line (independent of the vertices positions), 
$\hat C(\omega)$ is built as a sum of edges, i.e. 3rd kind cycles, and it could have boundaries at vertices, i.e. $\partial \hat C(\omega)$ can be a divisor at the vertices:
\beq
\partial \hat C(\omega) = \sum_v \partial_v \hat C(\omega)\ [v].
\eeq
This would imply that $\hat B(\hat C(\omega))$ would have poles with residues at $v$ equal to:
\beq
\partial_v \hat C(\omega) = \Res_v \hat B(\hat C(\omega)) = \Res_v \omega = 0.
\eeq
This proves that $\hat C(\omega)$ has no boundary at the vertices, in other words that $\hat C(\omega)$ is invariant under homotopic deformations of the graph in particular infinitesimal deformations of the $o_i$s and all vertices of $\Gamma$.

It remains to show that $\hat C(\omega)$ is invariant under topological changes of graphs.
It suffices to consider flop transitions:
\begin{center}
\includegraphics[scale=0.4]{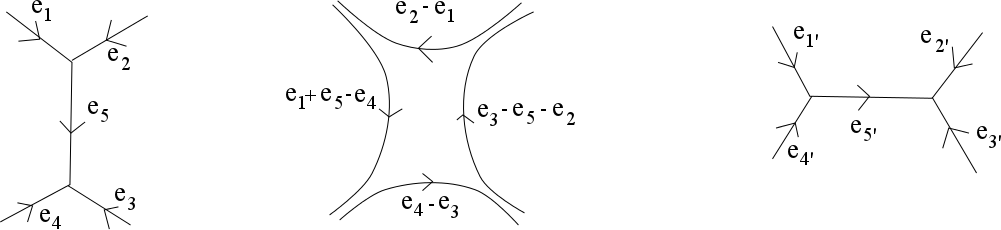}
\end{center}
We have (using $e_5^\perp = e_1^\perp+e_2^\perp=-e_3^\perp-e_4^\perp$):
\begin{multline}
(\oint_{e_1^\perp} \tilde\omega) e_1
+ (\oint_{e_2^\perp} \tilde\omega) e_2
+ (\oint_{e_5^\perp} \tilde\omega) e_5
+ (\oint_{e_3^\perp} \tilde\omega) e_3
+ (\oint_{e_4^\perp} \tilde\omega) e_4 \\
=  (\oint_{e_1^\perp} \tilde\omega) (e_1+e_5-e_4)
 - (\oint_{e_2^\perp} \tilde\omega) (e_3-e_5-e_2)
+ (\oint_{e_1^\perp} \tilde\omega+\oint_{e_4^\perp} \tilde\omega) (e_4-e_3) \\
=  (\oint_{e_1^\perp} \tilde\omega) (e'_1-e'_4)
 - (\oint_{e_2^\perp} \tilde\omega) (e'_3-e'_2)
+ (\oint_{e_1^\perp} \tilde\omega+\oint_{e_4^\perp} \tilde\omega) (e'_4+e'_5-e'_3) \\
= (\oint_{e_1^\perp} \tilde\omega) e'_1
+ (\oint_{e_2^\perp} \tilde\omega) e'_2
+ (-\oint_{e_2^\perp}-\oint_{e_1^\perp}-\oint_{e_4^\perp} \tilde\omega) e'_3
+ (\oint_{e_4^\perp} \tilde\omega) e'_4
+ (\oint_{e_1^\perp} \tilde\omega+\oint_{e_4^\perp} \tilde\omega) e'_5 \\
= (\oint_{e_1'^\perp} \tilde\omega) e'_1
+ (\oint_{e_2'^\perp} \tilde\omega) e'_2
+ (\oint_{e_5'^\perp} \tilde\omega) e'_5
+ (\oint_{e_3'^\perp} \tilde\omega) e'_3
+ (\oint_{e_4'^\perp} \tilde\omega) e'_4 
\end{multline}
which is thus invariant under a flop transition.

This shows that $\hat C(\omega)$ is independent of a choice of fundamental domain.

\eproof

\section{Proof of Corollary \ref{cor:LagrangeBPi}}
\label{app:LagrangeBPi}

\proof
If $\gamma_1$ and $\gamma_2$ both belong to $\Ker\hat B$, \eqref{eq:Binter} gives $\gamma_1\cap \gamma_2=0$.

Now let $\omega_1$ and $\omega_2$ be meromorphic forms,
and let
\beq
\tilde \omega_1 = \omega_1 - \sum_{p=\text{poles of }\omega_1} \sum_{k=0}^{\deg_p-1 \omega_1} t_{p,k} \hat B(\bcycle_{p,k})
\qquad , \quad
t_{p,k} = \frac{1}{2\pi\ii} \oint_{\acycle_{p,k}} \omega_1
\eeq
and similarly
\beq
\tilde \omega_2 = \omega_2 - \sum_{p=\text{poles of }\omega_2} \sum_{k=0}^{\deg_p-1 \omega_2} t'_{p,k} \hat B(\bcycle_{p,k})
\qquad , \quad
t'_{p,k} = \frac{1}{2\pi\ii} \oint_{\acycle_{p,k}} \omega_1.
\eeq
$\tilde \omega_1$ and $\tilde \omega_2$ have no pole on $\curve$.
By definition we have $\Pi(\bcycle_{p,k})=\bcycle_{p,k}$ and thus 
\beq
\hat C(\omega_{1}) = \hat C(\tilde \omega_{1}) + \sum_{p=\text{poles of }\omega_1} \sum_{k=0}^{\deg_p-1 \omega_1} t_{p,k} \bcycle_{p,k}
\eeq
\beq
\hat C(\omega_{2}) = \hat C(\tilde \omega_{2}) + \sum_{p=\text{poles of }\omega_2} \sum_{k=0}^{\deg_p-1 \omega_2} t'_{p,k} \bcycle_{p,k}.
\eeq
Since the cycles $\bcycle_{p,k}\cap \bcycle_{p',k'}$ don't intersect, and don't intersect the 1st kind part, we have
\beq
2\pi\ii \hat C(\omega_1) \cap \hat C(\omega_2) = 2\pi\ii \hat C(\tilde \omega_1) \cap \hat C(\tilde \omega_2) 
= \int_{\hat C(\tilde \omega_1)} \tilde \omega_2 - \int_{\hat C(\tilde \omega_2)} \tilde \omega_1.
\eeq
We use the same method as in appendix \ref{app:defChat}: the Riemann bilinear identity.
choose a fundamental domain, and define in it some functions $f_1$, $f_2$ such that $df_1=\tilde \omega_1$ and $df_2=\tilde \omega_2$.
We obtain
\beq
0=\sum_{p=\text{poles of}\,\tilde\omega_2} \Res_{p} f_1 \ \tilde \omega_2 
= \int_{\hat C(\tilde\omega_1)} \tilde\omega_2
\eeq
and
\beq
0=\sum_{q=\text{poles of}\,\tilde\omega_1} \Res_{q} f_2 \ \tilde\omega_1 
= \int_{\hat C(\tilde\omega_2)} \tilde\omega_1
\eeq
Therefore
\beq
\hat C(\omega_1)\cap\hat C(\omega_2)=0.
\eeq

\eproof

\section{Proof of lemma \ref{Lem:Qpositive}}
\label{app:Qpositive}

\bl
If $\Pi(\gamma)\neq 0$:
\beq
2\pi\ii  \ Q(\Pi(\gamma) ,\overline{\Pi (\gamma)}) >0
\eeq
\el

\proof

Let $\omega\neq 0$ be a meromorphic 1-form, and $\gamma=\hat C(\omega)=\Pi(\gamma)$.
Let
\beq
\tilde \omega = \omega - \sum_{p=\text{poles of }\omega} \sum_{k=0}^{\deg_p-1 \omega_1} t_{p,k} \hat B(\bcycle_{p,k})
\qquad , \quad
t_{p,k} = \frac{1}{2\pi\ii} \oint_{\acycle_{p,k}} \omega,
\eeq
which is a holomorphic 1-form on $\curve$.
Let $\curve_0$ be a fundamental domain of $\curve$ and $f$ such that $df=\tilde\omega$.
We have
\bea
0 & < & \int_\curve |\tilde{\omega}|^2 \cr
 &  &= \int_{\partial\curve_0}  \tilde\omega \ \bar f \cr
 &  &= \int_{\overline{\hat C(\tilde\omega)}}  \tilde\omega  \cr
 &  &= 2\pi\ii \mathcal Q(\overline{\hat C(\tilde\omega)},{\hat C(\tilde\omega)})  \cr
\eea
Then, using that $\overline{\bcycle_{p,k}}=\bcycle_{p,k}$ and $\bcycle_{p,k}\cap \bcycle_{p',k'}=0$ and the $\bcycle_{p,k}$--cycles don't intersect the 1st kind part, we derive the result.

\eproof

%
%
%
%
%
%

\section{Proof of lemma \ref{lemma:deltaX}}
\label{app:DeltaX}

\proof
Since $\acycle''_i\in\Ker \hat B$ we have
\bea
0 &=& \int_{\Delta_z\acycle''_i} B + \int_{\acycle''_i} \Delta_z B   \cr
 &=& \int_{\Delta_z\acycle''_i} B + \int_{\acycle''_i} \omega_{0,3}(z,.,.)   \cr
 &=& \int_{\Delta_z\acycle''_i} B 
\eea
so that $\Delta_z \acycle''_i=- \sum_{j} (\Delta_z X_{i,j}) \ \bcycle_j$ should be a linear combination of $\acycle''_j$:
\beq
- \sum_{j} (\Delta_z X_{i,j}) \ \bcycle_j = \sum_j C_{i,j} \acycle_j.
\eeq
Taking the intersection with $\bcycle_i$ we get $0=C_{i,j}$, and thus 
\beq
\Delta_z X=0.
\eeq
This implies that $\Delta_z \acycle''_i=0$, and since $\Delta_z \y = B$ and since $\acycle''_i\in \Ker\hat B$ we have
\beq
\Delta_z \left( \oint_{\acycle''_i}\y \right) = 0.
\eeq

Similarly, since $\acycle''_i\in\Ker \hat B$ we have
\bea
- \oint_{\partial_{\gamma_1\boxtimes \gamma_2}\acycle''_i} B &=&   \frac12 \left( \oint_{\acycle''_i}\oint_{\gamma_1} B \ \hat B(\gamma_2) + \oint_{\acycle''_i}\oint_{\gamma_2} B \ \hat B(\gamma_1)\right)  \cr
\sum_{j} \partial_{\gamma_1\boxtimes \gamma_2} X_{i,j} \oint_{\bcycle_j} B &=&   \frac12 \left( \oint_{\acycle''_i}\oint_{\gamma_1} B \ \hat B(\gamma_2) + \oint_{\acycle''_i}\oint_{\gamma_2} B \ \hat B(\gamma_1)\right)  .
\eea
Taking the intersection with $\acycle''_j$ gives
\bea
\partial_{\gamma_1\boxtimes \gamma_2} X_{i,j}  &=&   \frac12 \left( \oint_{\acycle''_i}\oint_{\gamma_1} B \ \acycle''_j\cap \hat B(\gamma_2) + \oint_{\acycle''_i}\oint_{\gamma_2} B \ \acycle''_j\cap \hat B(\gamma_1)\right)  .
\eea

\eproof

\end{document}